\documentclass[10pt,journal,compsoc]{IEEEtran}

\ifCLASSOPTIONcompsoc
  \usepackage[nocompress]{cite}
\else
  \usepackage{cite}
\fi

\ifCLASSINFOpdf
\else
\fi

\usepackage{cite}
\usepackage{amsmath,amssymb,amsfonts}
\usepackage{graphicx}
\usepackage{hyperref}
\usepackage{textcomp}
\usepackage[table,xcdraw]{xcolor}
\def\BibTeX{{\rm B\kern-.05em{\sc i\kern-.025em b}\kern-.08em
    T\kern-.1667em\lower.7ex\hbox{E}\kern-.125emX}}

\usepackage[most]{tcolorbox} %
\usepackage{xspace} %
\usepackage[normalem]{ulem}
\usepackage{booktabs} %
\usepackage{multirow} %
\usepackage{multicol} %
\usepackage{subcaption} 
\usepackage[noend]{algpseudocode}
\usepackage{enumitem}
\usepackage{adjustbox}
\usepackage{wrapfig}
\usepackage[ampersand]{easylist}
\usepackage[most]{tcolorbox}
\newtheorem{definition}{Definition}
\newtheorem{exmp}{\underline{Example}}

\newboolean{showcomments}

\setboolean{showcomments}{true}

\ifthenelse{\boolean{showcomments}}
  {\newcommand{\nb}[2]{
    \fbox{\bfseries\sffamily\scriptsize#1}
    {\sf\small$\blacktriangleright$\textit{#2}$\blacktriangleleft$}
   }
   
  }
  {\newcommand{\nb}[2]{}
   
  }

\definecolor{MidnightBlue}{HTML}{006895}

\newcommand*\circled[1]{\tikz[baseline=(char.base)]{
            \node[shape=circle,draw,inner sep=0.5pt] (char) {#1};}}

\newcommand{\ie}{\textit{i.e.,}\xspace}
\newcommand{\eg}{\textit{e.g.,}\xspace}
\newcommand{\etc}{\textit{etc.}\xspace}
\newcommand{\etal}{et al.\xspace}

\newcommand{\docode}{\textit{do$_{code}$}\xspace}

\newcommand{\nlms}{NCMs\xspace}
\newcommand{\nlm}{NCM\xspace}

\newcommand{\scm}{SCM\xspace}

\newcommand{\datainterI}{\textit{ProgramRepair}\xspace}
\newcommand{\datainterII}{\textit{SyntacticDifferences}\xspace}
\newcommand{\datainterIII}{\textit{UnCommenting}\xspace}
\newcommand{\datainterIV}{\textit{ASTNodeTypes}\xspace}

\newcommand{\modelinterI}{\textit{NumberLayers}\xspace}
\newcommand{\modelinterII}{\textit{NumberUnits}\xspace}

\newcommand{\ntp}{NTP\xspace}

\newcommand{\assoJS}{JS Dist.\xspace}
\newcommand{\assoPR}{Pearson\xspace}

\newcommand{\training}{\textit{CodeSearchNet}\xspace}
\newcommand{\BuggyTB}{\textit{BuggyTB}\xspace}
\newcommand{\CommentsTB}{\textit{CommentsTB}\xspace}
\newcommand{\BigCloneIITB}{\textit{BigClone2TB}\xspace}
\newcommand{\BigCloneIIITB}{\textit{BigClone3TB}\xspace}
\newcommand{\BigCloneTB}{\textit{BigCloneTB}\xspace}
\newcommand{\Galeras}{\textit{Galeras}\xspace}

\newcommand{\blocks}{\texttt{\small[blocks]}\xspace}
\newcommand{\tests}{\texttt{\small[tests]}\xspace}
\newcommand{\oop}{\texttt{\small[oop]}\xspace}

\newcommand{\exceptions}{\texttt{\small[exceptions]}\xspace}
\newcommand{\datatype}{\texttt{\small[datatype]}\xspace}

\newcommand{\operators}{\texttt{\small[operators]}\xspace}
\newcommand{\conditionals}{\texttt{\small[conditionals]}\xspace}
\newcommand{\extra}{\texttt{\small[extraTokens]}\xspace}

\newcommand{\jaccard}{\texttt{\small<jaccard>}\xspace}
\newcommand{\levenshtein}{\texttt{\small<levenshtein>}\xspace}
\newcommand{\soren}{\texttt{\small<sorensen-dice>}\xspace}

\newcommand{\node}{\texttt{\small[node]}\xspace}
\newcommand{\forinclause}{\texttt{\small[for\_in\_clause]}\xspace}

\newcommand{\rnn}{RNN$_{1,1024}$\xspace}
\newcommand{\gru}{GRU$_{1,1024}$\xspace}
\newcommand{\grui}{GRU$_{2,1024}$\xspace}
\newcommand{\gruii}{GRU$_{3,1024}$\xspace}
\newcommand{\gruiii}{GRU$_{1,512}$\xspace}
\newcommand{\gruiv}{GRU$_{1,2048}$\xspace}

\newcommand{\tf}{GPT-2$_{6,12}$\xspace}
\newcommand{\tfi}{GPT-2$_{12,12}$\xspace}
\newcommand{\tfii}{GPT-2$_{24,12}$\xspace}

\newcommand{\bert}{BERT$_{12,12}$\xspace}

\newcommand{\secref}[1]{Sec.~\ref{#1}\xspace}
\newcommand{\figref}[1]{Fig.~\ref{#1}\xspace}
\newcommand{\tabref}[1]{Table~\ref{#1}\xspace}

\newtcolorbox{boxK}{
    fontupper = \small,
    sharpish corners, %
    boxrule = 0pt,
    toprule = 4.5pt, %
    enhanced,
    fuzzy shadow = {0pt}{-2pt}{-0.5pt}{0.5pt}{black!35} %
}

\begin{document}
\bstctlcite{IEEEexample:BSTcontrol} %
\title{Toward a Theory of Causation for Interpreting Neural Code Models}

\author{David N.~Palacio,~\IEEEmembership{Student Member,~IEEE,}
        Alejandro Velasco,~\IEEEmembership{Graduate Student Member,~IEEE,}
        Nathan Cooper,~\IEEEmembership{Member,~IEEE,}
        Alvaro Rodriguez,~\IEEEmembership{}
        Kevin Moran,~\IEEEmembership{Member,~IEEE,}
        Denys Poshyvanyk,~\IEEEmembership{Fellow,~IEEE}%
\IEEEcompsocitemizethanks{\IEEEcompsocthanksitem David Nader Palacio, Alejandro Velasco, Nathan Cooper, and Denys Poshyvanyk are with the Department
of Computer Science, William \& Mary, Williamsburg,
VA, 23185.\protect\\
E-mail: danaderpalacio,svelascodimate,nacooper,dposhyvanyk\{@wm.edu\}
\IEEEcompsocthanksitem Kevin Moran is with the Department
of Computer Science, George Mason University, Fairfax,
VA, 22030.\protect\\
E-mail: kpmoran@gmu.edu
\IEEEcompsocthanksitem Alvaro Rodriguez is with the Department of Computer Science, Universidad Nacional de Colombia, Colombia, Bogota.
\protect\\
E-mail: aldrodriguezca@unal.edu.co}%
\thanks{Manuscript received January 16th, 2023;}}

\markboth{Accepted to Appear in IEEE Journal of Transactions on Software Engineering}%
{Palacio \MakeLowercase{\textit{et al.}}: Bare Demo of IEEEtran.cls for Computer Society Journals}

\IEEEtitleabstractindextext{%

\begin{abstract}
Neural Language Models of Code, or Neural Code Models (NCMs),  are rapidly progressing from research prototypes to commercial developer tools. 
As such, understanding the capabilities and limitations of such models is becoming critical. However, the abilities of these models are typically measured using automated metrics that often only reveal a portion of their real-world performance. While, in general, the performance of \nlms appears promising, currently much is unknown about how such models arrive at decisions. To this end, this paper introduces \docode, a post hoc interpretability method specific to \nlms that is capable of explaining model predictions. \docode is based upon causal inference to enable programming language-oriented explanations. While the theoretical underpinnings of \docode are extensible to exploring different model properties, we provide a concrete instantiation that aims to mitigate the impact of \textit{spurious correlations} by grounding explanations of model behavior in properties of programming languages. To demonstrate the practical benefit of \docode, we illustrate the insights that our framework can provide by performing a case study on two popular deep learning architectures and ten \nlms. The results of this case study illustrate that our studied \nlms are sensitive to changes in code syntax. All our \nlms, except for the BERT-like model, statistically learn to predict tokens related to blocks of code (\eg brackets, parenthesis, semicolon) with less confounding bias as compared to other programming language constructs. These insights demonstrate the potential of \docode as a useful method to detect and facilitate the elimination of confounding bias in \nlms.

\end{abstract}

}

\maketitle

\IEEEdisplaynontitleabstractindextext

\IEEEpeerreviewmaketitle

\IEEEraisesectionheading{\section{Introduction}\label{sec:introduction} }

\IEEEPARstart{T}{he} combination of large amounts of freely available code-related data, which can be mined from open source repositories, and ever-more sophisticated deep learning architectures for code, which we refer to as Neural Code Models (\nlms), have fueled the development of software engineering (SE) tools with increasing effectiveness. \nlms have (seemingly) illustrated promising performance across a range of different SE tasks~\cite{Watson:ICSE20,White:MSR15,ciniselli2021empirical,Mastropaolo2021StudyingTasks, Tufano.MSR.2018, SANER.2019, Tufano:icse2021}. In particular, \textit{code generation} has been an important area of SE research for decades, enabling tools for downstream tasks such as code completion~\cite{MSR-Completion}, program repair~\cite{Chen2019sequencer}, and test case generation~\cite{Watson:ICSE20}. In addition, industry interest in leveraging Large Language Models (LLMs), a scalable version of \nlms, has also grown as evidenced by tools such as Microsoft's IntelliCode \cite{intellicode}, Tabnine \cite{tabnine}, OpenAI's Codex \cite{openai_codex}, and GitHub's Copilot \cite{github_copilot}. Given the prior popularity of code completion engines within IDEs~\cite{murphy2006ide} and investment in commercial tools, \nlms for code generation will almost certainly be used to help build production software systems in the near future if they are not being used already. 

Recently, there has been increased interest in evaluating \nlms for code generation. A recent study from Chen \etal \cite{chen2021evaluating} illustrates that certain issues, such as alignment failures and biases, do exist for large-scale \nlms. Most of the conclusions from Chen~\etal's study were uncovered through manual analysis, \eg through sourcing counterexamples, making it difficult to rigorously quantify or to systematically apply such an analysis to research prototypes~\cite{wu2019errudite}. Given the rising profile and role that \nlms for code generation play in SE and the current limitations of adopted evaluation techniques, it is clear that \textit{new methods are needed to provide deeper insight into \nlms' prediction performance}.

Much of the quality assessment work on \nlms has primarily concentrated on accuracy-based metrics (\eg Accuracy, BLEU, METEOR, ROUGE) as opposed to multi-metric evaluations (\eg robustness, fairness, bias, efficiency). Moreover, skepticism within the NLP research community is growing regarding the efficacy of current accuracy-based metrics, as these metrics tend to overestimate model performance  ~\cite{ribeiro2020checklist, rei2020comet, kocmi2021ship}. Even benchmarks that span multiple tasks and metrics have been shown to lack robustness, leading to incorrect assumptions of model comparisons~\cite{dehghani2021benchmark}. Notable work has called for a more systematic approach that aims to understand a given model's behavior according to its linguistic capabilities~\cite{ribeiro2020checklist}, while others have suggested the need for holistic evaluations of Language Models~\cite{liang_holistic_2022}. 

In addition to limitations with current methods of model quality assessment, some of the most popular \nlms have been adapted from the field of Natural Language Processing (NLP), and thus may inherit the various limitations often associated with such models --- including biases, memorization, and issues with data inefficiency, to name a few~\cite{bender2021parrots}. However, perhaps the most problematic aspect of current neural language models is the fact that they are incapable of explaining their reasoning \cite{doshi-velez_towards_2017,Doshi-Velez2018ConsiderationsLearning,molnar_interpretable_2020,lipton_mythos_2017}. \nlms and, in general, \textit{deep learning architectures} seemingly trade effectiveness for transparency, as the same complexity that allows for impressive learning and generalization leads models to operate in a \textit{black-box} fashion. That is, we are uncertain how neural models --- including \nlms --- \textit{arrive at decisions}; a phenomenon described as \textit{incompleteness in problem formalization} \cite{Doshi-Velez2017TowardsLearning}. Such incompleteness manifests as an inability to explain models' predictions in human-understandable terms. If neural models fail at justifying their outputs, \textit{Can we trust these models? Will these models work in deployment? How brittle are neural models in practical software engineering settings?} Researchers and practitioners require neural models to be robust not only at making predictions but also in the \textbf{interpretability} of those predictions \cite{lipton_mythos_2017}. Despite the increasing popularity and apparent effectiveness of code generation tools based on \nlms, there is still much that is unknown about their behavior.

\textbf{A Motivating Example for  Interpretability.} Consider the scenario in which we \textit{observe} that a given \nlm is underperforming when predicting code from code inputs that are buggy. That is, we observe a negative correlation between \textit{buggy code} in the prompt and \textit{accuracy} after an initial exploratory analysis on this \nlm. However, a simple observation of a correlation between \textit{buggy code} and \textit{accuracy} is insufficient evidence to explain the \nlms prediction performance in this scenario, as different \textit{properties} of the buggy input code could be the \textit{cause} of the observation. To assert that \textit{input code being buggy} is an explanation or \textit{interpretation} for having a worsened accuracy, we must first establish a \textbf{causal relationship} between these two random variables. 

\begin{wrapfigure}{r}{0.15\textwidth}
  \begin{center}
    \includegraphics[width=0.15\textwidth]{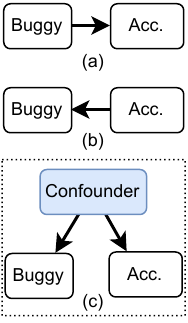}
  \end{center}
  \caption{Common Cause Principle in \nlms \cite{Scholkopf2022}}
   \label{fig:example}
\end{wrapfigure}

Statistical dependencies are induced by an underlying causal process (\ie Reichenbach's common cause principle \cite{Scholkopf2022}). In consequence, the correlation between \textit{buggy code} and \textit{accuracy} has three possible causal explanations\footnote{Pearl poses an additional causal explanation based on conditional independence in colliders used for the ``back-door criterion'' \cite{Pearl2016Causality}.}  depicted as causal graphs in \figref{fig:example}: (a) \textit{accuracy} is caused by \textit{buggy code}, (b) \textit{buggy code} is caused by \textit{accuracy}, and (c) the reason for a correlation is a third variable. This variable, known as a confounder, causally influences \textit{buggy code} and \textit{accuracy}. Based on our \textit{domain knowledge}, we may posit that option (c) is most likely to represent our causal assumptions, as many factors such as the \textit{number of tokens or subwords} can affect both variables of interest. 

If a causal connection indeed exists between \textit{buggy code} and \textit{accuracy}, we can claim that the setting \textit{buggy code} is an \textit{interpretation} for code predictions. However, in the above example, we would essentially be guessing, and as such the question remains: \textit{how can we quantify the causal effect of buggy code on \nlm's accuracy after controlling for the influence of hidden confounders?} Unfortunately, when attempting to understand the prediction performance of \nlms, no causal interpretability formalism is available to articulate or even answer the previous questions. 

In this paper, we cast this problem of achieving a more complete understanding of Neural Code Models as a \textit{Causal Interpretability} task and posit that we can leverage the \textit{theory of causation} as a mechanism to explain \nlms prediction performance. This mechanism includes a formalism to articulate and answer causal queries.  We hypothesize that this mechanism can serve as a useful verification tool to fulfill desiderata that we might want of \nlms. These desiderata, which are originally suggested in NLP literature, comprise notions such as facilitating debugging, detecting biases, providing recourse, and eventually, increasing the reliability and trust of \nlms \cite{molnar2019interpret,Doshi-Velez2017TowardsLearning,lo_trustworthy_2023}. As such, this paper introduces \docode, a novel global post hoc interpretability method specifically designed for understanding the effectiveness of \nlms using causal queries. Furthermore, \docode intends to establish a robust and adaptable methodology for \textit{interpreting predictions} of \nlms in contrast to simply \textit{measuring the accuracy} of these same \nlms.

\docode asks causal questions as to why prediction performance is affected by software engineering-based \textit{interventions}. These interventions are generally data or model properties (\eg Bugginess of Code, Number of Inline Comments, Number of Model Layers) that a domain expert might think are affecting a given \nlm. More specifically, \docode consists of four major conceptual steps, (1) \textbf{modeling} a \textit{structural causal graph}, (2) \textbf{identifying} causal estimand, (3) \textbf{estimating} causal effects, and (4) {\textbf{validating} the causal process by refuting the obtained effect estimate and vetting the graph creation}. Through the introduction of this interpretability method, we aim to help SE researchers and practitioners by allowing them to understand the potential limitations of a given model, work towards improving models and datasets based on these limitations, and ultimately make more informed decisions about how to build automated developer tools given a more holistic understanding of \textit{\textbf{what}} \nlms are predicting and \textit{\textbf{why}} the predictions are being made.

To showcase the types of insights that \docode can uncover, we perform a comprehensive study on seven practical \textit{interpretability scenarios} (see \secref{sec:case_studies}) on different variations of popular deep learning architectures for the task of code generation, namely RNNs~\cite{RNNs} and Transformers \cite{vaswani2017transformers} trained on the CodeSearchNet dataset~\cite{husain2019codesearchnet}. We instantiated our study (see \figref{fig:case_studies}) using configuration criteria such as type of architecture (\eg RNN, GPT, BERT), evaluation dataset (\eg codexglue, codesearchnet, galeras), intervention modality (\eg binary, linear), hyperparameter interventions (\eg layers, units), data interventions (\eg bug fixing, inline comments, semantic preserving), potential outcomes (\eg cross-entropy, next token predictions, Jaccard), causal inference metric (\eg Pearson, Jensen Shannon, Average Treatment Effect (ATE)), refutation testing (\eg placebo, unobserved common cause, random common cause), and syntactic clustering. This syntactic clustering comprises ten keyword-based code categories derived from Java and eleven grammar-based categories derived from Python (see \figref{fig:syntactic_aggregations}). 

Our case study design resulted in several notable findings illustrating the usefulness of our interpretability technique. Our most relevant findings are as follows:
\begin{itemize}
    \item The presence or absence of buggy code does not appear to causally influence (or explain) the prediction performance of our \nlms even under measured high correlation;
    \item The presence of Type II Clones in training data impacts (or causally explains) the effectiveness of \nlms in terms of cross-entropy. Our observed correlations and measured causal effects suggest an influence of interventions related to the insertion or removal of Type II clones in training data for different programmatic constructs, including: identifiers, literals, types, white spaces, layout, and comments;
    \item We observed strong correlations between the number of layers and model performance in terms of Next Token Prediction (NTP), which might suggest a causal interpretation of code predictions. However, \docode revealed that the reported Average Treatment Effects were close to zero for such phenomena -- demonstrating the presence of confounding bias suggesting multiple effects are at play; 
    \item  Finally, the intervention of masking random tokens has more impact on code predictions than masking grammar-based categories. This suggests that our BERT-like models do not entirely capture the structural information of programming languages.
\end{itemize}

\textbf{Our contributions.} Previous findings suggest the need for integrating causal interpretability in post-hoc analysis of neural language models.  We hope to provide researchers with an approach for integrating causal analysis into their research. To that end, in addition to our case study showing the power such an analysis can provide, we have additionally created a checklist that summarizes the steps required to apply Causal Interpretability to Neural Code Models (see \figref{fig:checklist}). In summary, this paper makes the following contributions: 

\textbf{1.} A formal four-step pipeline for interpreting \nlms that enables causal analyses of different settings for language model use;
\textbf{2.} A comprehensive study on seven practical scenarios rooted in software engineering practice in which \docode is applied to popular deep learning models for code generation;
\textbf{3.} A set of findings for these models that help to explain their behavior in different settings challenges some current notions of code generation models;
\textbf{4.} A checklist to help researchers apply Causal Interpretability to Neural Code Models;
\textbf{5.} And a replication package~\cite{icodegen} to both encourage the use of our technique in future work on developing and evaluating code generation models and for replicating our experiments.

\section{Background \& Related Work}
\label{sec:background}

\begin{figure}[h]
		\centering
		\includegraphics[width=0.45\textwidth]{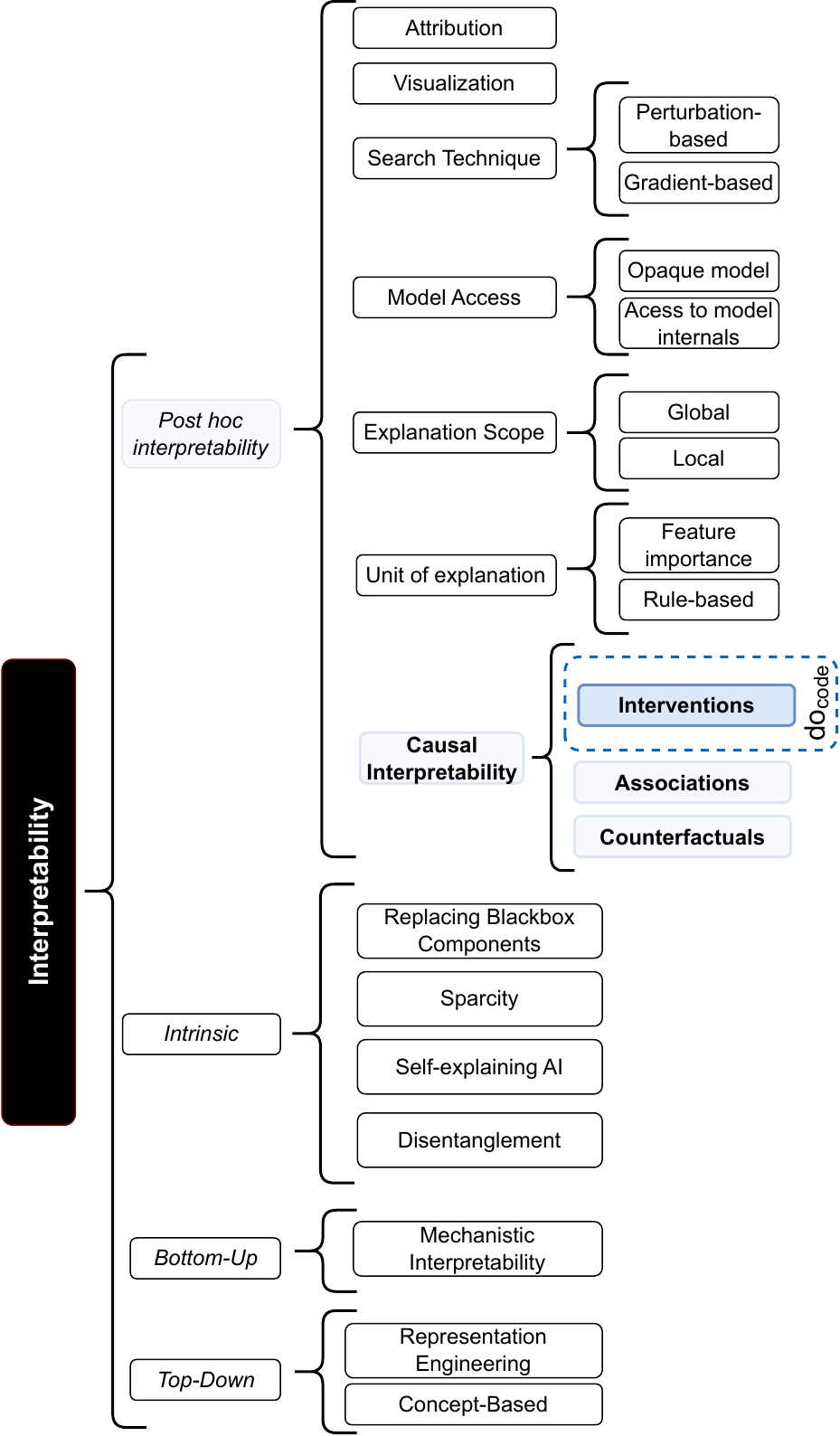}
		\caption{A classification of key methods in interpretability including \textit{Causal Interpretability}.}
        \label{fig:methods}
\end{figure}

\textit{Interpretable Machine Learning} is a research field aimed at understanding how opaque models give rise to predictions. This process is centered around human understanding. Although the research community has not achieved a consensus on a precise definition of interpretability, researchers usually refer to this field either as ``Interpretable Machine Learning'' or ``Explainable AI'' \cite{molnar_interpretable_2020,Doshi-Velez2018ConsiderationsLearning}. The main goal of the field is to create methods that explain models' reasoning and then verify whether such reasoning is sound \cite{Doshi-Velez2017TowardsLearning}. These interpretability methods can be classified into different criteria depending on the researcher's concentration. The most common criteria are post hoc vs. intrinsic, model-specific vs. model-agnostic, global vs. local explanation scope, feature importance vs. rule-based unit of explanation, and causal levels \cite{molnar_interpretable_2020,han_which_2022}. We depict in \figref{fig:methods} a summary of the most relevant methods published thus far. \figref{fig:methods} also positions our method \docode within the scope of \textit{Causal Interpretability}, which is based on the concept of Ladder of Causation by Pearl introduced in \secref{sec:motivation} \cite{Pearl2018Causality}.   

\begin{boxK}
\textit{Causal Interpretability} is a post hoc global approach by which Neural Code Models (\nlms) are interpreted or explained from a causal assumption encoded in a Structural Causal Model (SCM). By using the formalism of Pearl's Ladder of Causation, researchers can \textit{estimate} a quantifiable \textit{causal effect} of proposed interventions.
\end{boxK}

\textbf{Applications of \nlms in SE:} \nlms in SE have a rich history, stemming from the seminal work by Hindle \etal who proposed the concept of the \textit{naturalness} of software using \textit{n}-gram models~\cite{hindle2012natural}. %
Then, with the rise of Deep Learning, researchers began exploring the possibility of adapting NLMs to code, as initially demonstrated by White \etal~\cite{White:MSR15}. Since then, \nlms have been used for a number of SE tasks such as code completion \cite{Nguyen2013ASS, Raychev2014CodeCW, tu2014local, Hellendoorn2017AreCode, Karampatsis2019,  Karampatsis2020Open-VocabularyAbstract, ciniselli2021empirical, chen2021evaluating} and translation \cite{Chen2019sequencer, Roziere2020transcoder, Hu2018comment, Mastropaolo2021StudyingTasks, Tufano:2018:EIL:3238147.3240732, Tufano:tosem2019, Tufano:icse2019}. 
Researchers have also investigated various representations of \nlms for code~\cite{White2016clones} as well as graph-based representations based on ASTs~\cite{allamanis2018learning} and program dependency graphs~\cite{Nguyen:ICSE15}. It has also been shown that using a \nlm for code representation and then fine-tuning for specific SE tasks can achieve \textit{state-of-the-art} performance across tasks \cite{Hussain2020DeepTL, feng2020codebert, guo2021graphcodebert}.

\noindent\textbf{Interpretability of \nlms:} A growing body of research has investigated the potential of evaluating and understanding LMs using various techniques, which can be classified into two basic groups: (i) \textit{intrinsic techniques} such as probing for specific linguistic characteristics \cite{tenney2019bert} and examining neuron activations~\cite{dai2021knowledge}; and (ii) \textit{post hoc techniques} such as benchmarks \cite{wang2018glue, rajpurkar2018squad, husain2019codesearchnet, lu2021codexglue, chen2021evaluating}, attention-based analysis \cite{liu_reliability_2023,mohammadkhani_explainable_2023}, and intervention analyses \cite{khandelwal2018sharp, prabhakaran2019perturbation, ribeiro2020checklist, rabin2021generalizability}. Karpathy \etal were among the first to interpret NLMs for code using general Next Token Predictions~\cite{karpathy2015understand} as an interpretability method as well as investigating individual neurons. Our work extends Karpathy \etal's interpretability analysis using causal inference and a more statistically rigorous methodology that grounds its evaluation in SE-specific features and settings. 

Complementary to our work, the work by Polyjuice paper \cite{wu-etal-2021-polyjuice} aims to automatically generate perturbations to language, to see how a given model performs in different scenarios. Our work is differentiated by the fact that we both define a mapping of source code tokens to categories and define a methodology for linking causal relationships of model changes to these various categories. As such, our work functions alongside techniques such as Polyjuice, providing deeper insights into why model performance varies across different perturbations. Our work is also complementary to the work by Cito \etal, \cite{cito2022}  on Counterfactual explanations for models of source code. This work takes a significant step toward defining perturbations of code for counterfactual explanations of generative models. Our technique in conjunction with Cito et al.'s approach can provide deeper explanations as to \textit{why} a given \nlm's performance changed for a given perturbation.

Our method is different from the work on adversarial robustness via one key point, similar to the work of Cito \etal \textit{Our different testbeds are not intentionally designed to fool a model.} On the contrary, our different testbeds represent completely natural distributions of code tokens collected at scale. As such, our research has a completely different aim as compared to the work on adversarial robustness. Other related works from Explainable Artificial Intelligence (XAI) have focused on the usage of causal theory applied to generate post hoc explanations about the logical structure of a neural network \cite{pmlr-v162-hu22b} or explore the causal relationships between the explanation and the prediction \cite{karimi_relationship_2023}. Both causal XAI methods differ from our work since we employ the notion of Pearl's Ladder of Causation directly on the inputs and outputs of \nlms.

\section{Why do we Need Causal Interpretability for Deep Learning Models applied to Software Engineering?}\label{sec:motivation}

Our research leverages Pearl's theory of Causal Inference and grounds it in interpreting Neural Code Models (\nlms). Given the wealth of metrics and benchmarks that exist for \nlms it is natural to ask \textit{why should we study causation in Deep Learning for Software Engineering?} Causation has two main goals in science (i) discovering causal variables and (ii) assessing counterfactual interventions \cite{Pearl2018Causality}. The field of \textit{Deep Learning for Software Engineering (DL4SE)} can take advantage of the latter when dealing with uncertainty and \textit{confounding bias} of \nlms. Estimating counterfactual interventions is a powerful tool to generate explanations of the model's performance. Our method, \docode, can be applied to a wide range of SE models for detecting and eliminating confounding bias. We do not intend to pose \docode as a tool to enhance prediction performance but as an adaptable approach to analyze why \nlms obtain their predictions. However, quantifying the effects of interventions requires establishing a causal structure underlying the target data. 

Randomized controlled experiments were the general choice to explore causality before Pearl's definitions of graphical models. Nonetheless, it is not practical to force developers to perform interventions (such as removing comments from a training corpus) or even train hundreds of \nlms to test various treatments. Instead the $do-operator$ and causal graphs (\ie Structural Causal Models) are better tools for performing causal estimations from observational data. Reconstructing such graphical representations is challenging since it not only requires formalizing causation in the field of SE (\ie defining potential outcomes, common causes, and treatments) but also tracing and connecting software data to causal models (see the pipeline in \secref{sec:overview}). In addition, formulating interventions is not an easy process. We must hypothesize feasible transformations or interventions that can occur in code to simulate real-world settings for \nlms, a concept that we synthesize as \textbf{The Causal Interpretability Hypothesis} (see \secref{sec:hypo}). To that end, we have proposed a pipeline to help aid in adapting the process of causal inference to the interpretation of Neural Code Models (\nlms). This pipeline has been inspired by Pearl's notion of the \textit{Ladder of Causation}, introduced below, and the \textit{doWhy} library \cite{dowhy}.

\subsection{Pearl's Ladder of Causation}

According to Pearl \& Mackenzie \cite{Pearl2018Causality}, Causal Inference (CI) seeks answers to questions of \textbf{association} (\textit{what is?}), counterfactual \textbf{interventions} (\textit{what if?}), and pure \textbf{counterfactuals} (\textit{why?}). The authors introduce the concept of \textit{Ladder of Causation} to match distinct levels of cognitive ability with concrete actions: seeing~(\textit{level one}), doing~(\textit{level two}), and imagining~(\textit{level three}). Our proposed analysis is primarily concerned with levels one \& two. Particularly, our method \docode is an extension of the intervention level.  

\begin{figure}[h]
		\centering
		\includegraphics[width=0.45\textwidth]{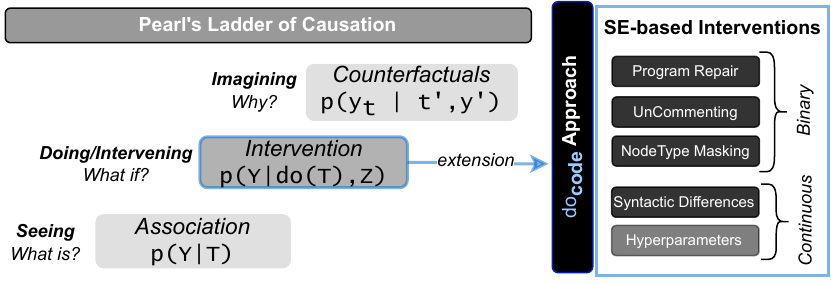}
		\caption{Ladder of Causation: \docode is an extension of the intervention level.}
        \label{fig:ladder}
        \vspace{-1em}
\end{figure}

\textbf{Causal Association.} In level one causation, $p(Y|T)$ is estimated by using typical correlation methods (\eg Pearson, Spearman, or Covariance) in addition to functional associations such as $y=g(t)$, which can be predicted with regressions or ML methods. For binary treatments similar to the ones we used in $T_{[data]}$ (\eg Buggy/Fixed and Commented/Uncommented), we opt to employ Pearson correlations $\rho_{YT}$ and Jensen-Shannon distance as association estimand for level one causation. 

\textbf{Causal Intervention.} If we want to go beyond \textit{``what is''} type questions, we must move past simple correlations and associations. This requires the $do-operator$ found in level two, causation $p(y|do(t))$. It is relevant to identify cases of \textit{spurious correlations} (\ie \textit{Confounding Bias}) or cases where $p(Y|T)\neq p(Y|do(T))$. Typically, \textit{association is not causation} due to the influence of a common cause or confounding variable $Z$. Such a variable is the one that is being controlled for or adjusted employing the concept of \textit{adjustment formula} in Eq.~\ref{eqn:do-all-lines}. Nonetheless, we can still compute the correlation $\rho_{YZ}$ to assess the variables affecting potential outcomes. 

\begin{figure}[t]
		\centering
		\includegraphics[width=0.4\textwidth]{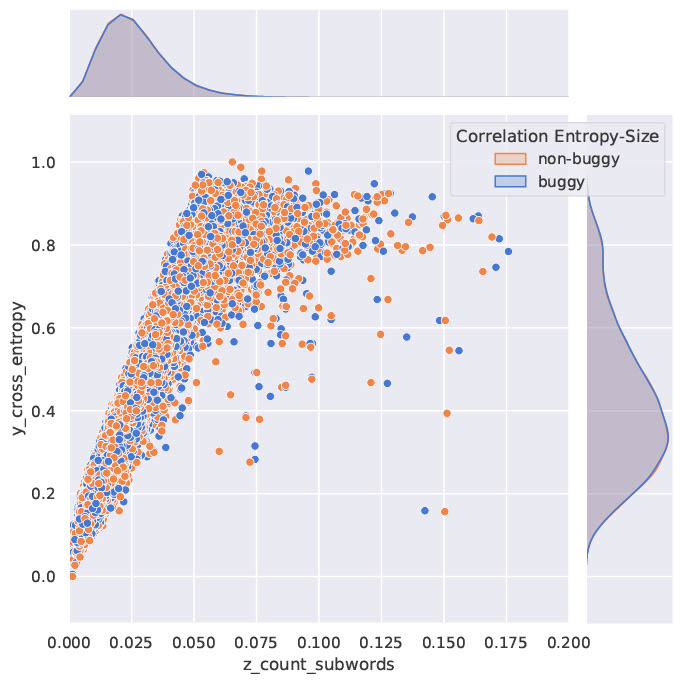}
		\caption{ \textit{Spurious Correlation} between the \textit{Number of Subwords} common cause and Cross-Entropy values ($p(Y|Z)\approx0.87$) for the \datainterI intervention generated from \tf. } 
        \vspace{-0.5cm}
        \label{fig:covariate}
\end{figure}

\textbf{An Example of Confounding Bias.} Consider an intervention that simulates the software engineering task of \datainterI where a treatment $T$ is predicting tokens following either Buggy/Fixed code, $Y$ is the cross-entropy of each method of a dataset of both buggy and fixed code (which we introduce in detail later in the paper), and $Z$ is the \textit{Number of Subwords} for each method. Using \docode we find there exists a spurious correlation after estimating the association distribution $p(Y|T)\approx0.67$ with \textit{Jensen-Shannon Distance} (see Def.~\ref{def:js}) and intervention distribution $p(Y|do(T))\approx-2E-4$ with \textit{Average Treatment Effect} (ATE) (see Def.~\ref{def:ate}). One possible explanation is that the common cause $Z$ (the number of subwords) confounds the relationship between the treatment and the outcome. Fig.~\ref{fig:covariate} depicts the influence of $Z$ on the potential outcome $Y$ for BuggyCode ($p(Y|Z,T=Buggy)\approx0.87$) and FixedCode ($p(Y|Z,T=Fixed)\approx0.86$). Blue and orange points in the plot are code snippets from the dataset. These points are equally distributed, which suggests that the \datainterI intervention has a negligible impact on model effectiveness as measured by cross-entropy.

\subsection{Software Engineering-Based Interventions}
In order to enable our analysis based upon Pearl's Ladder of Causation, we need to design \textit{interventions}, or changes in the input data distribution that represent a meaningful concept (\ie commented vs. uncommented code), to attribute cause from the changes in the model's performance in predicting code tokens in different settings. That is, if we observe that a model is less effective at predicting operators after changing ``test'' data sequences to which the model is applied, we may be able to confirm that the change \textit{caused} this drop in effectiveness if we properly control for confounders. We design \docode's interventions based on the fact that \nlms are often not applied to the same types of code corpora upon which they are trained. For instance, if a model trained on a well-commented dataset is applied to predict segments of poorly commented code, this could potentially impact performance. As such we define parallel code corpora which contain programming language-specific changes across the datasets, and specifically introduce four different initial interventions depicted on the right side of \figref{fig:ladder}. 

We define \textit{SE-based Interventions} to better understand model performance across different settings. We formulate these settings as parallel code corpora with differing specified semantic properties. For instance, a testbed aimed at simulating a debugging environment may consist of two parallel corpora: the buggy code, and the (corresponding) fixed code. Therefore, these datasets describe some high-level SE properties, which we employ in \docode's causal analysis. We define four types of SE application settings, adapted from both our prior work and community datasets: (i) buggy/non-buggy~\cite{Tufano2019LearningBug-Fixes}, (ii) commented/non-commented~\cite{lu2021codexglue}, (iii) type II, and (iv) type III clone pairs' differences~\cite{Svajlenko2015EvaluatingBigCloneBench}, and Abstract Syntax Trees node masking \cite{daniel23}. \docode is extensible, meaning that researchers can define their own code corpora interventions. In addition, we have also defined \textit{Model Hyper-parameter Interventions} to extend the causal analysis beyond data or code corpora perturbations (see \secref{sec:case_studies}).

\subsection{The Causal Interpretability Hypothesis}
\label{sec:hypo}

The following hypothesis aims to answer the question of why causal interpretability is necessary for the field of Deep Learning for Software Engineering (DL4SE). We claim that the prediction generated by any \nlm can be causally explained by employing the notion of the Ladder of Causation. Our proposed method \docode uses the process of causal inference to enable practitioners to answer their particular causal queries.

\begin{boxK}
\textbf{Hypothesis:} \docode is a causal interpretability method that aims to make DL4SE systems (\ie Neural Code Models) and their decision-making process understandable for researchers and practitioners. 
\end{boxK}

\section{An Overview of the \docode Approach}\label{sec:overview}

\begin{figure*}[ht]
		\centering
		\includegraphics[width=1\textwidth]{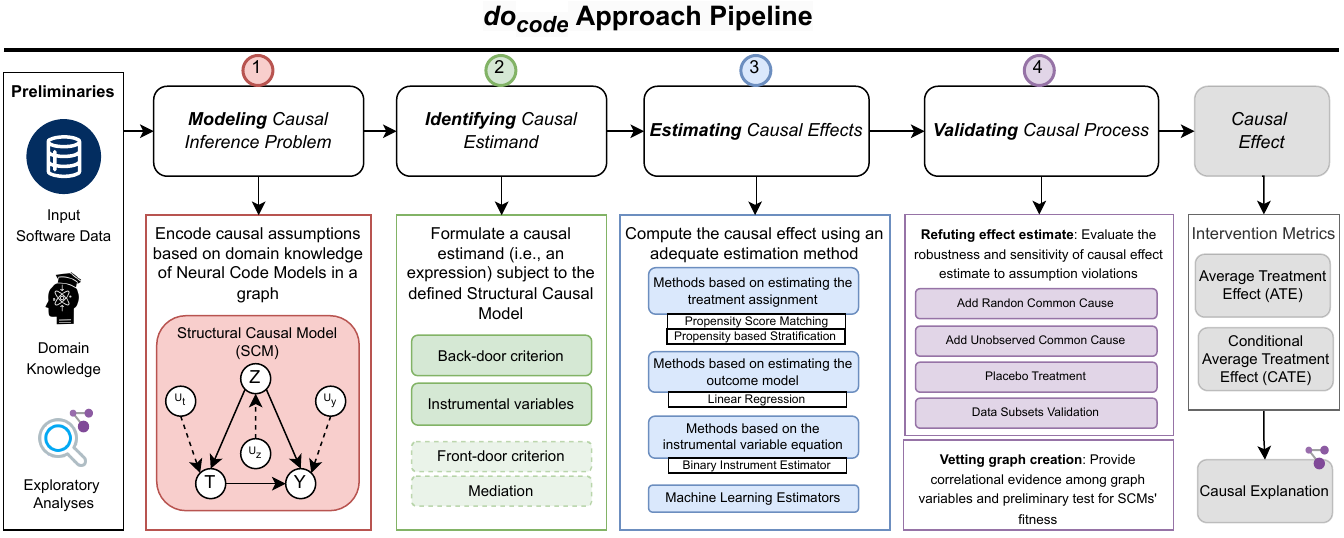}
		\caption{Overview of the \docode Approach: Each numeral represents a step in the process of generating causal interpretations. First, a Structural Causal Model (\scm) that frames the explanation hypothesis is formulated. Second, \docode executes graph surgery on the \scm to isolate a targeted estimand of the causal effect. Third, the causal effect is assessed based on the targeted estimand. Finally, the estimated effect and the \scm are scrutinized through refutation techniques and exploratory analysis to confirm their validity.}
        \label{fig:pipeline}
\end{figure*}

\docode comprises a set of statistical and causal inference methods to \textit{generate (and evaluate) post hoc interpretations} of \nlms. In the simplest of terms, our proposed interpretability approach asks \textit{Why does a \nlm make a given code prediction?} -- and provides a framework for answering this broad question by \circled{1} \textbf{modeling} the inference problem to encode causal assumptions, which rely on \textit{domain knowledge} of \nlms and \textit{observable software data}, in a graph representation; \circled{2} \textbf{identifying} the causal estimand based on the previous graph representation; \circled{3} \textbf{estimating} the causal effect based on probabilistic and machine learning methods that operate on observable software data; and { \circled{4} \textbf{validating} the causal process by, firstly, refuting obtained estimate using different sensitivity and robustness techniques (\eg placebo, random common cause, data subsets validation); and secondly, vetting the creation of the graph using correlational analyses on SCMs' variables. } \figref{fig:pipeline} depicts the four steps that consolidate our post hoc interpretability approach (\ie after models have been trained). We provide a high-level explanation of how each step of \docode pipeline functions (modeling: \secref{sec:modeling}, identifying: \secref{sec:identifying}, estimating: \secref{sec:estimating}, and validating: \secref{sec:refuting}) before describing the case study design in detail in \secref{sec:case_studies}. 

In addition to the previous pipeline and to help bridge the gap between a given \nlms low-level code prediction and human-understandable categories, we propose a \textbf{syntax clustering} criterion that aims to group well-known categories of programming languages from individual code tokens. This criterion is not explicitly defined in the pipeline but it is required for posterior evaluations. {Syntax clustering is introduced in \secref{sec:case_studies} and validated in \secref{sec:stx_val}.}

While our syntax clustering provides the building blocks for explaining global model behavior when applied to predict different types of software data, it may be difficult to determine whether any correlations in performance are \textit{causal}, \ie definitely resulting from the intervention, or \textit{spurious}, \ie potentially caused by confounding factors. To enable an analysis of whether a change in model performance is truly causal or not, allowing for the generation of \textit{accurate explanations}, we propose an analysis based on causal inference which we describe below.

{\textbf{Preliminaries.} \textit{Software Data}, \textit{Domain Knowledge}, and \textit{Exploratory Causal Analysis} are indispensable elements before starting the \docode pipeline. We refer to \textit{Software Data} as any artifact generated by a software engineering-based process (\eg source code, requirements, documentation). Because we concentrated on Neural Code Models, our target data is human-generated code for causal interpretability. \textit{Domain knowledge} enables software researchers to discern whether the causal assumptions about data are plausible on scientific grounds. This ability to qualitatively encode causal assumptions in a graphical model is known as \textit{Transparency}, depicting the process in which researchers recognize cause-and-effect relationships in a domain \cite{pearl_seven_2019}. Although \textit{Software Data} and \textit{Domain Knowledge} are sufficient elements to create a graphical form of perceived causal assumptions, we scrutinized the graph creation process using statistical tools. These tools aim at detecting correlational evidence among SCMs' variables to support the qualitative graph encoding. We present the methodology and results of these correlations in \textit{Exploratory Causal Analyses} (see \secref{sec:eda}). }

\textbf{Step One: Modeling Causal Inference Problem.} A variable $T$ is a \textit{cause} of $Y$ if the variable $Y$ depends on $T$ to determine its value. However, there are also variables $U$ that represent \textit{unmodeled} factors \cite{Scholkopf2022}. This causal relationship can be precisely formulated via a causal model that uses directed acyclic graphs (DAGs) to describe direct parent-child relationships, instead of probabilistic dependencies, among random variables such as $T$ and $Y$. It also allows us to introduce a graphical definition of causation  \cite{Scholkopf2022,Pearl2009Causality}. Therefore, Structural Causal Models (SCMs) are graphical models responsible for enabling us to quantitatively estimate the results of an action or intervention simulated in the graph. In other words, SCMs provide a framework for \textit{counterfactual reasoning}. 

\textbf{Step Two: Identifying Causal Estimand.} In this step, \docode formulates a causal estimand, (\ie mathematical expression) subject to the previously defined Structural Causal Model (SCM). {\docode employs graph-based criteria and \textit{the adjustment formula} for identifying an expected causal effect. These graph-based criteria determine which set of confounders $Z$ in the SCM should be conditioned on when seeking the causal relationship between $T$ and $Y$. The doWhy library is used to \textit{explore} the identification of different causal estimands with criteria such as back-door criterion, front-door criterion, instrumental variables, and mediation analysis \cite{dowhy}}. 

\textbf{Step Three: Estimating Causal Effects.} After obtaining an expression using any identification criteria, our method uses a suitable or proper estimation method to compute the causal effect. The estimation method depends on the nature of the SCM's variables (\eg binary, discreet, continuous). The library doWhy supports methods based on estimating the treatment assignment, the outcome model, the instrumental variable equation, and machine learning estimators \cite{dowhy}. {Our study concentrates on estimating \textit{propensity scores} for binary treatments and \textit{linear regression} for continuous treatments.} 

{\textbf{Step Four: Validating Causal Process.} An important step in causal analysis is validating our causal process. This validation comprises two parts: \textit{refuting effect estimate } and \textit{vetting the causal graph}. To perform the first part, we employ \textit{refutation methods} that calculate the robustness of the causal estimate by making the assumptions \textit{falsifiable}. These methods test the robustness of our assumptions through various types of perturbations on the graph or software data to see the impact on our Average Treatment Effect ($ATE$) estimation. There are multiple refutation methods, but in this paper, we focus on four: adding a random common cause or covariate $\mathcal{R}_1$, adding an \textit{unobserved} common cause or covariate $\mathcal{R}_2$, replacing the treatment with a random variable or placebo $\mathcal{R}_3$, and removing a random subset of the data $\mathcal{R}_4$. Depending on the method, the resulting value should either be the same as the original $ATE$ (\ie \textit{invariant transformation}) or close to zero (\ie \textit{nullifying transformation}). To perform the second part, we inspect the causal graph at two different times: 1) \textit{graph creation}, which indicates supporting the rationale behind the construction of the graph using correlational analyses on SCMs' variables, and 2) \textit{graph correctness}, which applies the concept of \textit{Testability} after executing an identification method \cite{Pearl2018Causality}.
}

\textbf{Enabling Causal Interpretability for NCMs.} Consider the scenario in which we want to understand how a given model operates between buggy and fixed code. To do this \docode constructs a SCM composed of an intervention variable $T=Buggy$ and potential outcome $Y=Acc$ (representing a measure of model effectiveness or performance in terms of our defined code categories). In addition, we identify confounding factors $Z$ for interventions and potential outcomes. Next, we want to perform the intervention wherein the model is applied to fixed code instead of buggy code. \docode then constructs a parallel graph for this intervention. Using notions from causal inference theory (\ie modeling, identifying, estimating, refuting), \docode is then able to determine how the intervention affected model performance in terms of code categories, and whether the change in performance is a true causal effect, or spurious, resulting from effects of covariates.

While the above scenario illustrates the intuition of our causal analysis, it is important to define how causal effects, and ultimately interpretations, will be generated. First, we need methods by which we can measure model performance. To do this we use Cross-Entropy loss, or the difference in distribution of the model's prediction and the ground truth for given token sequences, and average Next Token Predictions, or probabilities assigned by models to individual tokens, across token sequences as these measures are relevant values produced at inference time that reflect the prediction effectiveness. By relating these values to SE-based interventions $T$, we can gain an understanding of how well a studied model is \textit{generating code} under these treatments. To calculate causal effects for binary treatments similar to the ones we used in $T_{[data]}$ (\eg Buggy/Fixed and Commented/Uncommented), we first calculate association correlations (Def.~\ref{def:js}) and then causal effects (Def.~\ref{def:ate}).

If we want to estimate the extent buggy/non-buggy SE-based interventions affect the accuracy of a \nlm, we need before to \textit{identify} the effect of the treatment using the \textit{adjustment formula} of causal inference (see Def.~\ref{def:effect}). After the causal effect is calculated, we aim to generate \textbf{causal explanations} from our analysis via structured templates that relate the code token category to the intervention: \eg \textit{\texttt{\small [code token category]} performed worse by \texttt{\small [change in performance]}, due to a change in model application from \texttt{\small [intervention]} to \texttt{\small [intervention]}, with a causal analysis Average Treatment Effect of \texttt{\small [ATE value]}}.

\section{Step One: Modeling Causal Problem}
\label{sec:modeling}

\begin{figure}[ht]
		\centering
		\includegraphics[width=0.45\textwidth]{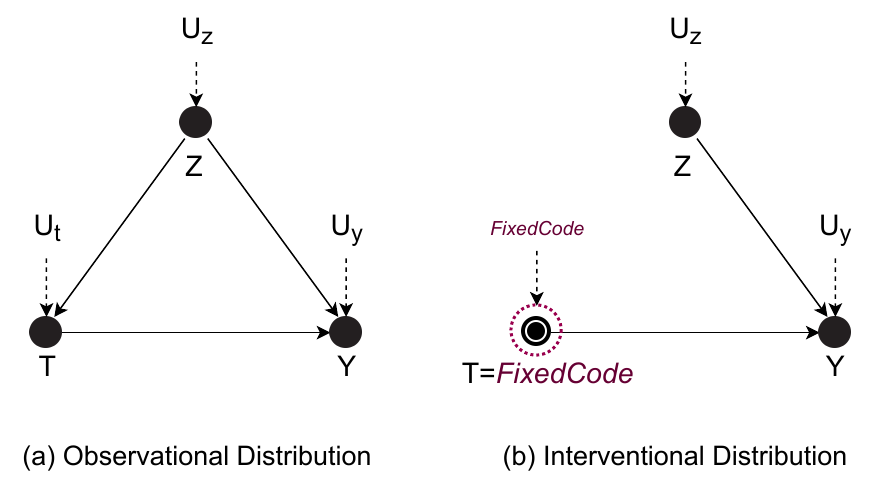}
		\caption{(a) \textit{Structural Causal Model} representing cause-effect relationships of program repair in \nlms. (b) The \textit{SCM} after intervening the treatment with \textit{Fixed Code}.}
        \label{fig:scm}
\end{figure}

In the first pipeline step, assumptions about relationships among data are defined in a Structural Causal Model (SCM) similar to Fig.~\ref{fig:scm}. This SCM is later modified to compute the interventional distribution $p(Y|do(T))$ in step two. Causal assumptions must be made explicit in the graph, which entails defining the interventions $T$ (\ie binary: buggy/non-buggy, discrete: layer modifications, {continuous: syntactic differences}), potential outcomes $Y$ (\ie Cross-Entropy or NTPs performance), and the confounders (or common causes) $Z$. For all SCMs proposed in this study, we assumed that confounders are SE quality metrics since they have the potential to influence models' code predictions, \ie a \nlm may be influenced by more or less for loops, as well as influence interventions, \ie more code is correlated with more bugs \cite{4027145}. 

{Before interpreting the prediction performance of a given Neural Code Model (\nlm) using causal explanations, we introduce the formalism of a \nlm as defined below.
\begin{definition}
\label{def:ncm}
\textbf{Neural Code Model (\nlm)}: A \nlm is a probability distribution $P(w_t | w_1, w_2, \ldots, w_{t-1})$ in which the output $w_t$ at time step $t$, given the input sequence $(w_1,w_2,..,w_{t-1})$, is inferred through the conditional probability $ P(w_t | h_t)$  \cite{Raychev2014CodeCW,White:MSR15}. The hidden state $h_t$ encapsulates the properties of the preceding context. The training corpora of \nlms primarily contain source code. Consequently, input sequences are split into subwords of code (\ie tokens $w$ from a vocabulary $\mathcal{V}$). Furthermore, \nlms are typically pre-trained for code generation, and fine-tuned to perform specific SE downstream tasks such as code summarization, test case generation, and bug fixing.
\end{definition}
}

The following example showcases a practical scenario in which code predictions can be causally explained by SE-based interventions on program repair.

\begin{exmp}
\label{exmp:association}
Consider a Bayesian network created to explain code predictions of a given \nlm, depicted as an arrow that links buggy input code to the model's accuracy $Buggy\to Acc$. Accuracy is a measure of agreement between current vs. expected predictions. We can use accuracy as a measure that contains and represents information about code predictions. The variables $Buggy$ and $Acc$ are dependent. These variables embody a \textit{program repair intervention}. Therefore, if we want to define the joint distribution $p(Buggy,Acc)$ to represent the network, we must specify by Bayes' rule using the prior $p(Buggy)$ and the conditional probability $p(Acc|Buggy)$. However, this joint distribution can be computed in the opposite direction $Acc\to Buggy$ using the same Bayes' rule for a prior $p(Acc)$ and conditional $p(Buggy|Acc)$. The fact that the joint distribution can be represented with both networks is non-intuitive since we know from experience that the model performance cannot give rise to a bug in the original input. In other words, the relationship between these variables is asymmetric or \textit{causal}. Hence, we expect that buggy snippets affect the prediction performance of \nlms, not the other way around.
\end{exmp}

A first attempt to address the relationship in Ex.~\ref{exmp:association} would be computing a correlation coefficient $\rho_{TY}\approx p(Y|T)=p(Acc|Buggy)$, where $T$ is a binary \textit{intervention} that represents the \textit{debugging} process and $Y$ is a \textit{potential outcome} that corresponds to the prediction performance of the model. This coefficient, however, is still symmetric: if $T$ is correlated with $Y$, then $Y$ is equally correlated with $T$. \textit{Causal networks} allow us to model causal asymmetries where directionality goes beyond probabilistic dependence. These causal models represent the mechanism by which data \textit{were generated} \cite{Pearl2016Causality}. Instead of testing whether $Buggy$ and $Acc$ are conditionally dependent, \textit{causation} asks which variable \textit{responds} to the other one: $Buggy$ to $Acc$ or $Acc$ to $Buggy$? \cite{Pearl2016Causality}. Therefore, we can formally introduce a definition of causation:

\begin{definition}
\label{def:causation}
  \textbf{Causation.} A variable $T$ is a \textit{cause} of $Y$ if the variable $Y$ depends on $T$ to determine its value. Formally, the value of $Y$ was \textit{assigned} based on what is known about $T$. In other words, the value of $Y$ is determined by a \textit{\textbf{structural equation}} $Y=f_y(T,U_y)$ and the arrow $T\to Y$. The $U$ variables in these equations represent \textit{unmodeled} variables that are exogenous to the causal network but disturb the functional relationship between the outcome and its treatment \cite{Scholkopf2022}.    
\end{definition}

Similarly, we can define a structural function for the treatment $T=f_t(U_t)$ that depends only on $U$ disturbances assuming that no \textit{common causes} exists between potential outcomes and interventions. A common cause (or confounder) is a random variable $Z$ that causally influences two variables that are initially perceived as statically dependent ($T \not\!\perp\!\!\!\perp Y$). However, this dependency can be explained by the underlying influence of $Z$ on the effects, making the effects conditionally independent ($T \perp\!\!\!\perp Y | Z$). Therefore, there exist more complex causal relationships between treatments and outcomes that we can model with structural equations that correspond to a \textit{\textbf{Structural Causal Model}} (SCM). We introduce a definition for Structural Causal Models as graphical models that capture causal assumptions:

\begin{definition}
\label{def:scm}
  \textbf{Structural Causal Models.} These directed acyclic graphs (DAGs) describe direct parent-child relationships, instead of probabilistic dependencies, among random variables $X_i$. The value $x_i$ of each variable $X_i$ is defined by the structural equations $x_i = f_i(PA_i,U_i)$ where $PA_i = {X_j: X_j \to X_i}$ denotes the set of parents or direct causes of $X_i$. This model allows us to introduce a graphical definition of causation  \cite{Scholkopf2022,Pearl2009Causality}.
\end{definition}

In summary, this step consists of \textit{setting down assumptions} about the causal relationships of software data employed to interpret \nlms. SCMs help us to describe relevant features of software data and how these features interact with each other. In the following subsection, we define each component of the Structural Causal Models (\ie interventions, potential outcomes, and common causes or confounders).

\subsection{SE-Based Counterfactual Interventions} \label{sec:seintervention}
\nlms are notorious for behaving differently across distinct datasets \cite{abu-mastafa}. For example, if a model trained on a well-commented dataset is applied to predict segments of poorly commented code, this mismatch could potentially impact its performance. As such, we assert that observing model performance across datasets with different characteristics can aid in understandability and interpretability. Hence, we defined \textit{SE-Based Counterfactual Interventions} $T$ to better understand model performance across different settings. We formulate these interventions based on domain knowledge from observable testbeds (\ie datasets) organized in sample pairs \textbf{treatment} $T=0$ (\ie BuggyCode) and \textbf{control} $T=1$ (\ie FixedCode). Note that we define testbeds according to different applications often described in SE research. The general process comprises of identification of some specific \textit{intervention} (\ie program repair) and 2) construction or collection of the necessary data via mining repositories or other means that contain these observed interventions. More complex interventions will likely be more challenging to prepare. 

In \docode, \textit{counterfactual interventions} produce explanations motivated by both semantic perturbations $T_{[data]}$ to our SE Application Settings (\eg Program Repair ``Buggy/Fixed Code'', (Un)commented Code, Syntactic Differences) and model hyper-parameter variations $T_{[hyp]}$ on \nlms (\eg layers, units, or heads). Although hyperparameter variations are NOT data perturbations based on SE settings, we include them to extend the analysis of possible causes of models' predictions beyond data interventions.  

\subsection{Potential Outcomes / Code Predictions} 
\begin{figure}[ht]
		\centering
		\includegraphics[width=0.483\textwidth]{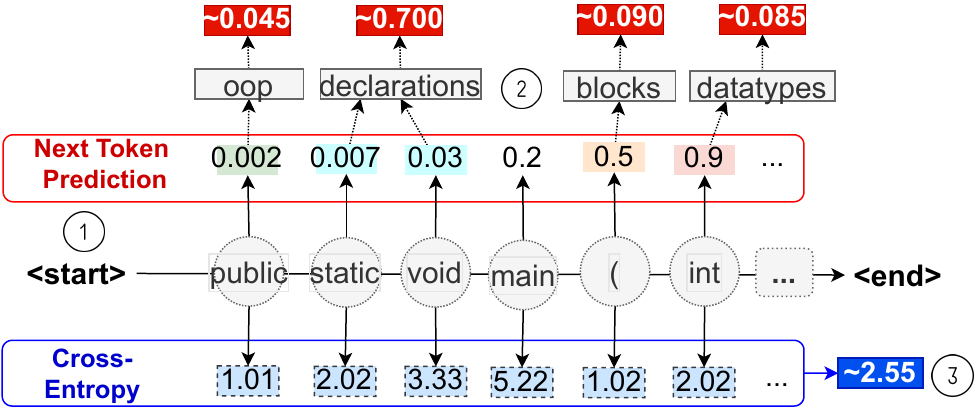}
		\caption{Potential Outcomes are Code Prediction of  \nlms: Cross-Entropy $(Y_g)$, Next Token Predictions $(Y_l)$, or Distance metrics $(Y_d)$.}
        \label{fig:performance}
\end{figure}

Cross-Entropy ({Fig.~\ref{fig:performance}-\circled{3}}), Next Token Prediction ({Fig.~\ref{fig:performance}-\circled{2}}), and Distant Metrics (\ie Jaccard, Levenshtein, and Sorence-Dicen) are relevant values produced at inference time that reflect the effectiveness of a model at \textit{predicting code}. By relating these values to counterfactual interventions $T$ (\ie program repair), we can gain an understanding of how well a studied \nlm is \textit{generating code} under these SE-based interventions. 

\textbf{Cross-Entropy.} \label{out:cross_entropy}
We refer to Cross-Entropy loss as a measure of a model's \textit{coarse-grained performance} $Y_g: w \to - \sum_{t \in |w|} P(w_t | d_t) \log Q(w_t | w_{<t})$ as these losses capture the overall performance of a \nlm over an entire sequence of tokens $w$. Due to the discrete nature of the data, the expression $P(w_t | w_{t-1:1} )$ can be estimated using a classifier. The classifier, in our particular case, is a \nlm \cite{Bengio2003AModel}. Hence, rather than using \textit{n}-grams or Markov Models to approximate $P(w_t | w_{t-1:1})$ \cite{Karampatsis2020Open-VocabularyAbstract}, it is convenient to use a latent model $P(w_t | w_{t-1:1} ) \approx P(w_t | d_t )$, where $d_t$ is known as a \textit{hidden state} that embeds the sequence information from past observations up to the time step $t$. Depending on \textit{how} the sequence is processed, the hidden state $d_t$ can be computed using either an autoregressive network (\ie such as a Transformer ($GPT$)~\cite{vaswani2017transformers}) or a Recurrent Neural Network ($RNN$). 

\textbf{Next Token Prediction (\ntp).} \label{out:ntp} 
Conversely, \ntp values signal \textit{fine-grained performance} $Y_l:w_{<t} \to P(w_t | w_{t-1:1})$ within token-level contexts. NTPs capture local predictions for individual tokens that are affected by complex interactions in \nlms and are equivalent to the estimated predicted value (or softmax probability) $\sigma(k)_t$ for each token. Bear in mind that the size of the vector $\sigma(k)_t$ is the vocabulary $|\mathcal{V}|$, in which $k$ represents the non-normalized log probabilities for each output token $t$. NTPs capture the value of the expected token $w_t$ instead of the maximum value estimated in the vector $\sigma(k)_t$. 

\textbf{Distance Predictions.}\label{out:distance}
Similarity distance scores play a crucial role in assessing the model's \textit{distance performance}, defined by the expression $Y_d: s' \to \nabla(s, s')$. Function $\nabla$ represents the similarity coefficient used for the pairwise comparison of two finite sample sets, $s$ and $s'$. For instance, if set $A$ denotes Node Types in the AST (Abstract Syntax Tree) of a ground-truth code sample, and set $B$ represents Node Types in the AST of a predicted code sample, we can quantify their similarity using metrics such as the Jaccard Index, Levenshtein distance, and Sorensen-Dice coefficient. The resulting similarity scores, which range from $0$ to $1$, effectively measure the degree of similarity between the sets, with $0$ indicating no similarity and $1$ meaning that the sets are identical.

\subsection{Common Causes or SE-based Confounders}\label{sec:confounders}
In past work, several factors such as code duplication~\cite{Allamanis19} have been illustrated to affect model predictions. As such, we derive a list of potential SE-based confounders that could influence a model's prediction of our clustered syntax categories beyond our interventions defined in the previous section. Our initial set of confounders (also called common causes) include McCabe's complexity, LoC (Lines of code), number of: 
\begin{multicols}{2}
    \begin{itemize}
        \item \texttt{\small returns}
        \item \texttt{\small loops}
        \item \texttt{\small comparisons}
        \item \texttt{\small try/catches}
        \item \texttt{\small parenthesized expressions}
        \item \texttt{\small string literals}
        \item \texttt{\small variables}
        \item \texttt{\small max nested blocks}
        \item \texttt{\small anonymous classes}
        \item \texttt{\small inner classes}
        \item \texttt{\small lambda expressions}
        \item \texttt{\small unique words}
        \item \texttt{\small log statements}
        \item \texttt{\small modifiers}
    \end{itemize}
\end{multicols}

Practitioners and researchers can extend the search of potential SE-based confounders based on their domain knowledge, empirical analysis on \nlms, or observations of \nlms' behavior in production.

\section{Step Two: Identifying Causal Estimand}
\label{sec:identifying}

In the second pipeline step, once the SCM (similar to \figref{fig:scm}) is constructed, {\docode must validate under which conditions the structure of the created graph is sufficient for estimating a causal effect from software data. Graph-based identification criteria are \textit{test} mechanisms or tools employed in Causal Inference to determine whether a causal effect can be estimated for any two variables $T$ and $Y$. \docode supports graph-based criteria such as back-door criterion, front-door criterion, instrumental variables, and mediation. The application of these criteria depends on the configuration and shape of the SCM. To formulate a correct estimand, \docode requires both an \textit{adjustment formula} (see Eq.~\ref{eq:effect}) and a \textit{graph-based criterion}. 

While we discuss identification criteria such as front-door and mediation as potential options for graph-based identification criteria, for our case study -in particular- we use the combination of the \textit{back-door criterion}, \textit{instrumental variables}, and the \textit{adjustment formula}, as these cover sufficient interpretability scenarios for a causal graph with the standard shape. The standard shape of proposed SCMs includes data-based $T_{[data]}$ and parameter-based $T_{[hyp]}$ interventions, SE-based confounders $Z \in SE_{metrics}$, potential outcomes $Y_l, Y_g, Y_d$, and (if applicable) instruments $I$. 

The back-door criterion identifies sets of confounders that should be adjusted/controlled for when we estimate causal effects from software data, while the instrumental criterion comprises variables with a direct effect only on the treatments. The mathematical details of how back-door or any other identification criteria operate are out of the scope of this paper. Detailed information about graph-based criteria can be found in academic manuals \cite{Pearl2016Causality,Pearl2009Causality, Scholkopf2022}. Nonetheless, the \textit{adjustment formula} for computing causal effects by simulating interventions (\ie do-calculus \cite{Pearl2018Causality}) is discussed in this section. 
}

Structural Causal Models (SCMs) are stable mechanisms that remain invariant to local changes unlike probabilities computed by Bayesian networks \cite{Pearl2016Causality}. This characteristic helps us to estimate quantitatively the results of an intervention in the graph without actually performing it in controlled settings (\ie randomized experiments). The mathematical tool employed to perform these interventions is the $do(\cdot)-operator$ \cite{Pearl2009Causality}. For instance, if we want to estimate how fixed code affects the performance of a \nlm, we need to compute the action $p(Acc|do(Buggy=False))$ (Eq.~\ref{eqn:do-1}). These actions are \textit{interventional} distributions since we \textbf{set} the value of $Buggy$ to $False$. Note that this interventional distribution is \textbf{not} the same as the distribution $p(Acc|Buggy=False)$ (Eq.~\ref{eqn:do-2}). This latter distribution is \textit{observational} as we are \textit{conditioning} the performance on the value of the $Buggy$ variable. Intervening on a variable in a SCM means fixing its value and, therefore, changing the value of other variables of the network as a result. Conversely, conditioning on a variable means narrowing the cases that the outcome takes once we assign a value to the intervention. 

\begin{exmp} 
\label{exmp:scm}
Consider Fig.~\ref{fig:scm} a generalization of the buggy influence on a deep model's performance. The first graph is a SCM composed of an intervention variable $T=Buggy$ and potential outcome $Y=Acc$. In addition, we identified some common causes $Z=SE_{Metrics}$ for interventions and potential outcomes. These common causes (or confounders) can be mapped to Software Engineering quality metrics (\eg Lines of Code, McCabe Complexity, Size of Methods). We want to perform the intervention $do(Buggy=False)$, which is the same as $do(T=FixedCode)$. The second graph depicts this program's repair intervention. 
\end{exmp}

Note that fixing the value of $T$ makes the SCM change by eliminating the effect or influence arrow of the confounder $Z$ to the intervention. The disturbance $U_t$ is also eliminated. This elimination process of input arrows to fixed variables is formally known as \textit{graph surgery}. Both $do(\cdot)-operator$ and graph surgery allow us to untangle causal relationships from mere correlations \cite{Pearl2009Causality}. The law of total probabilities and invariance principle are required to compute the observational and interventional distributions \cite{Scholkopf2022}:

\begin{subequations}
\begin{align}
p(Y|do(t=FixedCode)) &=\sum_{z \in _{metrics}}p(Y|t,z)p(z)\label{eqn:do-1} \\
p(Y|t=FixedCode) &= \sum_{z \in _{metrics}}p(Y|t,z)p(z|t) \label{eqn:do-2} 
\end{align}
\label{eqn:do-all-lines}
\end{subequations}

Note that Eq.~\ref{eqn:do-1} differs from  Eq.~\ref{eqn:do-2}: the prior $p(z)$ in contrast to $p(z|t)$, which is precisely the link that is eliminated in the SCM. Eq.~\ref{eqn:do-1} is formally known as the \textbf{\textit{adjustment formula}}. This formula is one of the building blocks in causal inference since it helps us to adjust common causes or control for confounders to allow the estimation of \textit{causal/treatment effects} \cite{Pearl2009Causality,Pearl2016Causality}.

\begin{definition}\label{def:effect}
\textbf{Treatment Effects.} Given a Structural Causal Model where a set of variables $PA$ denotes the parents of $T$, the treatment effect of T on the potential outcome Y is given by
\begin{subequations}
    \begin{align}
    p(Y=y|do(T=t)) &=  \label{eq:effect-1}\\
    \Sigma_z p(Y=y|T=t,PA=z)p(PA=z)  &= \label{eq:effect-2}\\
    \Sigma_z p(T=t,Y=y,PA=z)/p(T=t|PA=z) \label{eq:effect-3}
    \end{align}
\label{eq:effect}
\end{subequations} 
  
\end{definition}

In our initial causal statement, we generally accept that buggy code \textit{causes} a \nlm to predict poorly. Although the causal statement is true, it is not guaranteed that ``every buggy snippet'' is certain to make a model predict poorly. Therefore, causal relationships are \textbf{uncertain}. This uncertainty is captured by employing conditional probabilities described in Eq.~\ref{eq:effect-2}. Note Eq.~\ref{eq:effect-2} is a generalization of the adjustment formula Eq.~\ref{eqn:do-1}. We connect observational data with our interventional distribution to compute treatment effects. A standard way of connecting data with the interventional distribution is by employing a summation described in Eq.~\ref{eq:effect-3}. Note Eq.~\ref{eq:effect-3} is obtained with the application of Bayes' rule and algebraic manipulation once we multiply and divide Eq.~\ref{eq:effect-2} by the term $p(T=t|PA=z)$. This term is a conditional probability known as the \textit{propensity score}. This propensity score and the joint probability of all the nodes are distributions that can be \textit{obtained directly from data} \cite{Pearl2016Causality}. We explain this connection for code generation in \secref{sec:estimating}.

\section{Step Three: Estimating Causal Effects}
\label{sec:estimating}

In the third pipeline step, \docode estimates the causal effect using statistical and ML methods based on the adjustment formula from the previous step. \docode computes \textit{Propensity Score Matching} for binary SE interventions (\ie Buggy/Fixed) and \textit{Linear Regressions} for SE discrete interventions (\eg layers, units, or heads). We refer interested readers to the \textit{doWhy} documentation for full estimation methods details \cite{dowhy}. For completeness, we will solely show how to estimate a causal effect assuming a binary intervention as an example. We can start by explaining the notion of \textit{Average Treatment Effect (ATE)}. ATE is simply the average score of all treatment effects (see Def.~\ref{def:effect}) computed for a population. In our case, an individual of the population is just a code snippet $x$. 

\begin{definition}
\label{def:ate}
\textbf{Average Treatment Effect (ATE)} Defining the first intervention as $do(T=1)$ and the second by $do(T=0)$, the Average Treatment Effect is the population average of the difference of causal effects of each code snippet $x$.  
\begin{subequations}
    \begin{align}
     ATE = \mathbb{E}_{x\sim p(x)}[Y=y|x,do(T=t)]  &= \label{eq:ate-1}\\
     \mathbb{E}_{x\sim p(x)}[ \mathbb{E}[Y|x,do(T=1)] - \mathbb{E}[Y|x,do(T=0)] ] &= \label{eq:ate-2}\\
     \mathbb{E}_{x\sim p(x)}[ \mathbb{E}[Y^{1}|x,T=1] - \mathbb{E}[Y^{0}|x,T=0] ] &= \label{eq:ate-3}
    \end{align}
\label{eq:ate}
\end{subequations} 
\end{definition}

The previous Eq.~\ref{eq:ate} shows the formal definition of an ATE. We can derive the final expression by applying the law of total expectations and the ignorability assumption $Y \perp\!\!\!\perp  Z|T$, where the potential outcomes $Y$ are independent of intervention assignments conditioned on covariates $Z$ \cite{Pearl2009Causality}. That is, the effects of the hidden confounders $Z$ and missing data are ignored. In Eq.~\ref{eq:ate}, the term $\mathbb{E}[Y^1|x,T=1]$ represents the expected value of a potential outcome under an observable intervention (\ie FixedCode). Similarly, the term $\mathbb{E}[Y^0|x,T=0]$ represents an expected value of a potential outcome under an observable intervention (\ie BuggyCode). Both terms are quantities that can be \textit{estimated from data}. Covariate adjustment (in Eq.~\ref{eqn:do-1}), propensity score (in Eq.~\ref{eq:effect-3}), and linear regression are some of the estimation methods that we employ to approximate $ATEs$. Their usage depends upon the type of the intervention variable (\ie binary, discrete, or continuous) and causal graph assumptions.

{
\section{Step Four: Validating Causal Process}
\label{sec:refuting}
Assumptions encoded in causal graphs are justified by observations of a \textit{data generating process}. Therefore, testing for the quality of the causal graph fitting the data would be the main validity issue in the fourth pipeline step. We ask the question \textit{How can we assess the {validity} of the underlying causal process?} To answer the question, firstly, we must assess whether the estimated causal effect from step three is not significantly altered after assumption violations (\ie refutation methods). Secondly, we must conduct exploratory causal analyses to scrutinize how strong the correlations are among SCMs' variables. These exploratory analyses are a byproduct of graph validity to provide more evidence that supports causal assumptions on the top of the \textit{testability} with refutations and identification criteria. }

\subsection{Refuting Effect Estimate}
The obtained causal effect can be validated using \textit{refutation methods} that calculate the robustness and sensitivity of the estimate. {Refutation methods came from the idea that hypotheses respecting a \textit{data generating process} can be tested with conditions in which the hypotheses would be false. Structural Causal Models (SCMs) encode assumptions or hypotheses regarding a \textit{data generating process}. Consequently, SCMs are \textit{falsifiable} by modifying the model or the data to disprove initial assumptions.}

In essence, the refutation methods apply random perturbations to the original Structural Causal Model (SCM) to test for the robustness of the estimated causal effect ($ATE$). {Refutation methods inform researchers and analysts about the validity of the graph by applying \textit{invariant} or \textit{nullifying} transformations. \textit{Invariant} transformations modify the data so that the estimated value should not change; otherwise, the causal model would fail the test. For instance, a refutation method tests a causal graph for confounders when this method affects, adds, or perturbs common causes. As such, if the obtained refutation value for common causes is not robust (\ie different from the original $ATE$), it suggests that we should update our initial assumptions about confounders encoded in the graph. On the other hand, \textit{nullifying} transformations modify the data so that the estimated effect should be zero; otherwise, the causal model would fail the test too.} 

\begin{wrapfigure}{r}{0.15\textwidth}
  \begin{center}
    \includegraphics[width=0.15\textwidth]{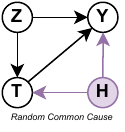}
  \end{center}
  \caption{{\scm transformation by adding a random common cause.}}
   \label{fig:refutation_unobserved_common_cause}
\end{wrapfigure}

Although researchers can design their tailored methods to test distinct parts of the causal graph (\eg treatments, confounders, hidden causes, datasets), we chose four baseline methods for our study:

\textbf{Adding a \textit{random} common cause or covariate $\mathcal{R}_1$}. We evaluate if the estimation method changes its estimate after introducing an independent random variable as a confounder from the dataset. {This refutation involves computing new causal estimates by simulating a common cause with randomly generated data and reporting the average value. A stable causal effect should remain unaffected by this change, maintaining its original estimate $p(Y|do(T)) = p(Y|do(T), H)$. \figref{fig:refutation_unobserved_common_cause} depicts the modification.}

\textbf{Adding an \textit{unobserved} common cause or covariate $\mathcal{R}_2$}. { 
A new dataset is artificially created for a correlated common cause between treatment $T$ and outcome $Y$. Next, the effect of the unobserved common cause is recomputed. Considering that the backdoor criterion assumes that all common causes are observed, the new estimate should not be altered drastically after violating the criterion assumption.}

\textbf{Replacing the treatment with a random variable or \textit{placebo} $\mathcal{R}_3$}. We evaluate to what extent the estimated causal effect changes if we replace the true treatment variable $T$ with an independent random variable of the same nature (\ie placebo). {Specifically, we simulate datasets where the treatment is the placebo, hence the computed causal effects should have a distribution close to zero. By using this refutation method, we seek to test the reliability of the estimator.}

\textbf{Removing a \textit{random subset} of the data $\mathcal{R}_4$}. We evaluate if the estimation method changes its estimate significantly after replacing the given dataset with a randomly selected subset. {Through multiple simulations, we calculate the causal estimate using subsets of the original data and report the mean value. The initial causal estimate should remain unaffected by this modification.}

{In summary, for high robustness, we anticipate that $\mathcal{R}_1$, $\mathcal{R}_2$, and $\mathcal{R}_4$ will closely approximate the estimated $ATE$ (Eq.~\ref{eq:ate}). Conversely, the placebo $\mathcal{R}_3$ is expected to be zero.}

{
\subsection{Vetting Causal Graph}
Beyond falsifying the structure and data of the causal graph with the refutation methods, we can also vet the validity of the graph at two different times:

\textbf{Graph Creation Validity}. Before or after the causal structure is modeled from \textit{domain knowledge}, we can conduct exploratory data analyses to alleviate the absence of statistical information about the variables under study. While not explicitly required from Causal Inference theory, we measure the correlation between the confounders and the treatments in one analysis \textcircled{\scriptsize $Z$} $\rightarrow$ \textcircled{\scriptsize $T$}. Then, we measure the correlation between the confounders and the outcomes in a separate analysis \textcircled{\scriptsize $Z$} $\rightarrow$ \textcircled{\scriptsize $Y$}. The purpose of these correlations is to support domain knowledge observations quantitatively. 

\textbf{Graph Correctness Validity}. The correctness of the graph is related to the concept of \textit{testability} \cite{Pearl2018Causality} in which we test whether a Structural Causal Graph (\scm) could have generated a dataset. Identification criteria such as back-door or front-door are standard ways of evaluating the fitness of a model. In fact, the back-door criterion uses a graphical mechanism known as \textit{d-separation}, which facilitates the demonstration that causal models have testable implications in the data they generate \cite{pearl_seven_2019}. 
}

\section{Case Study Design}\label{sec:case_studies}

\begin{figure*}[h]
		\centering
		\includegraphics[width=1\textwidth]{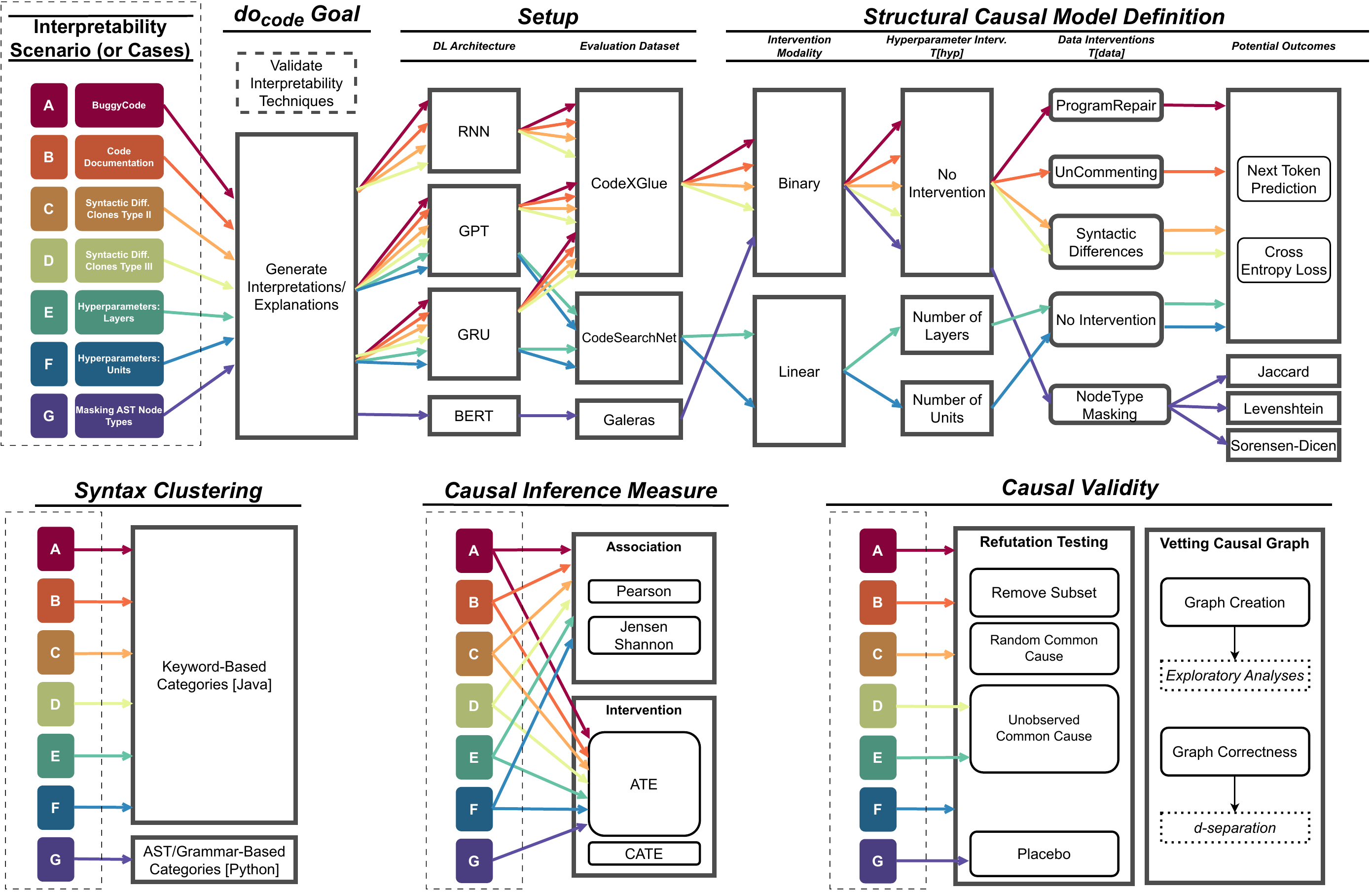}
		\caption{Configuration criteria for enabling causal interpretability. Seven scenarios are proposed for this case study. However, new interpretability cases can be formulated using a different permutation or extending the criteria.}
        \label{fig:case_studies}
\end{figure*}

In this section, we aim to illustrate how \docode is applied to \textit{enable causal interpretations} based on different interpretability scenarios. Fig. \ref{fig:case_studies} depicts an overview of seven scenarios (or cases from A to G) and six essential criteria that comprise the case study. Each scenario can be unfolded into the following criteria: (i) the goal of \docode (\ie generating interpretations or validating an interpretability technique), (ii) the setup (\ie the deep learning architecture under analysis and the evaluation dataset or testbed) (iii) the definition of the Structural Causal Model (\scm) from domain knowledge (\ie intervention modality, hyperparameter interventions, data interventions, and potential outcomes), (iv) a syntax clustering strategy for grouping potential outcomes based on token predictions, (v) the usage of causal inference measure whether the value is an association (\eg Pearson, Jensen Shannon) or an intervention (\eg ATE, CATE), and (vi) the refutation testing method employed to validate the robustness of the interpretations (\eg placebo, unobserved common cause). 

Note that practitioners and researchers should not necessarily have to stick to our seven cases, the configuration criteria can be extended to assess unexplored interpretability scenarios. Various permutations of the criteria outlined in \figref{fig:case_studies} can be formulated depending on the research goal of the causal analysis. The subsections below detail the criteria that we propose for this study.

\subsection{\docode Goal}
The goal of \docode is to enable \nlms' interpretations of code predictions by estimating the causal effects of specific SE-based interventions. On top of that, \docode also facilitates causal inference for \textit{validating} interpretability methods. Users of \docode should define their goals in alignment with this premise as these goals determine how the \scm is formulated. For example, \docode can be employed to interpret the effects of code smells on code predictions produced by given \nlm, the SE-based intervention would be a binary treatment of samples with and without smells. On the other hand, we decided not to introduce a scenario in which \docode validates an interpretability method as it is out of the scope of this study. However, we offer sufficient guidance on assembling the criteria for this purpose in this section.

\subsection{Setup}

Applying \docode requires selecting both the deep learning architecture (\ie \nlm from Def.~\ref{def:ncm}) and the evaluation dataset or testbed for our interpretability scenario.

{\textbf{About Deep Learning Architectures.} Because causal interpretability is agnostic to \nlm, \docode supports (but is not limited to) architectures such as RNN, GPT, GRU, and BERT. Recurrent Neural Networks (RRNs) update the hidden state $h_t$ using the current input and the previous hidden state, thus $h_t = f(h_{t-1}, w_t)$. RNNs can take the form of a Gated Recurrent Unit (GRU) \cite{cho_properties_2014}, which uses reset $r_t = \sigma(W_r \cdot [h_{t-1}, w_t])$, and update $u_t = \sigma(W_u \cdot [h_{t-1}, w_t])$ gates to control how it combines the previous hidden state $h_{t-1}$ with a new candidate state $h_t' = \tanh(W_h \cdot [r_t \odot h_{t-1}, w_t])$ to obtain the updated hidden state $h_t = (1 - u_t) \odot h_{t-1} + u_t \odot h_t'$. }

Furthermore, GPT and BERT are deep architectures known as Transformers \cite{vaswani2017transformers}. In transformer models, the hidden state $h_t$ in each time step is updated through a multi-head self-attention mechanism and a feed-forward network. In simpler terms, $h_t = \text{Attention}(Q_t, K_t, V_t) = \text{softmax}({{Q_iK_i^T}}/{\sqrt{d_k}})V_i$, where $Q_i, K_i$, and $V_i$ represent the queries, keys, and values for the time step $t$, and $d_k$ is the dimension of the key vectors. Transformers adopt the form of Encoder-based (\eg BERT) or Decoder-based (\eg GPT).

We employed RNNs and Transformers for all our scenarios as RNNs are widely used in SE \cite{watson2020dl4se}, while Transformers have gained popularity in the SE/NLP domain for their high performance ~\cite{Mastropaolo2021StudyingTasks}. In \figref{fig:case_studies}),  we opted to train the \nlms for cases $[A-F]$ to gain more control over the training data. This was particularly important given our need to manipulate the tokenizer, allowing us to assign special tokens to specific code categories (\ie syntax clustering). However, \docode is not confined to specialized models trained by the user as demonstrated in the $G$ case, in which we used an in-the-wild BERT model. Therefore, \docode operates effectively on already pre-trained \nlms as long as there exists a way to group fine-grained predictions (\ie BPE tokens) to human-understandable categories (\eg AST/Grammar-based or keyword-based).

Table~\ref{tab:training} provides a detailed overview of the training configurations for each scenario. In $[A-F]$ cases, the \nlms were trained using the Java portion of the \training dataset~\cite{husain2019codesearchnet}, which consists of a diverse array of methods (mts) from GitHub \cite{github}. We partitioned \training into training, validation, and test sets. For model development, we employed Tensorflow and Pytorch \cite{tensorflow2015-whitepaper, pytorch}, along with Huggingface's Transformers library \cite{wolf2020transformers}. To mitigate overfitting, training was terminated if cross-entropy did not improve by at least $1\text{e-2}$ over $5$ epochs. Inputs were standardized with a \textit{start of sentence} token and adjusted to a length of $300$. This process was executed on an Ubuntu 20.04 system with an AMD EPYC 7532 32-Core CPU, an A100 NVIDIA GPU with 40GB VRAM, and 1TB RAM. For the $G$ case, we used a pre-trained BERT model: codebert-base-mlm \cite{feng2020codebert}, trained specifically for a Masking Language Model (MLM) objective and also using the \training dataset.

\begin{table}[ht]
\centering
\caption{{\nlms training specifications for Recurrent Neural Networks (\ie Vanilla RNN, GRU), and Transformers (\ie GPT-2, BERT)}
\label{tab:training}
}
\scalebox{0.9}{%
\vspace{-0.2cm}

\begin{tabular}{lllll}
\multicolumn{5}{c}{\textit{\nlms Training}}                       \\
\textbf{RNNs} & \multicolumn{1}{c}{\textbf{Transformers}} &  & \multicolumn{1}{c}{\textbf{Hyper.}} & \multicolumn{1}{c}{\textbf{Val.}} \\ \cline{4-5} 
\textit{\nlm $_{lyr,unt}$} & \textit{\nlm $_{lyr,hds}$} &  & dropout RNN\cite{Karampatsis2020BigCode}   & 0.5     \\ \cline{1-2}
\rnn              & \tf               &  & dropout TF    & 0.l     \\
\gru              & \tfi              &  & optimizer\cite{Kingma2015AdamAM}     & adam    \\
\grui             & \tfii             &  & learning rate & 1e-3    \\
\gruii            & \bert             &  & beta1,beta2   & 0.9     \\
\gruiii           &                  &  & epsilon       & 1e-7    \\
\gruiv            &                  &  & epochs        & 64      \\
                 &                  &  & batch RNN,TG  & 512,128 \\ \cline{1-2} \cline{4-5} 
\end{tabular}

}

\end{table}

\textbf{About Evaluation Datasets.} To build the data intervention testbeds for our interpretability scenarios, we obtained \textit{Java} samples from \textit{CodeXGLUE} \cite{DBLP:journals/corr/abs-2102-04664} and \textit{Python} samples from \textit{Galeras} \cite{daniel23}. In  $[A-F]$ cases, parallel corpora were used to examine the impacts of buggy code (\BuggyTB: 64,722 methods or mts), code documentation (\CommentsTB: 6,664 mts), and syntactic alterations in semantically similar snippets, specifically focusing on type II (\BigCloneIITB: 666 mts) and type III (\BigCloneIIITB: 8,097 mts) clones from \BigCloneTB. For the $G$ case, we conducted binary interventions by masking tokens corresponding to each AST element in the \Galeras samples (8,299 mts), covering all node types in the Python grammar as defined by tree-sitter \cite{noauthor_tree-sitterintroduction_nodate}. In the control group, a comparable number of tokens were randomly masked in each sample.

\textbf{About Code Tokenization.} For all scenarios, Byte Pair Encoding (BPE) tokenization \cite{sennrich2015neural} was applied to the testbeds before processing them with the \nlms. Known for its efficacy in training \nlms on code, BPE significantly mitigates the \textit{out-of-vocabulary} problem \cite{Karampatsis2020BigCode}. For $[A-F]$ cases, we developed a BPE tokenizer trained on $10\%$ of our training data and with a vocabulary size of 10K. Conversely, for the $G$ case, we employed the pretrained BPE tokenizer from the BERT-selected model. Nonetheless, the BPE tokenization process sometimes resulted in sub-tokens that either combined multiple reserved keywords or split keywords across different tokens. This splitting issue presented a significant challenge for our interpretability analysis as our method relies on accurately grouping token predictions into semantic categories, a criterion that we named \textit{syntax clustering}. To address this, we developed clustering functions to ensure the tokens are correctly aligned with our defined categories, which will be discussed in \secref{sec:aggregations}.

\subsection{Structural Causal Model Definition}

\begin{figure*}[h]
		\centering
		\includegraphics[width=1\textwidth]{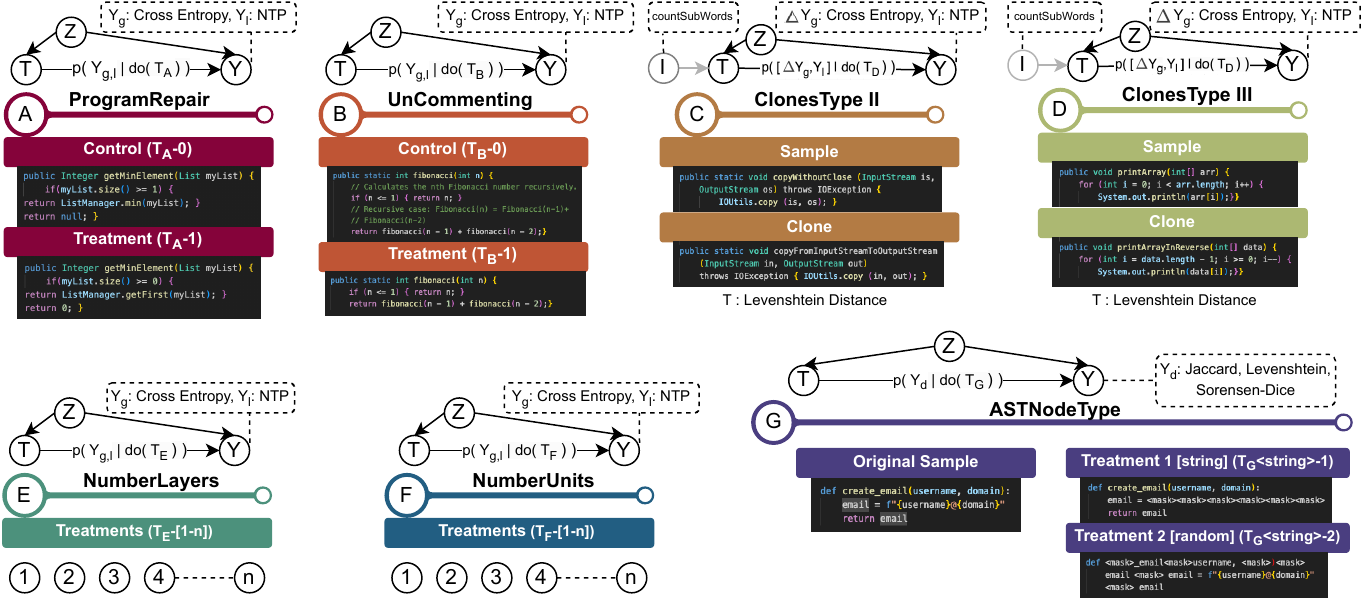}
		\caption{Structural Causal Models' Definition. Each \scm presents the causal estimand, the potential outcomes, the intervention type, and examples of interventions for each interpretability scenario.}
        \label{fig:causal_modes}
\end{figure*}

\docode users must design the Structural Causal Model (\scm) based on their domain knowledge and available data. This criterion involves choosing the intervention modality, specifying the type of the intervention, and computing the potential outcome. The SE-based interventions, contingent upon the objectives of \docode, primarily fall into two categories: Data Interventions $T_{[data]}$ and Hyper-parameter Interventions $T_{[hyp]}$. Data interventions occur within the testbed, whereas Hyper-parameter interventions involve model training parameters (\eg Learning rate, Batch Size, Number of Epochs, Number of Hidden Layers/Units, dropout rate), which are not embedded in the data per se. Moreover, potential outcomes cover a variety of measures such as Next Token Prediction (\ntp), Cross Entropy Loss, BLEU, CODEBLEU, and distance similarity scores including Jaccard, Levenshtein, and Sorensen-Dice. We aimed to estimate the causal effect of $T_{[data]}$ and $T_{[hyp]}$ interventions on different potential outcomes. \figref{fig:causal_modes} describes seven SCMs, the expected potential outcomes with their corresponding treatments $T_{A-G}$, and examples of interventions per scenario. {Additionally, \tabref{tab:method} depicts the interventions and datasets used for each scenario.}

\begin{table}[ht]
\centering
\caption{{Overview of SE-based Interventions Experiments.}
\label{tab:method}
}
\vspace{-0.2cm}
\scalebox{0.81}{%

\begin{tabular}{llllll}
\multicolumn{6}{c}{\multirow{2}{*}{\textit{Counterfactual Interventions}}}      \\
\multicolumn{6}{c}{}                                                            \\
\multicolumn{1}{c}{\textbf{Type}} &
  \multicolumn{1}{c}{\textbf{Interv.}} &
  \textbf{Case Id} &
  \multicolumn{1}{c}{\textbf{Intervention}} &
  \multicolumn{1}{c}{\textbf{Associated Dataset}} &
  \textbf{PL} \\ \hline
\multirow{5}{*}{\textit{$T_{[data]}$}}  & $T_A$ & A & \datainterI   & \BuggyTB\cite{Tufano2019LearningBug-Fixes}       & Java   \\
                                & $T_B$ & B & \datainterIII & \CommentsTB\cite{husain2019codesearchnet}    & Java   \\
                                & $T_C$ & C & \datainterII  & \BigCloneIITB\cite{Svajlenko2015EvaluatingBigCloneBench}  & Java   \\
                                & $T_D$ & D & \datainterII  & \BigCloneIIITB\cite{Svajlenko2015EvaluatingBigCloneBench} & Java   \\
                                & $T_G$ & G & \datainterIV  & \Galeras\cite{daniel23}      & Python \\ \hline
\multirow{2}{*}{\textit{$T_{[hyp]}$}} & $T_E$ & E & \modelinterI  & \training\cite{husain2019codesearchnet}      & Java   \\
                                & $T_F$ & F & \modelinterII & \training\cite{husain2019codesearchnet}      & Java   \\ \hline
\end{tabular}

}

\end{table}

\textit{For $A$ and $B$ cases}, we examined the impact of buggy code and inline comments on \nlm's predictions. The $A$ case assessed how the presence of bugs influences the cross-entropy and \ntp. Similarly, the $B$ case explored the effect of including inline comments on the model's performance.

\textit{For $C$ and $D$ cases}, we assessed how \nlms respond to minor and major syntactic variations in semantically equivalent code snippets, focusing on the effects of semantic-preserving changes on code generation. Due to the absence of a natural split in clone types in the \BigCloneTB dataset, a typical control and treatment approach for causal analysis was impractical. The selection between \textit{function$_1$} and \textit{function$_2$} methods, as categorized by \BigCloneTB, is arbitrary. Instead, we concentrated on how syntactic differences in methods performing identical functions impact our models. Hence, we used the differences between these function sets as our primary confounders and intervention. Specifically, we employed the Levenshtein Distance – a metric quantifying the necessary edit operations (insert, modify, remove) to transform one sequence into another – to simulate a 'refactoring' treatment. This approach approximated the number of edits needed to convert one method into another and allowed us to bypass the need for a natural split between \textit{function$_1$} and \textit{function$_2$}. {It is worth noting that the associated graphs include the instrumental variable \textit{countSubWords}. We found evidence of a subtle correlation between treatments and \textit{countSubWords} in \tabref{tab:distance_n_tokens}.}

\textit{For $E$ and $F$ cases}, as indicated in \tabref{tab:method}, our case study involved exploring two distinct deep learning architectures. We particularly examined how variations in hyper-parameters such as layers and units within each architecture influence the prediction performance.

\textit{For the $G$ case}, our goal was to generate interpretations for the prediction of AST node types using BERT. To achieve this, we essentially constructed the dataset by implementing two treatments: masking tokens corresponding to AST Node types (Treatment 1) and randomly masking an equivalent number of tokens (Treatment 2). We then calculated the normalized similarity distances (\ie Jaccard, Levenshtein, and Sorensen-Dice) between the AST of the predicted code and the ground truth samples, which served as the potential outcomes in the \scm.

\subsection{Syntax Clustering of Code Predictions}\label{sec:aggregations}

Consider the situation where a developer inserts a \fbox{\texttt{\small `('}} character after the \fbox{\texttt{\small `main'}} keyword in a function declaration in Java ({\circled{1}}-Fig.~\ref{fig:performance}). Inherently, a developer mentally rationalizes several things such as the concept of a function declaration and expected Java syntax. If a \nlm can make a similar prediction, it suggests to us that it has \textit{statistically learned} some understanding of the concept of a function declaration and corresponding syntax. Therefore, we assert that by associating human-interpretable categories (\eg Programming Languages Keywords or AST nodes) to model predictions and then analyzing the statistical properties of those predictions, we can begin to learn how well a given \nlm reflects human knowledge.

A major conjecture in interpretability research is that \nlms are more understandable when they \textit{reflect human knowledge} \cite{Kim2018InterpretabilityTCAV}. One way of determining whether a model reflects human knowledge is testing it to see whether or not it operates (or predicts) \textit{similar to how a human would operate}. \docode accomplishes this by grouping code token predictions of \nlms to human interpretable categories.

To help bridge the gap between a given \nlms token-level representation of code and human-understandable categories, \docode aggregates individual tokens to well-known syntactic categories from programming languages. Source code tokens can be clustered to any number of syntactic elements, and particularly for \docode, we focus on aggregating tokens to different syntactic categories, which \textit{\textbf{do not}} require manual labeling. This syntax clustering mitigates the cost involved with large-scale data labeling and still provides explanations rooted in categories with which most programmers and researchers are likely familiar.

The syntax clustering comprises high-level properties of code using a \textbf{clustering function} defined by $\phi_{\mathcal{H}}: \vec{w} \to \vec{h}$, in which the vector $\vec{w}$ corresponds to tokens from a vocabulary $\mathcal{V}$. Thus, each token in a sequence $w$ is assigned to a specific syntax-understandable category $h$. Note that \docode allows the definitions of any potential clustering function, users are not forced to use our clustering category system $\mathcal{H}$. The proposed categories in our system  $\mathcal{H}$ are classified into keyword-based or grammar-based. Below, we pose a separate clustering function for Java and another one for Python.

\begin{figure*}[h]
		\centering
		\includegraphics[width=1\textwidth]{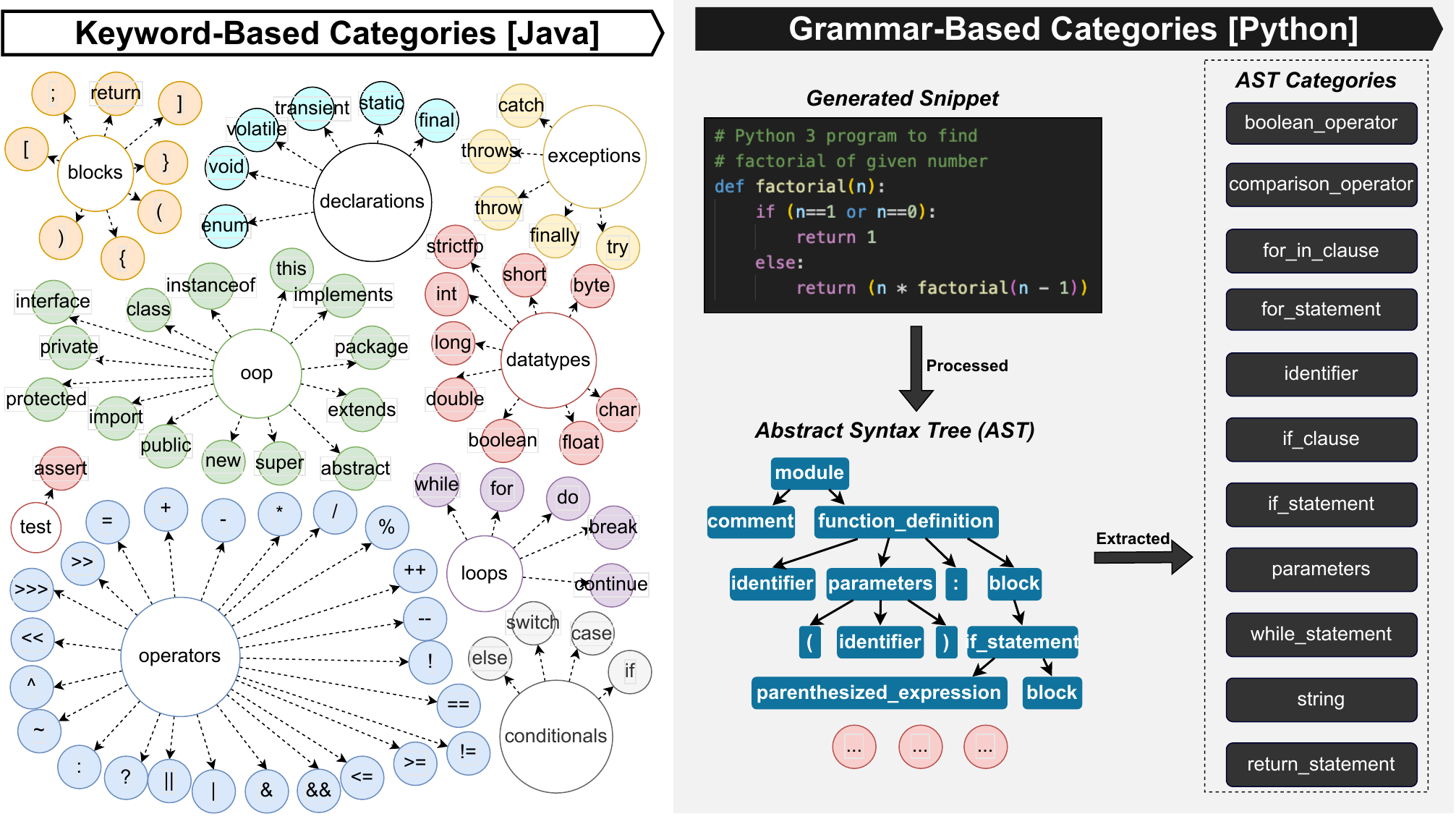}
		\caption{Code Syntax Clustering. Code predictions are grouped into Keyword-based or AST/Grammar-based categories.}
        \label{fig:syntactic_aggregations}
\end{figure*}

\subsubsection{Keyword-Based Aggregation}\label{sec:keyword_aggregation}

In programming languages, different types of tokens retain different semantic meanings. For instance \fbox{\texttt{\small `='}} and \fbox{\texttt{\small `<'}} are common \operators. Therefore, tokens can be grouped into semantically meaningful keyword-based \textit{categories} $\mathcal{H}$. We can establish a clustering function $\phi_{\mathcal{H}}: \vec{w} \to \vec{h}$, in which a token $w$ in a snippet $s$ is clustered to a corresponding keyword category $h$. We propose an initial set of nine categories for Java. Figure~\ref{fig:syntactic_aggregations} illustrates the categories and associated keyword tokens. For each category, we group keywords by functionality that were collected from the official Oracle Documentation and Manuals \cite{noauthor_java_nodate}. These keywords remain consistent across Java versions from Java 7 onwards. {In addition, we present an error analysis in \secref{sec:stx_val} to explore the validity of this clustering function.}

\subsubsection{AST/Grammar-Based Aggregation}\label{sec:grammar_aggregation}
Every expression in a programming language is determined by production rules defined in its Context-Free Grammar defined by the expression $\mathcal{H} = (\alpha, \lambda, \omega, \beta)$, in which $\alpha$ denotes the finite set of non-terminal nodes, $\lambda$ the finite set of terminal nodes, $\omega$ the finite set of production rules, and $\beta$ the start node. The set of production rules $\omega$ for any type of statement (\eg conditional, assignation, operator) is expressed in terms of the terminal and non-terminal nodes $(\alpha, \lambda)$. In programming languages, terminal and non-terminal nodes retain different semantic meanings. Following this, we can establish a clustering function $\phi_{\mathcal{H}}: \vec{w} \to \vec{h}$, where each token $w$ in a snippet $s$ is aggregated and assigned to a corresponding node $h$. We extracted a total of $196$ node types (\ie terminal and non-terminal), from tree-sitter Python's grammar \cite{noauthor_tree-sittertree-sitter-python_2023}. This set of categories was then used to group tokens from code snippets. Figure~\ref{fig:syntactic_aggregations} offers a visual example of this clustering.

\subsection{Causal Inference Measures}

Below, we explain association and intervention metrics that we estimate to validate the reliability of an interpretation. Association metrics serve as baselines that we need to confirm for confounding after computing intervention metrics, which represent actual causal effects.  

\begin{figure}[ht]
  \centering
  \begin{subfigure}[b]{0.46\linewidth}
    \includegraphics[width=\linewidth]{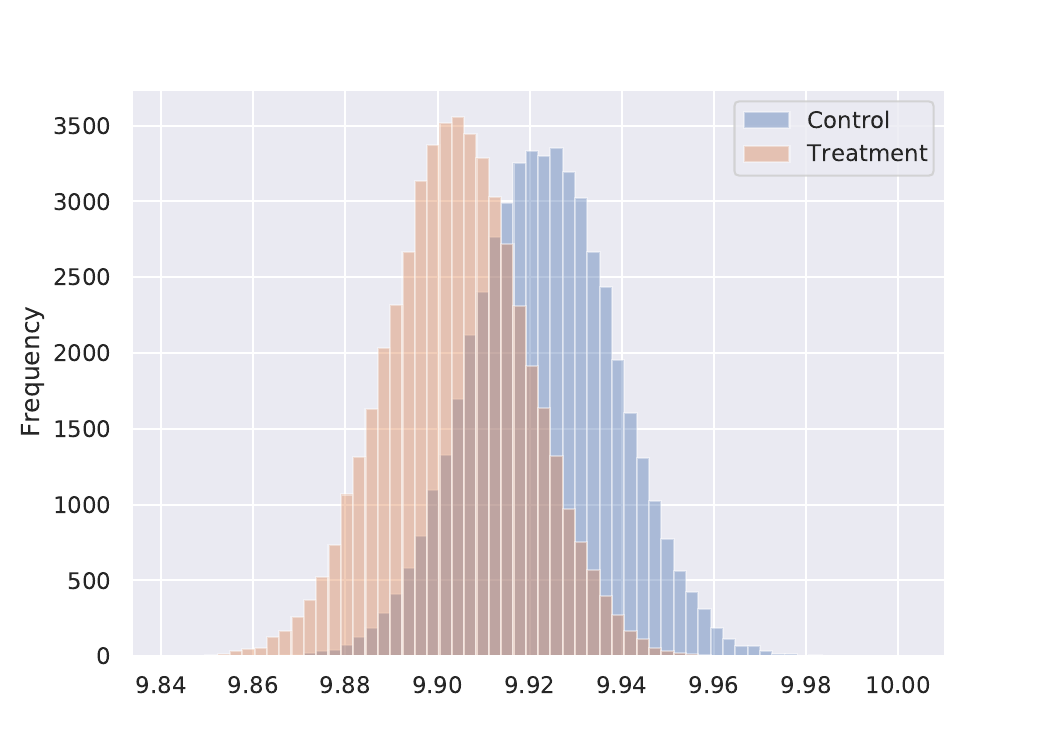}
  \end{subfigure}
  \begin{subfigure}[b]{0.46\linewidth}
    \includegraphics[width=\linewidth]{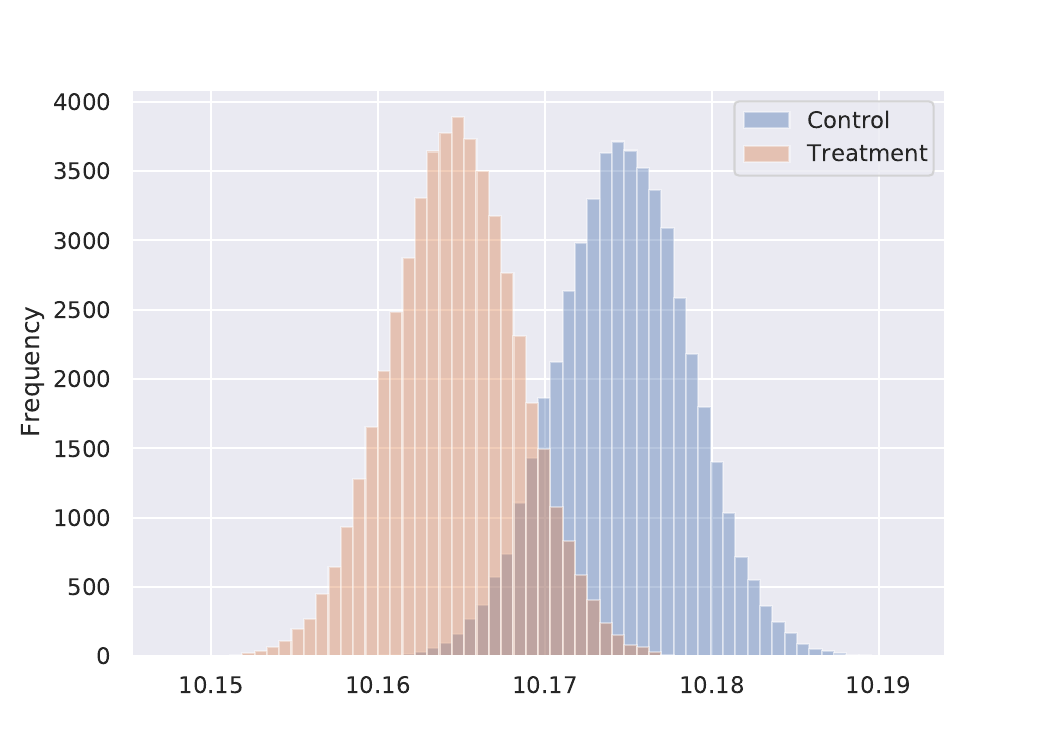}
  \end{subfigure}
  \begin{subfigure}[b]{0.46\linewidth}
    \includegraphics[width=\linewidth]{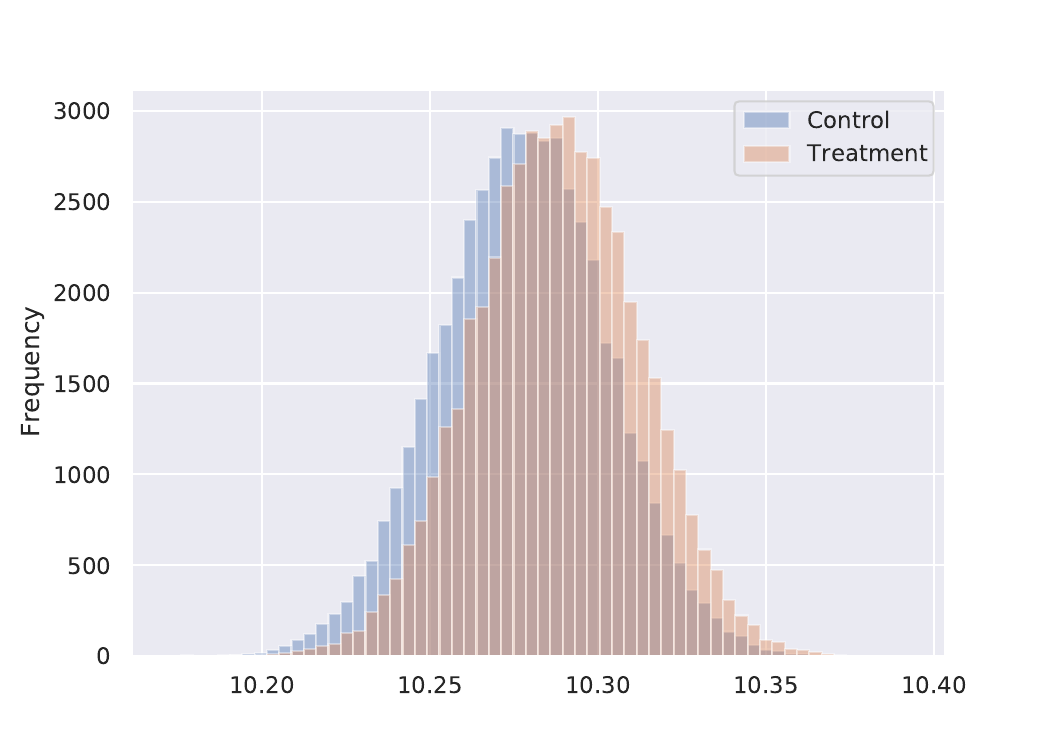}
  \end{subfigure}
  \begin{subfigure}[b]{0.46\linewidth}
    \includegraphics[width=\linewidth]{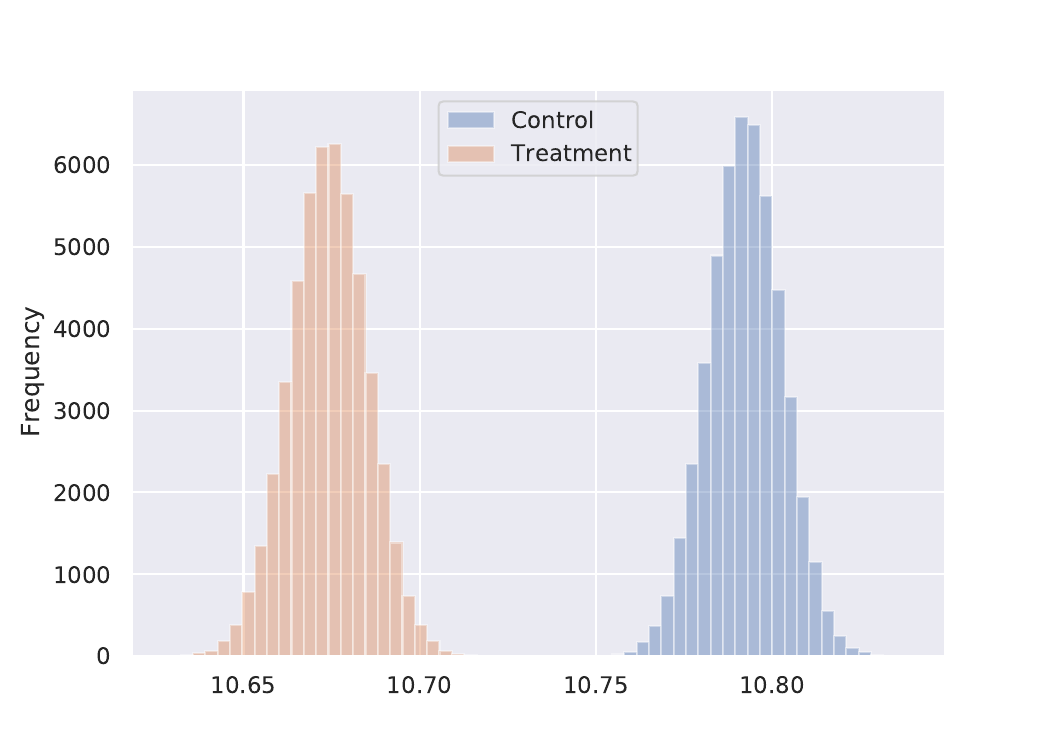}
  \end{subfigure}
  \vspace{-0.2cm}
  \caption{\datainterII Intervention (\BigCloneIIITB) for Global Performance: Bootstrapped Cross-Entropy. Top Left: \rnn ($JS=0.3$), Top Right: \gru ($JS=0.8$), Bot Left: \tf ($JS=0.6$), Bot Right: \tfi ($JS=1$).}
  \label{fig:jssimilarity}
  \vspace{-0.3cm}
\end{figure}

\textbf{Association Metrics.} For $[A-F]$ cases, we employed two methods to empirically estimate the association distributions $p(Y_g|T)$ and $p(Y_l|T)$. These two methods are the classic Pearson correlation and the Jensen-Shannon distance. For the latter, imagine we wish to understand the correlation between syntactic changes, \ie variable renaming, alterations in white space, \etc and a \nlms performance. One way we can study this is through computing the association of Cross-Entropy values $Y$ under two treatments, the first $T=0$, would be an unaltered code snippet and the second $T=1$ would be its Type III clone. Computing this association can be done using the Jensen Shannon distance $p(Y|T)\approx JS(Y^0,Y^1)$ as defined in Def.~\ref{def:js} for four models. Fig.~\ref{fig:jssimilarity} shows the distributions of $Y^0$ and $Y^1$ with their distances after applying bootstrapping as an example. 

\begin{definition}
\label{def:js}
\textbf{Jensen-Shannon Distance (JS).} The Jensen-Shannon divergence (JSD) overcomes the asymmetric computation of the KL divergence and provides a measure of the difference between distributions Eq.~\ref{eq:js_divergence}. The JS distance is the square of the JS divergence $p(Y|T)\approx JS(Y^{T=0},Y^{T=1}) = JSD(Y^{0},Y^{1})^2$. JS is proportional to the influence of $T$ on $Y$, which measure the separation of the distributions $Y^{0},Y^{1}$. The notation $Y^{T=0}$ refers to the potential outcomes observed under the treatment $T=0$. 

\begin{subequations}
    \begin{align}
     JSD(Y^{0}=y^0,Y^{1}=y^1) &= \label{eq:js_divergence-1}\\
     \frac{1}{2}\left[ D_{KL}\left(y^0||\frac{y^0+y^1}{2}\right) + D_{KL}\left(y^1||\frac{y^0+y^1}{2}\right)\right] &= \label{eq:js_divergence-2}
    \end{align}
\label{eq:js_divergence}
\end{subequations} 

\end{definition}

\textbf{Intervention Metrics.} Conversely, the interventional distributions $p(Y_g|do(T))$, $p(Y_l|do(T))$, and $p(Y_d|do(T))$ are estimated in terms of the Average Treatment Effect (ATE), as previously introduced in Eq~.\ref{eq:ate}. \docode can potentially estimate other types of causal effect metrics such as \textit{Conditional Average Treatment Effect (CATE)}.

\subsection{{Causal Validity}}

\textbf{{About Refutation Testing.}} After \docode estimates causal effects, these effects are validated for robustness. Ergo, \docode incorporates various refutation methods to evaluate the sensitivity of the causal estimations as we defined in \secref{sec:refuting}. We employ the following refutations: introducing a random common cause (\ie independent common causes added randomly should not impact the causal estimates), applying placebo treatments (\ie causal effect should go zero if the true treatment is replaced by an independent random variable), considering unobserved common causes (\ie causal estimates should not be too sensitive when we add an independent common cause correlated with the treatment and outcome), and performing data subset validations (\ie the causal effect should not change when the dataset is replaced by a random subset).

{\textbf{About Vetting Causal Graph.} We conducted exploratory analyses in \secref{sec:eda} that provide statistical evidence of the causal assumptions encoded in the graph. These analyses comprise measuring for correlations among the graph's variables. In addition, the \textit{testability} of each proposed causal graph occurs in the identification step (\secref{sec:identifying}). The identification step consists of finding an estimand for each proposed graph using the \textit{d-separation} criteria. These criteria help us to verify the fitness of the causal models in the datasets they generate. We can reject or accept the causal graph depending on the conditional independence tested from the data. All proposed SCMs in \figref{fig:causal_modes} were accepted.}

\subsection{Research Questions}

Inspired by the notion of Pearl's Ladder of Causation \cite{Pearl2016Causality,Sharma2021DoWhyAssumptions}, we framed two research questions to explore \docode's potential in enabling causal interpretability for \nlms. \texttt{\ref{rq:inference}} computes the feasibility of some SE-based Interventions and \texttt{\ref{rq:robustness}} performs a sensitivity analysis to validate the robustness of previously computed causal effects.

\begin{enumerate}[label=\textbf{RQ$_{\arabic*}$:}, ref=\textbf{RQ$_{\arabic*}$}, wide, labelindent=5pt]\setlength{\itemsep}{0.2em}
      \item \label{rq:inference} {\textbf{Causal Inference}: \textit{To what extent do SE interventions affect code prediction?}} 
      \item {\label{rq:robustness} {\textbf{Causal Validity}: \textit{How robust is the underlying causal graph under assumed SE interventions and code confounders?}}}
\end{enumerate}

\section{Results}\label{sec:results}
This section presents the results for both research questions per interpretability scenario introduced in \secref{sec:case_studies}.

{To answer \ref{rq:inference}, we begin by examining the correlations between potential outcomes and the interventions defined for each scenario. These correlations provide an early indication of how much the interventions might impact the outcomes. \tabref{tab:entropy_correlations} offers an overview of the correlation coefficients observed between the Cross-Entropy (\ie $Y_g$ \ref{out:cross_entropy}) and our proposed interventions. Additionally, \tabref{tab:semantic_correlations} details the Pearson correlations results for \ntp values (\ie $Y_l$ \ref{out:ntp}) for our \datainterII interventions, calculated with all the models we used in our experiments excluding \bert (used in scenario G).

Next, \docode generates causal explanations of code predictions for each scenario. Tables \ref{tab:results_A_global} and \ref{tab:results_B_global} display the computed causal effects for cross-entropy ($Y_g$). Similarly, Tables \ref{tab:resultsLocalJS} and \ref{tab:resultsLocalPearson} report both the causal inference metrics and correlation coefficients for each token category within $Y_l$.
\tabref{tab:resultsLocalPearsonMLM} presents both the correlation and causal effects computed for explaining the distance predictions (\ie $Y_d$ \ref{out:distance}) in scenario $G$.

We answer \ref{rq:robustness} by validating our causal effect estimates employing the four standard refutation methods outlined in Section \secref{sec:refuting}.
}

\begin{table*}[ht]
\centering
\caption{{\textit{Jensen-Shannon Dist.}, and \textit{Pearson Corr.} values obtained between Cross-Entropy $Y_g$ and Treatments in scenarios A,B,C,D and F } (\textbf{bold}:strong corr.)
\label{tab:entropy_correlations}
}
\scalebox{0.8}{%
\vspace{-0.2cm}
\begin{tabular}{lccccccclc}
\multicolumn{1}{c}{} &  &                       &  &              &  & \multicolumn{2}{c}{\datainterII}      &  &               \\
\multicolumn{1}{c}{} &  & \textit{\textbf{\datainterI} - $T_A$} &  & \textit{\textbf{\datainterIII} - $T_B$} &  & \textbf{{BigClone2TB} - $T_C$} & \textbf{{BigClone3TB} - $T_D$} &  & \textit{\textbf{\modelinterI} - $T_F$} \\
\nlm                  &  & \assoJS                &  & \assoJS       &  & \assoPR               & \assoPR        &  & \assoPR        \\ \cline{1-1} \cline{3-3} \cline{5-5} \cline{7-8} \cline{10-10} 
\rnn                  &  & \textbf{0.730$_{JS}$} &  & $0.180_{JS}$ &  & \textbf{0.45$_{PR}$} & $-0.056_{PR}$ &  & -             \\
\gru                  &  & $0.230_{JS}$          &  & $0.220_{JS}$ &  & $0.598_{PR}$         & $0.14_{PR}$   &  & $-0.093_{PR}$ \\
\tf                   &  & \textbf{0.670$_{JS}$} &  & $0.250_{JS}$ &  & $0.452_{PR}$         & $-0.14_{PR}$  &  & $-0.485_{PR}$ \\ \cline{1-1} \cline{3-3} \cline{5-5} \cline{7-8} \cline{10-10} 
\end{tabular}

}

\end{table*}

\begin{table}[ht]
\centering
\caption{{Average \textit{Pearson Corr.} of \datainterII interventions using all RNNs, GRUs, and GPTs.}
\label{tab:semantic_correlations}
}
\scalebox{0.9}{%
\vspace{-0.2cm}
\begin{tabular}{lllll}
                  &  & \multicolumn{3}{c}{\textbf{\datainterII}}                            \\
\textbf{\textit{Category}} &  & \textit{Clones Type II - $T_C$}        &  & \textit{Clones Type III - $T_D$}        \\ \cline{1-1} \cline{3-3} \cline{5-5} 
$[blocks]$            &  & 0.12 $\pm$ 0.039  &  & 0.214 $\pm$ 0.107  \\
$[exceptions]$        &  & 0.097 $\pm$ 0.068 &  & 0.036 $\pm$ 0.102  \\
$[top]$               &  & 0.388 $\pm$ 0.241 &  & 0.037 $\pm$ 0.023  \\
$[tests]$             &  & 1.0 $\pm$ 0.0     &  & nan $\pm$ nan      \\
$[declarations]$      &  & 0.192 $\pm$ 0.128 &  & 0.161 $\pm$ 0.135  \\
$[conditionals]$      &  & 0.078 $\pm$ 0.103 &  & 0.077 $\pm$ 0.144  \\
$[loops]$             &  & 0.183 $\pm$ 0.163 &  & 0.019 $\pm$ 0.094 \\
$[operators]$         &  & 0.316 $\pm$ 0.202 &  & 0.072 $\pm$ 0.05   \\
$[datatypes]$         &  & 0.016 $\pm$ 0.083 &  & -0.009 $\pm$ 0.064 \\
$[extra]$             &  & 0.163 $\pm$ 0.065 &  & 0.159 $\pm$ 0.14  
\end{tabular}

}

\end{table}

\begin{table*}[ht]
\centering
\caption{Causal Interventions $p(Y_g|do(T))$ of Cross-Entropy across models and datasets.\\(background:best effect)}
\label{tab:results_A_global}
\scalebox{0.74}{%

\begin{tabular}{llcccccccccc}
\multicolumn{1}{c}{\multirow{2}{*}{\textbf{\begin{tabular}[c]{@{}c@{}}\end{tabular}}}} &
   &
  \multicolumn{3}{c}{\multirow{2}{*}{\textbf{\begin{tabular}[c]{@{}c@{}}{\datainterI} - $T_A$\end{tabular}}}} &
  \multicolumn{1}{l}{} &
  \multicolumn{3}{c}{\multirow{2}{*}{\textbf{\begin{tabular}[c]{@{}c@{}}{\datainterIII} - $T_B$\end{tabular}}}} &
  \multicolumn{1}{l}{} &
  \multicolumn{2}{c}{\multirow{2}{*}{\textbf{\begin{tabular}[c]{@{}c@{}}{\modelinterI} - $T_E$\end{tabular}}}} \\
\multicolumn{1}{c}{} &
   &
  \multicolumn{3}{c}{} &
  \multicolumn{1}{l}{} &
  \multicolumn{3}{c}{} &
  \multicolumn{1}{l}{} &
  \multicolumn{2}{c}{} \\
\nlm &
   &
  \textit{\rnn} &
  \textit{\gru} &
  \textit{\tf} &
  \multicolumn{1}{l}{} &
  \textit{\rnn} &
  \textit{\gru} &
  \textit{\tf} &
  \multicolumn{1}{l}{} &
  \textit{\gru} &
  \textit{\tf} \\ \cline{1-1} \cline{3-5} \cline{7-9} \cline{11-12} 
\textit{Causal Eff. ATE} &
   &
  -0.0003 &
  -2.33E-05 &
  -0.0002 &
   &
  0.0023 &
  2.90E-05 &
  0.0026 &
   &
  -0.0058 &
  -0.0124 \\
\textit{Random Comm. Cause} &
   &
  -0.0003 &
  -2.45E-05 &
  -0.0002 &
   &
  0.0011 &
  -0.0004 &
  0.0015 &
   &
  -0.0058 &
  -0.0124 \\
\textit{Unobserved Comm. Cause} &
   &
  -0.0003 &
  1.54E-05 &
  -0.0001 &
   &
  0.0002 &
  -0.0001 &
  0.0007 &
   &
  -0.0050 &
  -0.0108 \\
\textit{Placebo} &
   &
  0.0001 &
  1.44E-05 &
  0.0001 &
   &
  0.0006 &
  -1.33E-05 &
  0.0006 &
   &
  -0.0001 &
  -2.77E-05 \\
\textit{Remove Subset} &
   &
  -0.0003 &
  -3.68E-05 &
  -0.0002 &
   &
  0.0012 &
  -0.0003 &
  0.0016 &
   &
  -0.0058 &
  -0.0124 \\ \cline{1-1} \cline{3-5} \cline{7-9} \cline{11-12} 
\end{tabular}
}
\end{table*}

\begin{table*}[]
\small
\centering
\caption{Causal Interventions $p(Y_g|do(T))$ of Cross-Entropy across Clone datasets.\\(background:best effect)}
\label{tab:results_B_global}
\scalebox{0.67}{

\begin{tabular}{llccccccccccccccc}
 &
   &
  \multicolumn{15}{c}{\textit{\textbf{\datainterII}}} \\
\multicolumn{1}{c}{\textbf{\begin{tabular}[c]{@{}c@{}}\end{tabular}}} &
   &
  \multicolumn{7}{c}{Backdoor Criterion} &
  \multicolumn{1}{l}{} &
  \multicolumn{7}{c}{Instrumental Variable $I_{[countSubWords]}$} \\
 &
   &
  \multicolumn{3}{c}{\textit{\textbf{BigClone2TB} - $T_C$}} &
   &
  \multicolumn{3}{c}{\textit{\textbf{BigClone3TB} - $T_D$}} &
   &
  \multicolumn{3}{c}{\textit{\textbf{BigClone2TB} - $T_C$}} &
   &
  \multicolumn{3}{c}{\textit{\textbf{BigClone3TB} - $T_D$}} \\
\nlm &
   &
  \textit{\rnn} &
  \textit{\gru} &
  \textit{\tf} &
   &
  \textit{\rnn} &
  \textit{\gru} &
  \textit{\tf} &
   &
  \rnn &
  \gru &
  \tf &
   &
  \rnn &
  \gru &
  \tf \\ \cline{1-1} \cline{3-5} \cline{7-9} \cline{11-13} \cline{15-17} 
\textit{Causal Eff. ATE} &
   &
  \cellcolor[HTML]{EFEFEF}0.6288 &
  \cellcolor[HTML]{EFEFEF}0.8713 &
  \cellcolor[HTML]{EFEFEF}0.5635 &
   &
  -0.1042 &
  0.1085 &
  -0.2739 &
   &
  0.00076 &
  0.0018 &
  0.0017 &
   &
  -0.18 &
  \cellcolor[HTML]{EFEFEF}0.72 &
  -0.25 \\
\textit{Random Comm. Cause} &
   &
  0.6297 &
  0.8720 &
  0.5651 &
   &
  -0.1043 &
  0.1084 &
  -0.2741 &
   &
  0.00076 &
  0.0018 &
  0.0017 &
   &
  -0.18 &
  0.72 &
  -0.25 \\
\textit{Unobserved Comm. Cause} &
   &
  0.2950 &
  0.4257 &
  0.2737 &
   &
  -0.0800 &
  0.0830 &
  -0.2168 &
   &
  0.00077 &
  0.0018 &
  0.0018 &
   &
  -0.18 &
  0.72 &
  -0.26 \\ \cline{1-1} \cline{3-5} \cline{7-9} \cline{11-13} \cline{15-17} 
\end{tabular}

}
\end{table*}

\begin{table*}[ht]
\footnotesize
\centering
\caption{\ntp Association Results $p(Y_l|T)$ are \textit{Jensen-Shannon Dist}. Causal Effects are ATEs $p(Y_l|do(T))$ \\(\textbf{bold}:strong corr., background:best effect)}
\vspace{-0.2cm}
\label{tab:resultsLocalJS}

\scalebox{0.74}{%

\begin{tabular}{llccccccccccccccc}
\multicolumn{1}{c}{\textbf{\begin{tabular}[c]{@{}c@{}}\end{tabular}}} &
   &
  \multicolumn{7}{c}{\textbf{\begin{tabular}[c]{@{}c@{}}{\datainterI} - $T_{A}$\end{tabular}}} &
  \multicolumn{1}{l}{} &
  \multicolumn{7}{c}{\textbf{\begin{tabular}[c]{@{}c@{}}{\datainterIII} - $T_{B}$\end{tabular}}} \\
\nlm &
   &
  \multicolumn{3}{c}{\textit{\gru}} &
  \multicolumn{1}{l}{} &
  \multicolumn{3}{c}{\textit{\tf}} &
  \multicolumn{1}{l}{} &
  \multicolumn{3}{c}{\textit{\gru}} &
  \multicolumn{1}{l}{} &
  \multicolumn{3}{c}{\textit{\tf}} \\
\textbf{Categories} &
   &
  \textit{\begin{tabular}[c]{@{}c@{}}Association\\ \assoJS\end{tabular}} &
  \textit{\begin{tabular}[c]{@{}c@{}}Causal Eff.\\ ATE\end{tabular}} &
  \textit{R1} &
   &
  \textit{\begin{tabular}[c]{@{}c@{}}Association\\ \assoJS\end{tabular}} &
  \textit{\begin{tabular}[c]{@{}c@{}}Causal Eff.\\ ATE\end{tabular}} &
  \textit{R1} &
   &
  \textit{\begin{tabular}[c]{@{}c@{}}Association\\ \assoJS\end{tabular}} &
  \textit{\begin{tabular}[c]{@{}c@{}}Causal Eff.\\ ATE\end{tabular}} &
  \textit{R1} &
   &
  \textit{\begin{tabular}[c]{@{}c@{}}Association\\ \assoJS\end{tabular}} &
  \textit{\begin{tabular}[c]{@{}c@{}}Causal Eff.\\ ATE\end{tabular}} &
  \textit{R1} \\ \cline{1-1} \cline{3-5} \cline{7-9} \cline{11-13} \cline{15-17} 
$[blocks]$ &
   &
  0.374 &
  \cellcolor[HTML]{EFEFEF}-0.0005 &
  -0.00052 &
   &
  0.794 &
  \cellcolor[HTML]{EFEFEF}-0.0001 &
  -0.000115 &
   &
  \textbf{0.867} &
  \cellcolor[HTML]{EFEFEF}0.0003 &
  0.000488 &
   &
  \textbf{0.948} &
  -0.0006 &
  -0.000352 \\
$[exceptions]$ &
   &
  \textbf{0.893} &
  -1.00E-06 &
  1.00E-06 &
   &
  0.349 &
  \cellcolor[HTML]{EFEFEF}-1.20E-05 &
  -1.10E-05 &
   &
  \textbf{0.835} &
  -4.80E-05 &
  -4.80E-05 &
   &
  \textbf{0.909} &
  \cellcolor[HTML]{EFEFEF}-1.20E-05 &
  -4.20E-05 \\
$[oop]$ &
   &
  \textbf{0.952} &
  \cellcolor[HTML]{EFEFEF}2.00E-05 &
  1.20E-05 &
   &
  \textbf{0.942} &
  \cellcolor[HTML]{EFEFEF}-1.30E-05 &
  -9.00E-06 &
   &
  \textbf{0.930} &
  -7.00E-06 &
  -4.70E-05 &
   &
  \textbf{0.910} &
  \cellcolor[HTML]{EFEFEF}-4.90E-05 &
  -2.60E-05 \\
$[tests]$ &
   &
  0.419 &
  9.00E-06 &
  8.00E-06 &
   &
  0.003 &
  1.00E-05 &
  9.00E-06 &
   &
  0.405 &
  1.90E-05 &
  4.00E-05 &
   &
  0.276 &
  \cellcolor[HTML]{EFEFEF}-1.70E-05 &
  2.00E-06 \\
$[declarations]$ &
   &
  \textbf{0.929} &
  4.00E-06 &
  4.00E-06 &
   &
  \textbf{0.946} &
  1.00E-06 &
  1.00E-06 &
   &
  \textbf{0.939} &
  -2.10E-05 &
  -0.000145 &
   &
  \textbf{0.852} &
  \cellcolor[HTML]{EFEFEF}4.60E-05 &
  6.00E-05 \\
$[conditionals]$ &
   &
  0.655 &
  \cellcolor[HTML]{EFEFEF}-3.90E-05 &
  -3.90E-05 &
   &
  \textbf{0.824} &
  -8.00E-06 &
  -1.00E-05 &
   &
  \textbf{0.889} &
  8.00E-06 &
  7.40E-05 &
   &
  0.179 &
  \cellcolor[HTML]{EFEFEF}-0.000596 &
  -0.000603 \\
$[loops]$ &
   &
  \textbf{0.870} &
  -2.00E-06 &
  -3.00E-06 &
   &
  0.677 &
  \cellcolor[HTML]{EFEFEF}2.10E-05 &
  1.90E-05 &
   &
  \textbf{0.973} &
  -1.70E-05 &
  1.50E-05 &
   &
  \textbf{0.915} &
  1.80E-05 &
  3.00E-06 \\
$[operators]$ &
   &
  \textbf{0.877} &
  -6.00E-06 &
  -5.00E-06 &
   &
  \textbf{0.961} &
  1.50E-05 &
  1.10E-05 &
   &
  0.722 &
  0.0002 &
  0.000345 &
   &
  0.002 &
  \cellcolor[HTML]{EFEFEF}0.0079 &
  0.008991 \\
$[datatypes]$ &
   &
  0.747 &
  -1.00E-05 &
  -1.00E-05 &
   &
  \textbf{0.832} &
  9.00E-06 &
  1.00E-05 &
   &
  0.251 &
  \cellcolor[HTML]{EFEFEF}0.0003 &
  0.000328 &
   &
  0.568 &
  0.0002 &
  0.000111 \\
$[extra]$ &
   &
  \textbf{0.831} &
  2.60E-05 &
  2.00E-05 &
   &
  \textbf{0.879} &
  5.70E-05 &
  6.00E-05 &
   &
  0.632 &
  0.0014 &
  0.001016 &
   &
  0.474 &
  0.0024 &
  0.002004 \\ \cline{1-1} \cline{3-5} \cline{7-9} \cline{11-13} \cline{15-17} 
\end{tabular}

}
\vspace{-0.1cm}

\end{table*}

\begin{table*}[ht]
\footnotesize
\centering
\caption{\ntp Association Results $p(Y_l|T)$ are \textit{Pearson Corr}. Causal Effects are ATEs $p(Y_l|do(T))$. {We use $-$ to indicate undetermined causal effects due to an unfeasible linear model.}(\textbf{bold}:strong corr., background:best effect)
}
\vspace{-0.2cm}
\label{tab:resultsLocalPearson}

\scalebox{0.74}{%

\begin{tabular}{llccclccccccclccc}
\textbf{\begin{tabular}[c]{@{}l@{}}\end{tabular}} &
   &
  \multicolumn{7}{c}{\textbf{\begin{tabular}[c]{@{}c@{}}{\datainterII} - $T_D$\end{tabular}}} &
  \multicolumn{1}{l}{} &
  \multicolumn{7}{c}{\textbf{\begin{tabular}[c]{@{}c@{}}{\modelinterI} - $T_E$\end{tabular}}} \\
\nlm &
   &
  \multicolumn{3}{c}{\textit{\gru}} &
   &
  \multicolumn{3}{c}{\textit{\tf}} &
   &
  \multicolumn{3}{c}{\textit{\gru}} &
   &
  \multicolumn{3}{c}{\textit{\tf}} \\
\textbf{Categories} &
   &
  \textit{\begin{tabular}[c]{@{}c@{}}Association\\ \assoPR\end{tabular}} &
  \textit{\begin{tabular}[c]{@{}c@{}}Causal Eff.\\ ATE\end{tabular}} &
  \textit{R1} &
   &
  \textit{\begin{tabular}[c]{@{}c@{}}Association\\ \assoPR\end{tabular}} &
  \textit{\begin{tabular}[c]{@{}c@{}}Causal Eff.\\ ATE\end{tabular}} &
  \textit{R1} &
   &
  \textit{\begin{tabular}[c]{@{}c@{}}Association\\ \assoPR\end{tabular}} &
  \textit{\begin{tabular}[c]{@{}c@{}}Causal Eff.\\ ATE\end{tabular}} &
  \textit{R1} &
   &
  \textit{\begin{tabular}[c]{@{}c@{}}Association\\ \assoPR\end{tabular}} &
  \textit{\begin{tabular}[c]{@{}c@{}}Causal Eff.\\ ATE\end{tabular}} &
  \textit{R1} \\ \cline{1-1} \cline{3-5} \cline{7-9} \cline{11-13} \cline{15-17} 
$[blocks]$ &
   &
  0.026 &
  -1.50E-05 &
  -1.50E-05 &
   &
  0.186 &
  -2.80E-05 &
  -2.80E-05 &
   &
  {\color[HTML]{333333} -0.102} &
  {\color[HTML]{333333} -0.010559} &
  {\color[HTML]{333333} -0.010559} &
   &
  {\color[HTML]{333333} \textbf{0.725}} &
  \cellcolor[HTML]{EFEFEF}{\color[HTML]{333333} 0.018004} &
  {\color[HTML]{333333} 0.018004} \\
$[exceptions]$ &
   &
  0.017 &
  - &
  - &
   &
  0.002 &
  - &
  - &
   &
  {\color[HTML]{333333} -0.070} &
  {\color[HTML]{333333} -} &
  {\color[HTML]{333333} -} &
   &
  {\color[HTML]{333333} 0.349} &
  {\color[HTML]{333333} -} &
  {\color[HTML]{333333} -} \\
$[oop]$ &
   &
  0.049 &
  - &
  - &
   &
  0.012 &
  - &
  - &
   &
  {\color[HTML]{333333} 0.019} &
  {\color[HTML]{333333} -} &
  {\color[HTML]{333333} -} &
   &
  {\color[HTML]{333333} 0.255} &
  {\color[HTML]{333333} -} &
  {\color[HTML]{333333} -} \\
$[tests]$ &
   &
  - &
  - &
  - &
   &
  - &
  - &
  - &
   &
  {\color[HTML]{333333} -0.130} &
  {\color[HTML]{333333} -} &
  {\color[HTML]{333333} -} &
   &
  {\color[HTML]{333333} 0.174} &
  {\color[HTML]{333333} -} &
  {\color[HTML]{333333} -} \\
$[declarations]$ &
   &
  0.375 &
  - &
  - &
   &
  0.034 &
  - &
  - &
   &
  {\color[HTML]{333333} -0.257} &
  {\color[HTML]{333333} -} &
  {\color[HTML]{333333} -} &
   &
  {\color[HTML]{333333} 0.405} &
  {\color[HTML]{333333} -} &
  {\color[HTML]{333333} -} \\
$[conditionals]$ &
   &
  0.274 &
  - &
  - &
   &
  -0.087 &
  - &
  - &
   &
  {\color[HTML]{333333} -0.009} &
  {\color[HTML]{333333} -} &
  {\color[HTML]{333333} -} &
   &
  {\color[HTML]{333333} \textbf{0.682}} &
  {\color[HTML]{333333} -} &
  {\color[HTML]{333333} -} \\
$[loops]$ &
   &
  0.024 &
  - &
  - &
   &
  0.111 &
  - &
  - &
   &
  {\color[HTML]{333333} 0.042} &
  {\color[HTML]{333333} -} &
  {\color[HTML]{333333} -} &
   &
  {\color[HTML]{333333} 0.275} &
  {\color[HTML]{333333} -} &
  {\color[HTML]{333333} -} \\
$[operators]$ &
   &
  0.099 &
  - &
  - &
   &
  0.062 &
  - &
  - &
   &
  {\color[HTML]{333333} -0.032} &
  {\color[HTML]{333333} -} &
  {\color[HTML]{333333} -} &
   &
  {\color[HTML]{333333} 0.389} &
  {\color[HTML]{333333} -} &
  {\color[HTML]{333333} -} \\
$[datatypes]$ &
   &
  0.037 &
  - &
  - &
   &
  -0.069 &
  - &
  - &
   &
  {\color[HTML]{333333} 0.002} &
  {\color[HTML]{333333} -} &
  {\color[HTML]{333333} -} &
   &
  {\color[HTML]{333333} 0.275} &
  {\color[HTML]{333333} -} &
  {\color[HTML]{333333} -} \\
$[extra]$ &
   &
  0.192 &
  -3.00E-05 &
  -3.00E-05 &
   &
  -0.017 &
  5.40E-05 &
  5.40E-05 &
   &
  {\color[HTML]{333333} 0.092} &
  \cellcolor[HTML]{EFEFEF}{\color[HTML]{333333} 0.014588} &
  {\color[HTML]{333333} 0.014588} &
   &
  {\color[HTML]{333333} \textbf{0.606}} &
  {\color[HTML]{333333} 0.014525} &
  {\color[HTML]{333333} 0.014525} \\ \cline{1-1} \cline{3-5} \cline{7-9} \cline{11-13} \cline{15-17} 
\end{tabular}

}
\vspace{-0.1cm}

\end{table*}

\begin{table*}[ht]
\centering
\caption{Masking AST Node Types. Distance Association Results $p(Y_d|T_{G})$ are \textit{Pearson Corr} and Causal Effects are ATEs $p(Y_d|do(T_{G}))$. (\textbf{bold}:strong corr., background:best effect)
}
\vspace{-0.2cm}
\label{tab:resultsLocalPearsonMLM}

\scalebox{0.74}{%

\begin{tabular}{llccccccccccc}
\textbf{\begin{tabular}[c]{@{}c@{}}\end{tabular}} &
   &
  \multicolumn{11}{c}{\textbf{\datainterIV} -  $T_G$} \\
\textit{Distance metric $Y_d$} &
   &
  \multicolumn{3}{c}{\textit{\jaccard}} &
  \multicolumn{1}{l}{} &
  \multicolumn{3}{c}{\levenshtein} &
  \multicolumn{1}{l}{} &
  \multicolumn{3}{c}{\soren} \\
\textbf{Categories} &
   &
  \textit{\begin{tabular}[c]{@{}c@{}}Association\\ assoPR\end{tabular}} &
  \textit{\begin{tabular}[c]{@{}c@{}}Causal Eff.\\ ATE\end{tabular}} &
  \textit{R4} &
  \multicolumn{1}{l}{} &
  \textit{\begin{tabular}[c]{@{}c@{}}Association\\ assoPR\end{tabular}} &
  \textit{\begin{tabular}[c]{@{}c@{}}Causal Eff.\\ ATE\end{tabular}} &
  \textit{R4} &
  \multicolumn{1}{l}{} &
  \textit{\begin{tabular}[c]{@{}c@{}}Association\\ assoPR\end{tabular}} &
  \textit{\begin{tabular}[c]{@{}c@{}}Causal Eff.\\ ATE\end{tabular}} &
  \textit{R4} \\ \cline{1-1} \cline{3-5} \cline{7-9} \cline{11-13} 
$[boolean\_operator]$ &
   &
  \textbf{-0.3294} &
  \cellcolor[HTML]{EFEFEF}-0.1508 &
  0.0052 &
   &
  -0.2845 &
  \cellcolor[HTML]{EFEFEF}-0.1363 &
  0.0147 &
   &
  -0.2809 &
  -0.0947 &
  0.0009 \\
$[comparison\_operator]$ &
   &
  -0.2648 &
  -0.0272 &
  -0.0066 &
   &
  -0.2071 &
  -0.0183 &
  -0.0031 &
   &
  -0.2228 &
  -0.0154 &
  -0.0019 \\
$[for\_in\_clause]$ &
   &
  \textbf{-0.4232} &
  -0.0532 &
  -0.0004 &
   &
  \textbf{-0.3724} &
  -0.0449 &
  0.0013 &
   &
  \textbf{-0.3568} &
  -0.0291 &
  0.0002 \\
$[for\_statement]$ &
   &
  \textbf{-0.4006} &
  -0.1011 &
  0.0341 &
   &
  -0.2525 &
  -0.0411 &
  -0.0124 &
   &
  \textbf{-0.3949} &
  -0.0832 &
  -0.0037 \\
$[identifier]$ &
   &
  0.0024 &
  -0.0752 &
  0.0230 &
   &
  -0.0375 &
  -0.0734 &
  -0.0017 &
   &
  0.0243 &
  -0.0387 &
  0.0068 \\
$[if\_clause]$ &
   &
  \textbf{-0.3904} &
  -0.0407 &
  0.0002 &
   &
  \textbf{-0.3666} &
  -0.0365 &
  5.9143 &
   &
  \textbf{-0.3395} &
  -0.0220 &
  0.0011 \\
$[if\_statement]$ &
   &
  \textbf{-0.3667} &
  \cellcolor[HTML]{EFEFEF}-0.1211 &
  -0.0328 &
   &
  -0.2409 &
  -0.0933 &
  -0.0130 &
   &
  \textbf{-0.3624} &
  -0.0954 &
  -0.0131 \\
$[parameters]$ &
   &
  \textbf{-0.3287} &
  -0.0481 &
  0.0071 &
   &
  \textbf{-0.3230} &
  -0.0457 &
  -0.0104 &
   &
  -0.2951 &
  -0.0294 &
  0.0032 \\
$[return\_statement]$ &
   &
  -0.2450 &
  \cellcolor[HTML]{EFEFEF}-0.1211 &
  0.0062 &
   &
  -0.2250 &
  \cellcolor[HTML]{EFEFEF}-0.1127 &
  -0.0158 &
   &
  -0.2261 &
  -0.0756 &
  -0.0083 \\
$[string]$ &
   &
  \textbf{-0.3651} &
  \cellcolor[HTML]{EFEFEF}-0.1683 &
  -0.0041 &
   &
  -0.2961 &
  \cellcolor[HTML]{EFEFEF}-0.1450 &
  0.0274 &
   &
  \textbf{-0.3199} &
  \cellcolor[HTML]{EFEFEF}-0.1161 &
  0.0036 \\
$[while\_statement]$ &
   &
  -0.2677 &
  -0.0964 &
  -0.0197 &
   &
  -0.1383 &
  -0.0091 &
  -0.0027 &
   &
  -0.2979 &
  0.0113 &
  0.9199 \\ \cline{1-1} \cline{3-5} \cline{7-9} \cline{11-13} 
\end{tabular}

}
\vspace{-0.1cm}

\end{table*}

\subsection{Interpretability Scenario $A$: BuggyCode}\label{sec:results_scenario_a}

\textbf{To what extent does \datainterI affect the predictions of \nlms?}
For the $A$ case, focusing on the \datainterI intervention, both \rnn and \tf models exhibit considerably high JS distances of $0.73$ and $0.67$, indicating a strong correlation between this intervention and the cross-entropy: $P(Y_g|T_{A})$. Additionally, we observed a similar pattern in \ntp outcomes for $Y_{l\blocks}$ with the \gru model, which had the highest correlation at $0.626$ for \datainterI.

However, when we controlled for SE covariates, the correlations initially observed turned out to be misleading for causation. Across the three models (\rnn, \gru, and \tf), the ATEs were very small, both in cross-entropy $p(Y_g|do(T_{A}))$ with values of $-3E-4$, $-2.33E-05$, and $-2E-4$, and in \ntp outcomes $p(Y_{l\blocks}|do(T_{A}))$ with a value of $-5.25E-4$. Having null effects after observing high correlations confirms the presence of confounding bias for this case. Therefore, the prediction performance is being affected by other confounders different from the actual buggy code intervention.

\textbf{How robust are the causal effect estimations?}
To verify the sensitivity of the ATEs for $Y_g$, we computed four robustness tests: $\mathcal{R}_1$, $\mathcal{R}_2$, $\mathcal{R}_3$, and $\mathcal{R}_4$. The results for $\mathcal{R}_1$, $\mathcal{R}_2$, and $\mathcal{R}_4$ aligned closely with the ATEs we found for the three models. Furthermore, the results for $\mathcal{R}_3$ were nearly zero. This indicates that the estimated causal effects were robust. Similarly, $\mathcal{R}_1$ values for $Y_l$ results were close to the obtained ATEs.

\begin{boxK}
\textbf{BuggyCode Finding $F_A$:} The presence or absence of buggy code does not appear to causally influence (or explain) the prediction performance of our \nlms even under measured high correlation.
\end{boxK}

\subsection{Interpretability Scenario $B$: Code Documentation}
\textbf{To what extent does \datainterIII (or inline comments) intervention affect the predictions of \nlms?}
In the \datainterIII interventions, there were no strong correlations in cross-entropy $P(Y_{g}|T_{B})$ for the \rnn, \gru, and \tf models, with values of $0.18$, $0.22$, and $0.25$ respectively. However, the \ntp outcomes showed significant associations in specific categories: for \gru, a strong correlation in \datatype ($P(Y_{l\datatype}|T_{B})=0.749$), and for \tf, notable correlations in \operators ($P(Y_{l\operators}|T_{B})=0.998$), \conditionals ($P(Y_{l\conditionals}|T_{B})=0.821$), and \tests ($P(Y_{l\tests}|T_{B})=0.724$).

Once we adjusted for covariates, the ATEs in both $p(Y_g|do(T_{B}))$ and $p(Y_l|do(T_{B}))$ were close to zero, showing a trend towards no causal effects. This suggests that actions like removing comments from code have little to no causal impact on cross-entropy or \ntp values.

\textbf{How robust are the causal effect estimations?}The obtained results across refutation tests indicated unstable causal effects. Specifically, the outcomes for $\mathcal{R}_1$, $\mathcal{R}_2$, and $\mathcal{R}_4$ did not align closely with the ATEs. This discrepancy could stem from several factors: 1) a lack of sufficient data samples, 2) inaccuracies in our causal diagram assumptions (\eg confounders, instrumental variables, and effect modifiers), and/or 3) the treatment is inadequate.

\begin{boxK}
\textbf{Code Documentation Finding $F_B$:} Despite observing strong correlations between the removal of comments and \ntp, we cannot causally interpret code predictions from inline comments since measured ATEs are not robust after refutations. This suggests that other hidden confounders are influencing the estimation of this causal effect.
\end{boxK}

\subsection{Interpretability Scenario $C$: Clones Type II}
\textbf{To what extent does \datainterII intervention (Clones Type II) affect the predictions of \nlms?}
In the \datainterII intervention for Type II clones, we observed a positive correlation between Levenshtein ``edit'' distance and the difference of cross-entropy values (\ie $p(\Delta Y_g|T_{C})$) across \rnn, \gru, and \tf. The correlation coefficients were $0.45$, $0.6$, and $0.452$ for each model respectively, as detailed in Table~\ref{tab:results_B_global}.

{Conversely, we found weak correlations in \ntp scores across all \nlms (excluding \bert) in the nine categories analyzed. For instance, $Y_l$ was correlated with \oop tokens and \operators with $p(Y_{l\oop}|T_{C})=0.388\pm0.241$ and $p(Y_{l\operators}|T_{C})=0.316\pm0.202$, as detailed in \tabref{tab:semantic_correlations}. The high standard deviation observed across the \nlms in these categories indicates a wide variance in what the models statistically learned.}

{The computed causal effects on $\Delta Y_g$, using backdoor criterion, showed an exceptionally high influence of clones Type II on the cross-entropy, with values of $0.63$, $0.87$, and $0.56$ for \rnn, \gru, and \tf. On the other hand, when we used $I_{[countSubWords]}$ as an instrumental variable for estimand identification, the observed causal effects were substantially lower (see \ref{tab:results_B_global}).}

{Since most of the ATEs for $Y_l$ could not be computed, using backdoor criterion or the instrumental variable, we omitted showcasing these results in \tabref{tab:resultsLocalPearson}. Formulating continuous interventions was hindered by the non-linear distribution of the data points. This can be explained due to the nature of Clone Types II, which oftentimes, maintains the same functionality while introducing minimal alterations in the structure of the source code. This led to a lower variability in the computed Levenshtein distances.}

\textbf{How robust are the causal effect estimations?} For testing the robustness of the ATEs, we could not compute the refutation methods  \textit{Placebo} and \textit{Remove Subset} due to limited data. However, the other two methods, \textit{Random Comm. Cause} and \textit{Unobserved Comm. Cause}, showed stability, reinforcing our confidence in the ATEs for global results.

\begin{boxK}
\textbf{Clones Type II Finding $F_C$:} The presence of Clones Type II impacts (or causally explains) the cross-entropy on \rnn, \gru, and \tf, as obtained correlations and causal effects suggest an influence of some interventions in identifiers, literals, types, white spaces, layout, and comments.
\end{boxK}

\subsection{Interpretability Scenario $D$: Clones Type III}
\textbf{To what extent does \datainterII intervention (Clones Type III) affect the predictions of \nlms?}
By contrast, our analysis showed a different pattern for Type III clones. We detected no strong correlations between the Levenshtein ``edit'' distance and the cross-entropy (\ie $p(\Delta Y_g|T_{D})$) in any of our \nlms. This is evident from the marginal correlation values for the three models—\rnn, \gru, and \tf: $-0.056$, $0.14$, and $-0.14$, respectively. {Additionally, we found even lower correlation values in \ntp scores (\ie $P(Y_l|T_{D})$) across all \nlms (excluding \bert), as detailed in \tabref{tab:semantic_correlations}. The highest coefficient is observed for \extra tokens, with a value of $p(Y_{l\extra}|T_{C})=0.159\pm0.14$.}

In our models, the causal effect on cross-entropy using the back-door criterion was generally small but showed interesting variations. For the \tf model, this effect was negative, with $p(Y_g|do(T_{D}))\approx-0.274$, whereas, for the \gru model, it was positive at $p(Y_g|do(T_{D}))\approx0.11$. {We observe a tendency to null causal effects when using $I_{[countsubWords]}$ as instrumental variable to identity the causal estimand.}

{Additionally, the causal effect on \ntp outcomes for \blocks was negligible $p(Y_{l\blocks}|do(T_{D}))=2.80E-05$. Unfortunately, most of the ATEs could not be computed for each keyword category since it was unfeasible to fit a linear model due to the shape of the data (\ie grouped predictions by syntax categories have similar values).} 

\textbf{How robust are the causal effect estimations?}
Similar to the findings with clone type II, testing the robustness of the ATEs presented challenges. Due to limited data, we were unable to apply $\mathcal{R}_3$ and $\mathcal{R}_4$. However, the use of the other two methods, $\mathcal{R}_1$ and $\mathcal{R}_2$, demonstrated stability in our results.

\begin{boxK}
\textbf{Clones Type III Finding $F_D$:} The presence of Clones Type III does not consistently impact (or causally explain) the cross-entropy and \ntp, as revealed by negative and positive ATEs obtained across the \nlms under analysis.
\end{boxK}

\subsection{Interpretability Scenario $E$: Layers}
\textbf{To what extent does \modelinterI intervention affect the predictions of \nlms?}
We observed that \modelinterI interventions are negatively correlated with cross-entropy. This trend is evident in the values of $p(Y_g|T_{E})$ for \gru and \tf models: $-0.093$ and $-0.485$ respectively. This suggests that as the number of layers increases, the cross-entropy tends to decrease. Conversely, we found strong correlations for \ntp outcomes across keyword-based categories ($p(Y_l|T_{E})$): the categories \blocks, \conditionals, and \extra showed notably high correlations, with values of $0.725$, $0.682$, and $0.606$ respectively.

Nonetheless, we detected confounding bias in almost all categories of our intervention analysis for \ntp outcomes. This was indicated by the extremely low ATEs, such as $p(Y_{\blocks}|do(T_{E}))=0.018$ and $p(Y_{l\extra}|do(T_{E}))=0.0145$. A plausible explanation for this phenomenon can be traced to SE metric confounders. For instance, the size of the methods appears to have a significant influence on both cross entropy and \ntp results across the \nlms.

\textbf{How robust are the causal effect estimations?}
To verify the sensitivity of the ATEs for $Y_g$, we estimated $\mathcal{R}_1$, $\mathcal{R}_2$, $\mathcal{R}_3$, and $\mathcal{R}_4$. The results for $\mathcal{R}_1$, $\mathcal{R}_2$, and $\mathcal{R}_4$ aligned closely with the ATEs we found for \gru and \tf. Furthermore, $\mathcal{R}_3$ results were nearly zero. Thus, the estimated causal effects were robust. Similarly, $\mathcal{R}_1$ values for $Y_l$ results were close to the obtained ATEs, reinforcing the robustness of our findings.

\begin{boxK}
\textbf{Layers Finding $F_E$:} Although it is observed strong correlations between the number of layers and \ntp, which might suggest a causal interpretation of code predictions, the reported ATEs close to zero demonstrate the presence of confounding bias.
\end{boxK}

\subsection{Interpretability Scenario $F$: Units}
\textbf{To what extent does \modelinterII intervention affect the predictions of \nlms?}

In a similar vein, \modelinterII interventions tend to be negatively correlated with the cross-entropy. The number of units showed a negative effect with the \gru model, with $p(Y_g|T_{F})\approx-0.084$.

Likewise, we found negative values for $P(Y_l|T_{F})$, suggesting that increasing the number of units negatively impacts the \ntp outcomes across all keyword-based categories. However, these values were relatively low, making it challenging to draw definitive conclusions. For example, the highest negative correlation observed was $P(Y_{l\exceptions}|T_{F})\approx -0.253$ for \ntp.

Negative correlations observed for the cross-entropy were consistent with the corresponding negative ATEs estimated. However, the negative correlation observed for \blocks was not in fact causal: $P(Y_{l\blocks}|do(T_{F}))\approx \text{-5E-06}$. Similarly, for the other categories, the ATEs were so minimal that they can be considered as causal null effects.

{We omit the results of \modelinterII intervention from Tables \ref{tab:results_A_global}, \ref{tab:results_B_global}, \ref{tab:resultsLocalJS} and \ref{tab:resultsLocalPearson}, since the causal effect estimates for $Y_g$ using \rnn and \tf, as well as most $Y_l$ categories, were either null or could not be determined due to the shape of the data (see online Appendix in \cite{icodegen}).}

\textbf{How robust are the causal effect estimations?} 
Just as with \modelinterI, we conducted robustness tests $\mathcal{R}_1$, $\mathcal{R}_2$, $\mathcal{R}_3$, and $\mathcal{R}_4$ for $Y_g$. The outcomes of $\mathcal{R}_1$, $\mathcal{R}_2$, and $\mathcal{R}_4$ closely matched the ATEs we observed in the \gru model. Furthermore, the $\mathcal{R}_3$ values were nearly zero, reinforcing the robustness of the estimated causal effects. Similarly, the $\mathcal{R}_1$ values for $Y_l$ are also closely aligned with the corresponding ATEs, further confirming the robustness of the results.

\begin{boxK}
\textbf{Units Finding $F_F$:} Intervening the number of units tends to be negatively correlated with the cross-entropy; in fact, measured causal effects on \ntp are null suggesting that this intervention does not explain code predictions.
\end{boxK}

\subsection{Interpretability Scenario $G$: Masking AST Nodes}

\textbf{To what extent does \datainterIV intervention affect the predictions of \nlms?}
We computed the distance outcomes $Y_d$ for some AST node types, as detailed in \secref{sec:grammar_aggregation}. However, in \tabref{tab:resultsLocalPearsonMLM}, we specifically focus on a subset of both terminal and non-terminal nodes. This subset was chosen for its familiarity with developers and includes elements such as conditional statements, identifiers, and repetition statements. The intervention demonstrated strong negative correlations between AST node types and the distance outcomes $Y_d$, consistent across all distance metrics. For example, $P(Y_{d\jaccard}|T_{G\forinclause})=-0.4232$, $P(Y_{d\levenshtein}|T_{G\forinclause})=-0.3724$, and $P(Y_{d\soren}|T_{G\forinclause})=-0.3568$.

The correlation results are further supported by the estimated ATEs $P(Y_d|do(T_{G\node}))$, demonstrating no confounding bias. Although the causal effects are lower than the estimated correlations, they reveal the presence of negative causation.

\textbf{How robust are the causal effect estimations?}
We calculated $\mathcal{R}_1$, and we obtained the same values for ATEs indicating that the causal effects are robust. Furthermore, we also calculated  $\mathcal{R}_4$, and the values were close to zero.

\begin{boxK}
\textbf{Masking AST Nodes Finding $F_G$:} The intervention of masking random tokens has more impact on code predictions than masking grammar-based categories. This suggests that \bert does not entirely capture the nodes' information of Abstract Syntax Trees (ASTs).
\end{boxK}

\section{Exploratory Causal Analyses}\label{sec:eda}

{This section introduces supplementary analyses to validate the assumptions made during the formulation of the SCMs in our interpretability scenarios, thereby facilitating the \textit{causal discovery} of our graphs. We exhaustively explored the datasets of interventions to support the encoding process from domain knowledge. We assess the validity of the causal graph encoding by exploring correlations among SE confounders (see \secref{sec:confounders}), potential outcomes, and interventions. In other words, by finding statistical dependencies among models' variables, we provide sufficient evidence for creating SCMs. 

It is worth clarifying that we employ the term \textit{covariates} to refer to a general set of Software Engineering metrics that describe code. Conversely, the term \textit{confounders} refers to a proven subset of covariates that influence treatments and outcomes. Therefore, some covariates might or might not be confounders. Because this exploration is usually conducted before confounder identification, preliminary correlational analyses are performed with a rudimentary list of SE covariates. 

Moreover, we introduce a descriptive \textit{error analysis} of the syntax clustering categories in \secref{sec:aggregations}. Specifically, we measured the average prediction by keyword-based aggregations for each Neural Code Model baseline. The error analysis aims to explore the feasibility of the proposed clustering function, which is fundamental for defining interpretable outputs across the \docode pipeline. }

\subsection{Statistical dependencies between confounders and potential outcomes  \textcircled{\scriptsize $Z$} $\rightarrow$ \textcircled{\scriptsize $Y$}}
\begin{table*}[ht]
\centering
\footnotesize
\caption{{Most correlated covariates influencing cross-entropy results (\ie $Y_g$) and \ntp scores (\ie $Y_l$) across \nlms and datasets (\textbf{bold}:strong corr.)}
\label{tab:yg_covariates}
}
\scalebox{0.79}{%
\vspace{-0.2cm}
\begin{tabular}{lllccclcccc}
\textbf{} &
  \textbf{} &
   &
  \multicolumn{3}{c}{\textit{Cross Entropy Loss}} &
   &
  \multicolumn{4}{c}{\textit{Next Token Prediction}} \\
\multirow{2}{*}{\textbf{Dataset}} &
  \multirow{2}{*}{\textbf{Model}} &
   &
  \textbf{\begin{tabular}[c]{@{}c@{}}Correlation with\\ {[}z\_count\_subwords{]}\end{tabular}} &
  \multicolumn{2}{c}{\textbf{\begin{tabular}[c]{@{}c@{}}Max correlation excluding\\ {[}z\_count\_subwords{]}\end{tabular}}} &
   &
  \multicolumn{2}{c}{\textbf{\begin{tabular}[c]{@{}c@{}}Max Correlation with\\ {[}z\_count\_subwords{]}\end{tabular}}} &
  \multicolumn{2}{c}{\textbf{\begin{tabular}[c]{@{}c@{}}Max correlation excluding\\ {[}z\_count\_subwords{]}\end{tabular}}} \\
 &
   &
   &
  $\rho$ &
  Covariate $[Z]$ &
  $\rho$ &
   &
  Potential Outcome $[Y_l]$ &
  $\rho$ &
  \textbf{Potential Outcome $[Y_l]$, Covariate $[Z]$} &
  $\rho$ \\ \cline{1-2} \cline{4-6} \cline{8-11} 
\BuggyTB &
  \rnn &
   &
  \textbf{0.87} &
  $Z_{[loc]}$ &
  0.25 &
   &
  $Y_{[loops]}$ &
  0.1 &
  $Y_{[blocks]}$, $Z_{[loc]}$ &
  -0.417 \\
\BuggyTB &
  \gru &
   &
  0.36 &
  $Z_{[parenthesizedExpsQty]}$ &
  -0.47 &
   &
  $Y_{[blocks]}$ &
  -0.072 &
  $Y_{[blocks]}$, $Z_{[parenthesizedExpsQty]}$ &
  \textbf{0.682} \\
\BuggyTB &
  \tf &
   &
  \textbf{0.88} &
  $Z_{[loc]}$ &
  0.27 &
   &
  $Y_{[tests]}$ &
  0.065 &
  $Y_{[blocks]}$, $Z_{[parenthesizedExpsQty]}$ &
  0.476 \\
\CommentsTB &
  \rnn &
   &
  0.41 &
  $Z_{[uniqueWordsQty]}$ &
  \textbf{0.54} &
   &
  $Y_{[tests]}$ &
  0.197 &
  $Y_{[tests]}$, $Z_{[uniqueWordsQty]}$ &
  0.3 \\
\CommentsTB &
  \gru &
   &
  0.15 &
  $Z_{[maxNestedBlocksQty]}$ &
  0.35 &
   &
  $Y_{[operators]}$ &
  0.082 &
  $Y_{[blocks]}$, $Z_{[parenthesizedExpsQty]}$ &
  0.152 \\
\CommentsTB &
  \tf &
   &
  0.39 &
  $Z_{[uniqueWordsQty]}$ &
  0.52 &
   &
  $Y_{[extraTokens]}$ &
  0.059 &
  $Y_{[declarations]}$, $Z_{staticMethodsQty}$ &
  -0.245 \\ \cline{1-2} \cline{4-6} \cline{8-11} 
\end{tabular}

}

\end{table*}

{
The accuracy of \nlms is significantly influenced by the context window size as these architectures rely heavily on contextual information to predict the next token \cite{dong_survey_2023}. Motivated by the previous premise, we found a strong correlation between the number of words $Z_{[countSubwords]}$ and the Cross-Entropy ($\rho = 0.87$) for the \rnn using the \BuggyTB dataset. A similar trend was observed using \tf with a correlation value of $\rho = 0.88$. For other architectures with different configurations (see \tabref{tab:yg_covariates}), the correlation still exists below $0.5$, indicating that while there is some evidence of dependencies, the impact of other potential covariates could be stronger. For instance, in the case of the \CommentsTB dataset, when using both \rnn and \tf, the most significant covariate impacting performance was the number of unique words $Z_{[\text{uniqueWordsQty}]}$, with correlation coefficients of $\rho = 0.54$ and $\rho = 0.52$, respectively.

More broadly, \tabref{tab:yg_covariates} showcases the Pearson correlation values ($\rho$) after exploring the dependencies among proposed covariates (see \secref{sec:confounders}), the \BuggyTB and \CommentsTB. These testbeds were used for the \datainterI intervention in scenario $A$ and the \datainterII intervention in scenario $B$.}

{Similarly, we computed correlation coefficients between our keyword-based categories and potential covariates. \tabref{tab:yg_covariates} reveals that the \ntp scores grouped by categories are not significantly correlated with $Z_{[countSubwords]}$, as evidenced by the low maximum coefficient values. Conversely, $Z_{[paranthesizedExpsQty]}$ seems to be appreciably correlated with the prediction of $[blocks]$ in the \BuggyTB dataset using \gru ($\rho = 0.682$), and \tf ($\rho = 0.476$). These findings seem to correspond with the grammar rules of Java since the parenthesized expressions are often used within blocks of code.}

\begin{boxK}
{\textbf{Finding \textcircled{\scriptsize $Z$} $\rightarrow$ \textcircled{\scriptsize $Y$} }: The statistical dependency between covariates and outcomes varies across different \nlms. The Cross-Entropy is highly correlated to the number of subwords for the \BuggyTB dataset. In contrast, the \CommentsTB dataset exhibits lower correlation values. Additionally, \ntp scores for $blocks$ tokens show a strong dependency on the number of parenthesized expressions for both binary treatments. }
\end{boxK}

\subsection{Statistical dependencies between confounders and interventions \textcircled{\scriptsize $Z$} $\rightarrow$ \textcircled{\scriptsize $T$}}

{In this analysis, we delve into statistical dependencies between covariates $Z$ and interventions. Specifically, \datainterI, \datainterIII, and \datainterII (Clones Type II and III) interventions. We assess and compare the distributions of covariates $Z$ across the treatment and control groups within each dataset using the Jensen-Shannon distance. Measured divergence and distance values do not exhibit substantial evidence of a significant separation between the number of tokens ($Z_{[countSubwords]}$) in the control and treatment distributions. Furthermore, all covariate distributions follow a similar tendency of lower distances between treatment and control groups (see \tabref{tab:distance_n_tokens}).
}

\begin{table}[ht]
\centering
\caption{{Differences between $Z_{\text{[countSubwords]}}$ Treatment and Control groups across multiple datasets using Jensen-Shannon divergences and distances.}
\label{tab:distance_n_tokens}
}
\scalebox{0.9}{%
\vspace{-0.2cm}
\begin{tabular}{llcc}
\textbf{}     &  & \multicolumn{2}{c}{\textit{Control vs Treatment}} \\
\textit{\textbf{Dataset}} &  & \multicolumn{1}{l}{\textit{\textbf{$\Delta$JS Divergence}}} & \multicolumn{1}{l}{\textit{\textbf{$\Delta$JS distance}}} \\ \cline{1-1} \cline{3-4} 
\CommentsTB    &  & 0.005                    & 0.07                   \\
\BuggyTB       &  & 0.004                    & 0.06                   \\
\BigCloneIITB  &  & 0.004                    & 0.06                   \\
\BigCloneIIITB &  & 0.002                    & 0.04                   \\ \cline{1-1} \cline{3-4} 
\end{tabular}

}

\end{table}

\begin{boxK}
{\textbf{Findings \textcircled{\scriptsize $Z$} $\rightarrow$ \textcircled{\scriptsize $T$}}: A subtle separation between control and treatment distributions exists when segregating by any SE covariate $Z$.
}
\end{boxK}

\subsection{Code Syntax Clustering Error Analysis}
\label{sec:stx_val}

\begin{figure*}
  \centering
  \includegraphics[width=1\linewidth]{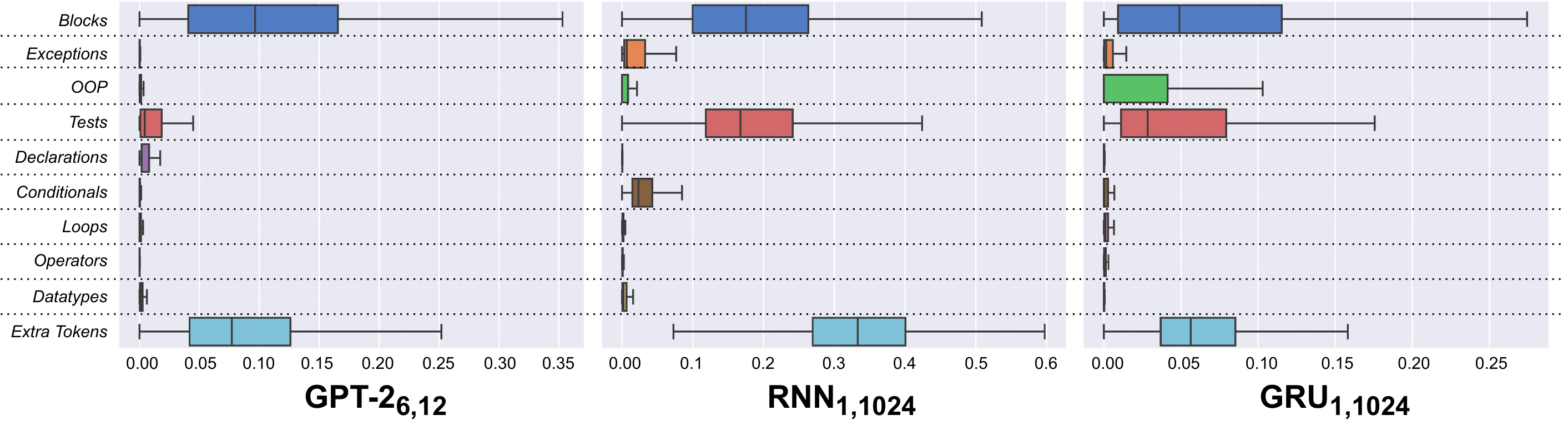}
  \caption{{Descriptive Error Analysis. Keyword-Based Categories probabilities (Normalized $Y_l$ - \ntp) for \tf \rnn and \gru on the \BuggyTB dataset. Higher values indicate less erroneous predictions.}}
  \vspace{-0.5cm}
  \label{fig:bug_rnn_tax_ntp}
\end{figure*}

\begin{figure*}
  \centering
  \includegraphics[width=1\linewidth]{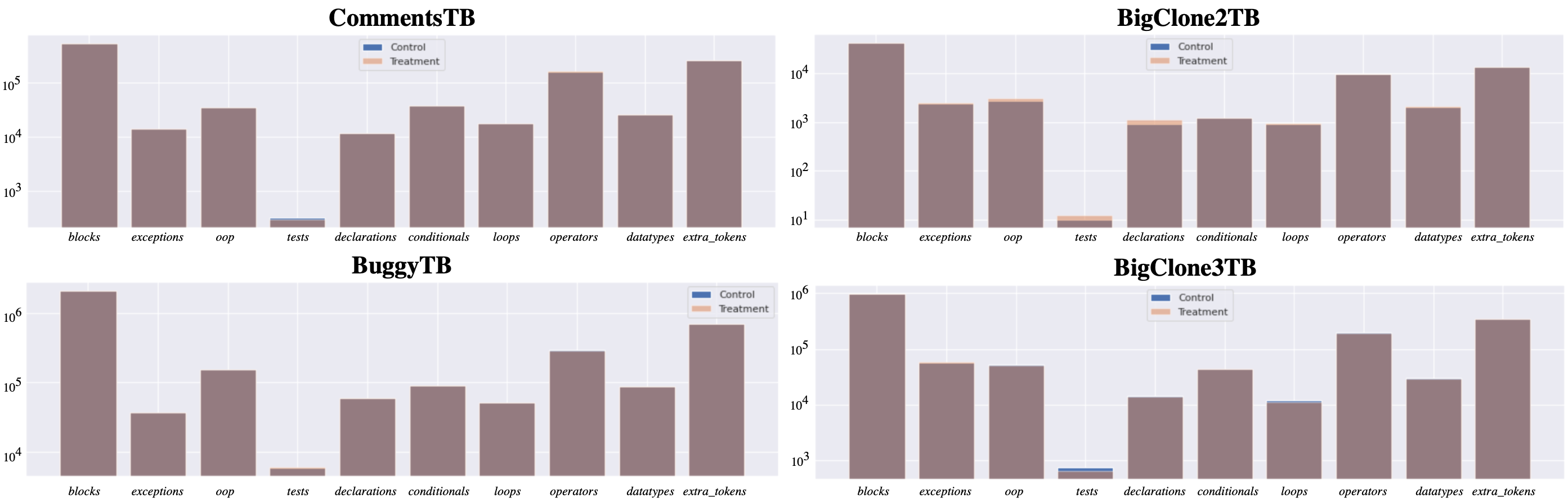}
  \caption{{Frequency analysis of keyword-based categories by \nlm.}}
  \vspace{-0.5cm}
  \label{fig:categories_frequency}
\end{figure*}

{
We conducted an exploratory analysis to evaluate and describe the behavior of \nlms predictions grouped by keyword-based categories. These categories are detailed in \figref{fig:syntactic_aggregations}. Specifically, we computed the normalized \ntp values of tokens for each sample in the dataset \BuggyTB, used in the \datainterI intervention for scenario $A$. Subsequently, we clustered token-predicted probabilities for each of the categories. The exploration aims to unveil the average prediction \textit{error} for a set of outcomes $Y_l$ for three \nlms: \tf, \rnn, and \gru. An \textit{error} is any keyword-based category whose average predicted probability is lower than $0.5$.

This exploration reveals that each keyword-based category follows a similar error tendency regardless of the \nlm type as illustrated in \figref{fig:bug_rnn_tax_ntp}. For example, the $[blocks]$ category in \tf exhibited the highest median value at $0.09$ followed by $[extraTokens]$ and $[tests]$ with a value of $0.06$ and $0.01$ respectively. Conversely, $[operators]$ has the lowest median value, and the interquartile range (\textit{iqr}) for $[extraTokens]$ is narrower than $[blocks]$. Similar distributions are observed in \rnn and \gru, where $[blocks]$, $[tests]$ and $[extraTokens]$ are among the categories with better prediction values. Moreover, a peculiar behavior is visible in the $[OOP]$ category for \gru, where the \ntp scores are slightly higher, as indicated by the third quartile approaching $0.05$. However, for half of the data points, the values remain close to the median, which is nearly zero, as highlighted by the second quartile.

More broadly, the highest normalized \ntp value observed among the categories falls below $0.5$, which suggests an absence of empirical evidence that supports a significant \textit{statistical understanding} related to these categories. Nonetheless, our error analysis entails a statistical technique demonstrating the feasibility of measuring code predictions by syntax categories.  

We also investigated the distribution of tokens by category within each Java dataset associated with $T_{[data]}$ interventions: \BuggyTB, \CommentsTB, \BigCloneIITB, and \BigCloneIIITB. Our findings, depicted in \figref{fig:categories_frequency}, reveal that tokens categorized under $[blocks]$ are the most frequent across four datasets. On the other hand, tokens associated with $[tests]$ are consistently the least frequent.
}

\begin{boxK}
{\textbf{Error Analysis Finding}: \tf, \rnn and \gru produce erroneous $Y_l$ predictions (\ie normalized \ntp $<0.5$) for all the syntax categories in the \BuggyTB dataset.}
\end{boxK}

\section{Discussion}
\label{sec:discussion}

The ultimate goal of interpretability in Deep Learning for Software Engineering is to generate explanations of \nlms' predictions. We achieved this goal by formulating causal interpretations of such predictions, quantified through performance values (\eg Cross-Entropy, Next Token Prediction, or Distance Metrics). {We demonstrate that regardless of conventional evaluations of \nlms (\eg BLEU, Accuracy, Perplexity), producing causal interpretations is vital for model understanding. For example, even if a model demonstrates a deficient BLEU score when its outputs are evaluated against the ground truth for a downstream task, we can still generate causal explanations of those outputs by defining an appropriate SCM.}

Our interpretability method, \docode, generates causal explanations (or interpretations) based on the idea of estimating \textit{the effect of \textbf{interventions}}. These interventions take place on the data inputs or parameters that configure \nlms. Furthermore, estimating their causal effects is grounded in Pearl's theory of causation \cite{Pearl2009Causality, Pearl2016Causality, Pearl2018Causality}. 

Below, we pose three aspects of the discussion: 1) some general insights from the case study, 2) a brief explanation of the guidelines to use \docode in practice, and 3) a list of challenges and opportunities practitioners might face when adapting \docode for their analyses.
 
\subsection{Insights From The Case Study}

In \docode, we examine model semantics by categorizing code tokens into different conceptual categories and examining both raw model performance according to these groups, and the causal relationships of these categories across treatments. Syntax can be measured by how often the model predicts a given token that is syntactically correct. While we did not directly report syntactic correctness, in our observations, our studied models rarely made syntactic errors. However, the local prediction performance of tokens across different semantic categories tends to vary quite a bit. This is partially explained by the prevalence of different token categories in the training set, but further work on model architectures that improve the performance of underperforming token groups could help to advance research on \nlms.

For each \nlm, a strong correlation across testbeds for a given code token category means that we observe the prediction performance to have \textit{changed} significantly across the SE-based treatments (\eg commented/uncommented code). This change could correspond to prediction performance for these token categories increasing or decreasing. However, because our approach uses a causal graph (\ie Structural Causal Model), we can determine the amount of causal effect between the correlated variables based on covariates, and determine whether or not they are \textbf{spurious correlations}, due to \textit{confounding bias} Fig.~\ref{fig:covariate}, or true causal relationships caused by the change in treatment. Based on our experiments, NCMs seem to have a stronger \textit{statistical understanding} of syntax and a more limited understanding of semantics, according to our definition of semantics. However, once we started exploring Masking Language Models such as BERT-like code models, we found the opposite: AST node tokens are not better predicted than a random and, therefore, unstructured set of tokens.

As the results suggest, our \nlms under study learn to predict tokens related to code blocks (\eg brackets, parentheses, semicolons) more effectively than most other code token types (\eg loops, conditionals, datatypes), we found that our \nlms are \textit{sensitive to seemingly subtle changes in code syntax}, reinforcing previous studies concluding the same~\cite{rabin2021generalizability}, and our models are only \textit{marginally impacted} by the presence of comments and bugs, which challenges findings from previous work~\cite{Baishakhi2016buggy}. Consequently, practitioners can identify which \textit{categories} of code tokens were important in the model’s decision-making process. In other words, \docode supports the identification of which 1) tokens, 2) layers, or 3) hyperparameters are impacting code predictions, which we describe below:

\textbf{Tokens.} In \secref{sec:aggregations}, we propose a \textit{syntax clustering} function for grouping the \nlms' code predictions into more understandable syntax categories (\ie keyword-based and AST/Grammar-based). Although the syntax clustering formalism was originally omitted from the \docode pipeline, we realized during the experimentation phase that a clustering strategy is vital as practitioners must make sense of generated subwords or fine-grained levels of code.

Our syntax clustering formalism stems from the need to make fine-grained code predictions more understandable to practitioners. As such, our approach enables determining which token categories a model can learn (or not) across different application settings (\eg SE-based interventions). {For example, given the normalized \ntp values computed for scenario \textit{A}, as depicted in \figref{fig:bug_rnn_tax_ntp}, one could conclude that block-category tokens are more effectively predicted by \tf, \rnn and \gru. Yet such \nlms typically struggled to predict operator-category tokens.} More importantly, we can observe causal relationships related to the importance of code tokens in decision-making \textit{changes} across treatments (\ie data interventions or model parameters alterations).

\textbf{Layers.} Our method supports interventions on the \textit{number of layers} for a given architecture. Therefore, \docode can identify if the layers, as a hyperparameter, influence the performance. Nonetheless, identifying specific layers to influence the performance would require going beyond the causal interpretability scope and exploring the inner mechanisms of the neural network \cite{molnar2019interpret}.

\textbf{Hyperparemeters.} $do_{code}$ supports the identification of hyperparameters that influence neural network performance by extending SE-based intervention definitions at the model level (see \secref{sec:seintervention}). Our case study supports two types of parameters: \textit{NumberOfLayers} and \textit{NumberOfUnits}.

{\textbf{About causal graph validity.} In our exploratory analyses, we conducted experiments accounting for other types of variables in the SCM beyond treatments, outcomes, and confounders. For example, we collected information on software-based effect modifiers (\ie variables affecting only outputs) and instrumental variables (\ie variables affecting only treatments). Particularly, for semantic preserving treatments, we made \textit{subword counting} an instrumental variable instead of a confounder. We found that these graph configurations were not robust under refutation methods. However, further research is required to construct proper tools to identify instrumental variables and modifiers.}

{\textbf{About Causal Transportability.} Although we concentrated on datasets and keyword-based categories for Java (\ie scenarios $A-F$) in our exploratory analyses, the assumptions we made to build the SCMs still hold for scenario $G$. This \textit{transportability} of assumptions across similar domains have been researched in Causal Inference \cite{pearl_transportability_nodate}. As such, the causal information learned from our experiments can be reused in analogous settings if there is homogeneity in the effect modifiers. Transportability allows us to maintain our assumptions across scenarios, as long as the underlying structure of the causal graph remains consistent, keeping the same type of treatments, potential outcomes, and confounding variables.

}

\subsection{\docode in practice}

While it may appear that \nlms have begun to achieve promising performance, it is insufficient to \textit{only} generate code predictions (\ie the \textit{\textbf{what}} of \nlms' decision). This current status quo, at best, provides an incomplete picture of the limitations and caveats of \nlms for code. Given the potential impact and consequence of these models and their resulting applications, there is a clear need to strive for a more complete understanding of how they function in practice. As such, we must push to understand how \nlms arrive at their predictions (\ie the \textit{\textbf{why}} of \nlms' decision) as shown in our proposed Structural Causal Models for each scenario. With the design and results of interpretability scenarios, we demonstrate that \docode comprises a causal explanation method that aims to make Deep Learning for Software Engineering \nlms, and their decision-making process, understandable to practitioners. 

\begin{figure*}[ht]
		\centering
		\includegraphics[width=0.99\textwidth]{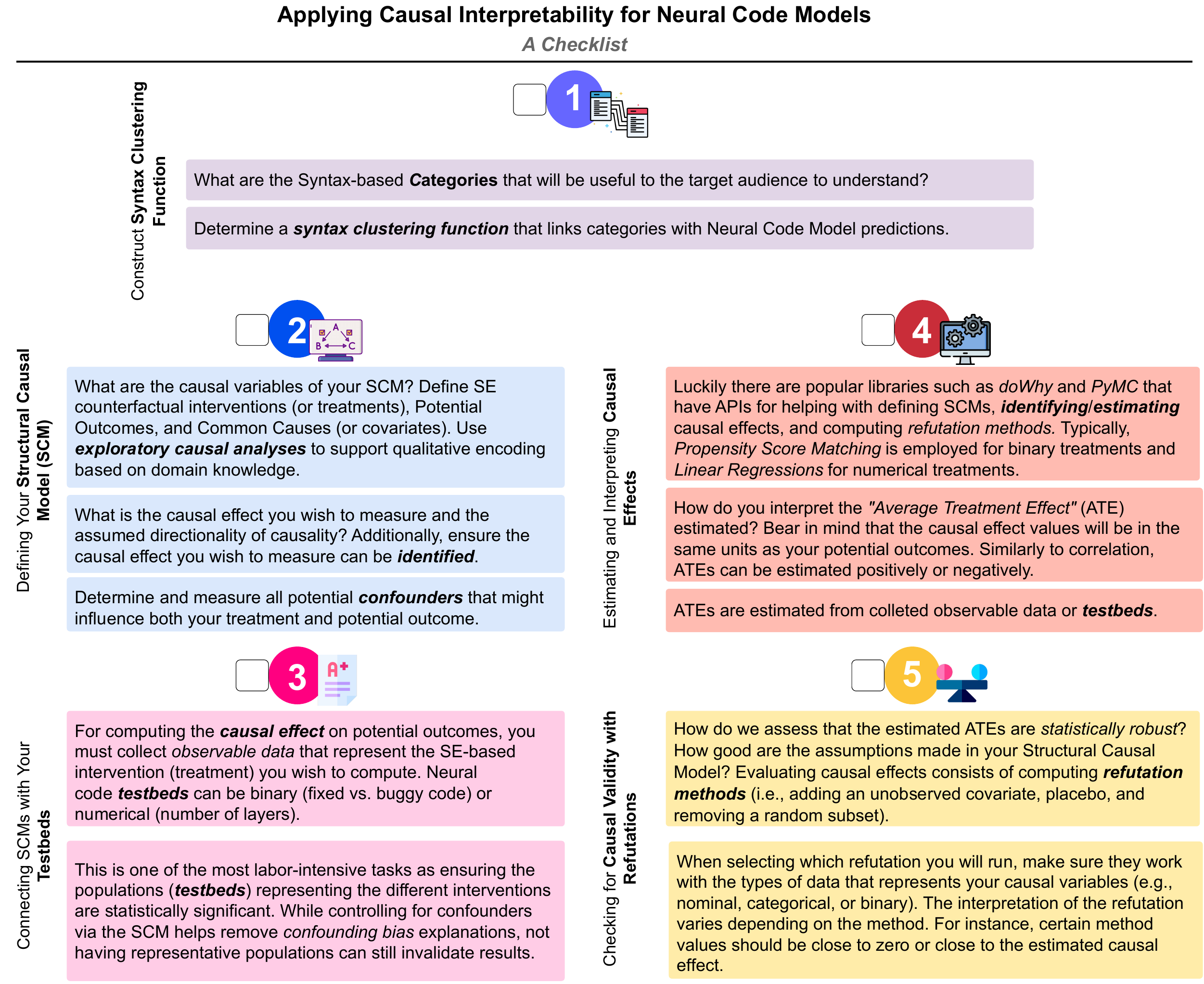}
		\caption{ Five Guidelines for Applying Causal Interpretability to Neural Code Models.} 
        \vspace{-0.5cm}
        \label{fig:checklist}
\end{figure*}

To that end, we have created a checklist that summarizes the general process researchers can use to apply causal interpretability to Neural Code Models \ref{fig:checklist}. For instance, one of the practical applications of \docode is facilitating the debugging process of \nlms. By debugging a \nlm we refer to the tasks of reducing the amount of \textit{confounding bias} between SE-interventions and code predictions. Our proposed guidelines below help to design a pipeline in which \docode facilitates the detection of this bias for a given setting.

As shown in \figref{fig:checklist}, the proposed checklist comprises five guidelines, which correspond to both the approach pipeline (see \secref{sec:overview}) and the definition of a syntax clustering function (see \secref{sec:aggregations}). In the first guideline, a researcher determines what would be understandable to the target audience and constructs a syntax clustering function. This clustering function translates the model's fine-grained code predictions to the target audience. Once this clustering is built, the researcher can move on to the second guideline that resolves around defining a Structural Causal Model (\scm). It is important to have this step after defining the clustering function so the interventions, outcomes, and confounders can be properly modeled. With the \scm at hand, in the third guideline, the researcher can now collect the data holding the observable SE-based interventions (\eg buggy vs. fixed code), used to estimate the causal effect of the treatment. Then, in the fourth guideline, existing popular libraries (\eg doWhy) can be used to estimate the causal effect. Lastly, in the fifth guideline, the researcher must check the original assumptions (\ie the confounders and \scm) using refutation methods. With this checklist, we hope to ease the complexity around causal analysis for researchers.

\subsection{Challenges \& Future Work}
In this section, we list some challenges $CHs$ that practitioners might face when adapting our method to their interpretability analyses.

\textbf{$CH_1$: Proposing a new syntax clustering criterion.} Our proposed categories, designed to group fine-grained code predictions into more understandable categories, are based entirely on definitions and guidelines extracted from Programming Languages (\ie Java and Python). This might represent a limitation if a practitioner requires a more in-depth analysis of the syntax interactions. \docode introduces a clustering formalism that can be extended and reformulated to specific interpretability needs. Different clustering functions give rise to different challenges. For instance, one BPE token may simultaneously belong to multiple categories and span across different AST nodes. Future research needs to address the following questions: How do practitioners deal with this overlapping? Do they enforce a singular classification for the BPE token, in a similar way to our keyword-based clustering? or do they allow for some statistically controlled overlaps as we did in AST/Grammar-based clustering?

\textbf{$CH_2$: Collecting data for formulating SE-based interventions.} \docode was not implemented to perform \textit{causal discovery}, unveiling the causes of code predictions directly from observations. In fact, \docode is restricted to estimate causal inference measures (\eg correlations, ATEs, CATEs) based on hypotheses and assumptions, from domain knowledge, embedded in Structural Causal Models. Therefore, users of \docode must define and formulate specific SE-based interventions depending on the available data, and also provide the possible confounding factors that can alter the effect on code predictions. Future work can be oriented to facilitate the definition of the Structural Causal Model into a more automatic approach. Such an automatic approach should include recent techniques on causal discovery.

\textbf{$CH_3$: Integrating \docode in Deep Learning for Software Engineering life-cycle.} To use \docode in practical settings, we must propose strategies for integrating interpretability approaches into Software Engineering pipelines that incorporate Deep Learning models. This will not only improve the reliability of the system but also facilitate the monitoring of critical information.

{\textbf{$CH_4$: Creating the Structural Causal Model.} The quality of the causal explanations generated by \docode heavily relies on the structure of the causal graph. Consequently, employing \docode in practical scenarios would require statistical evidence supporting each \scm's component. Although we provide evidence supporting these components for the proposed scenarios, it may not hold for every new setting. For instance, the confounders and code syntax clustering introduced in our study might differ for functional programming languages. Building a sound \scm would entail automatically identifying potential confounders and formulating assumptions from a rigorous causal discovery process not covered in our pipeline.}

In summary, our interpretability method \docode is based on the idea of outlining causal queries, given a \scm defined from the domain knowledge. These causal queries are obtained by estimating an interventional distribution, where the potential outcome is generally a prediction performance value of the Neural Code Model under study. Furthermore, the interventions are a set of software-based properties that help us construct explanations about the generative model. \docode has been implemented in an open-source library, which is available in our repository ~\cite{icodegen}.

\vspace{-0.5em}
\section{Conclusion}\label{sec:conclusion}
\vspace{-0.5em}
We presented \docode, an interpretability method for understanding \nlms for code. \docode combines rigorous statistical instruments and causal inference theory to give rise to contextualized SE explanations of \nlms using Structural Causal Models (SCMs). SCMs provide a more robust and interpretable framework for modeling complex systems such as Neural Code Models in software engineering. SCMs can help to better understand the underlying mechanisms and causal relationships between different variables, which can improve the accuracy and generalizability of deep learning models. In addition, SCMs can provide a more transparent and explainable approach to Deep Learning for Software Engineering, allowing for a better understanding of the decision-making process of the model and facilitating more effective detection of \textit{confounding bias} and error analysis. Additionally, we also carried out a case study evaluation using \docode on popular \nlms, namely, RNNs, GRUs, and Transformers, to interpret code predictions. \docode exposes erroneous syntax-based categories (\eg conditionals, or loops) by examining Average Treatment Effects and refutation methods. We hope our research opens the field for posterior empirical analysis in interpretability to empower researchers with statistical and causal inference methods.

\ifCLASSOPTIONcompsoc
\else
\fi

\section*{Acknowledgments}
  This research has been supported in part by the NSF CCF-2311469, CNS-2132281, CCF-2007246, CCF-1955853, CCF-2311468, and CCF-2132285. We also acknowledge support from Cisco Systems. Any opinions, findings, and conclusions expressed herein are the authors’ and do not necessarily reflect those of the sponsors.

\ifCLASSOPTIONcaptionsoff
  \newpage
\fi

\bibliographystyle{IEEEtran}
\bibliography{IEEEabrv,main}

\begin{IEEEbiography}[{\includegraphics[width=1in,height=1.25in,clip,keepaspectratio]{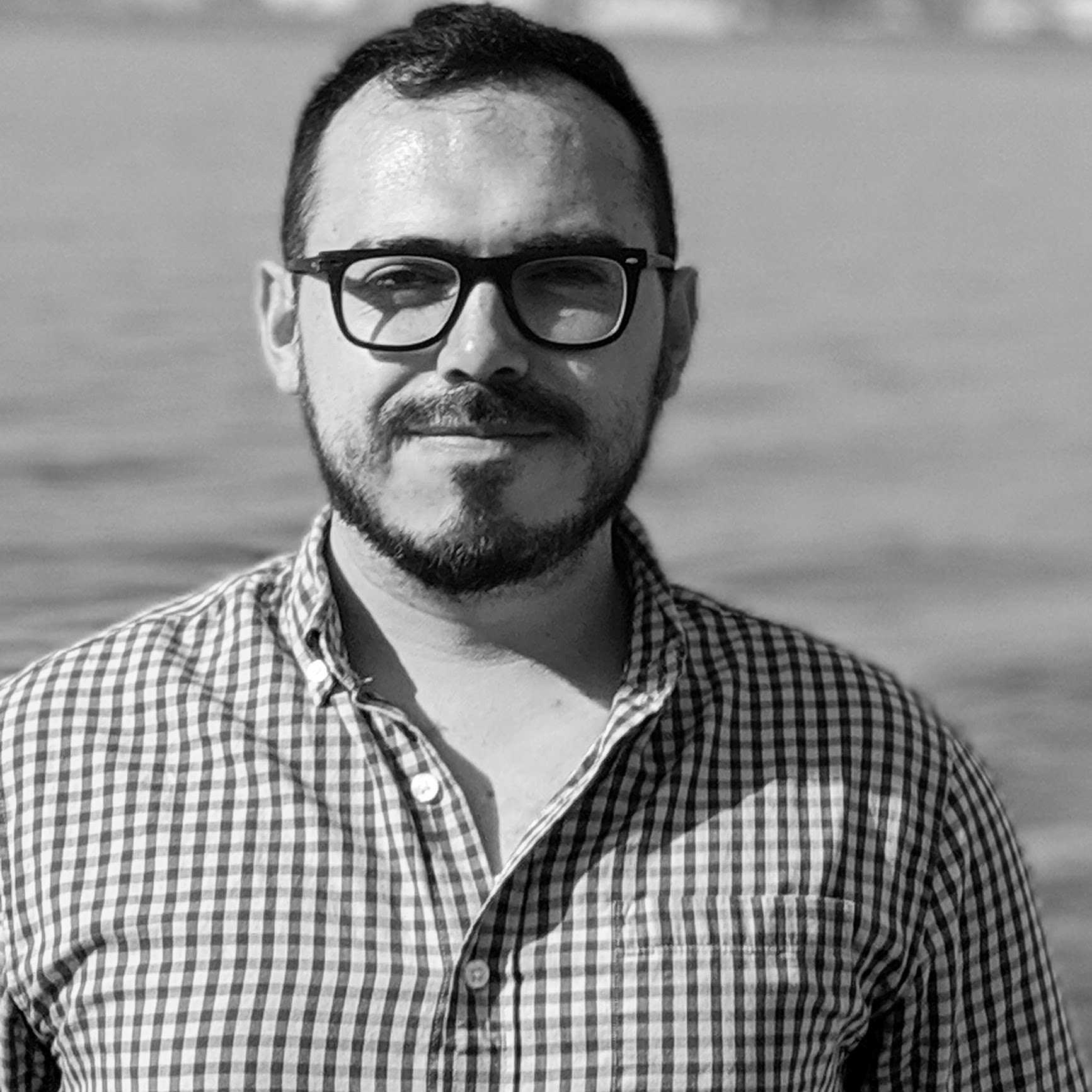}}]{David N. Palacio} is a Ph.D. Candidate in Computer Science at William \& Mary, where he is a member of the SEMERU Research Group supervised by Dr. Denys Poshyvanyk. He received his MSc. in Computer Engineering at Universidad Nacional de Colombia (UNAL), Colombia, 2017. His research is concentrated on interpretable methods for deep learning code generators, specifically, on using causal inference to explain deep software models. His fields of interest lie in complexity science, neuroevolution, causal inference, and interpretable machine learning for the study and automation of software engineer processes. More information is available at \url{https://danaderp.github.io/danaderp/}.
\end{IEEEbiography}

\begin{IEEEbiography}[{\includegraphics[width=1in,height=1.25in,clip,keepaspectratio]{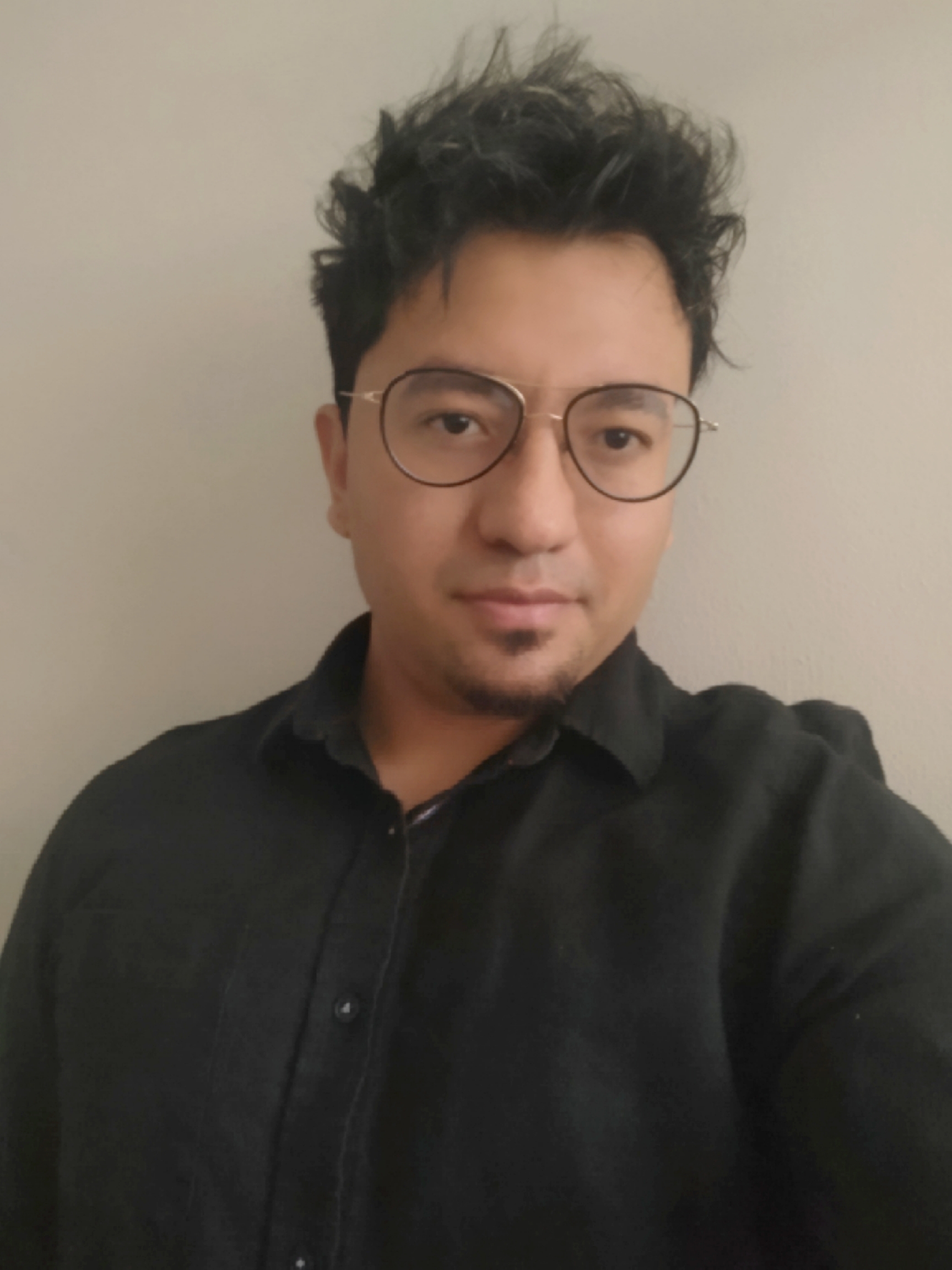}}]{Alejandro Velasco} is a Ph.D. student in Computer Science at William \& Mary, currently a member of SEMERU Research Group under the mentorship of Dr. Denys Poshyvanyk. He obtained his B.Sc and MSc. in Computer Engineering at Universidad Nacional de Colombia (UNAL). His research interests include software engineering, deep learning, interpretable machine learning, and software maintenance and evolution.  More information is available at  \url{https://svelascodimate.github.io}
\end{IEEEbiography}

\begin{IEEEbiography}[{\includegraphics[width=1in,height=1.25in,clip,keepaspectratio]{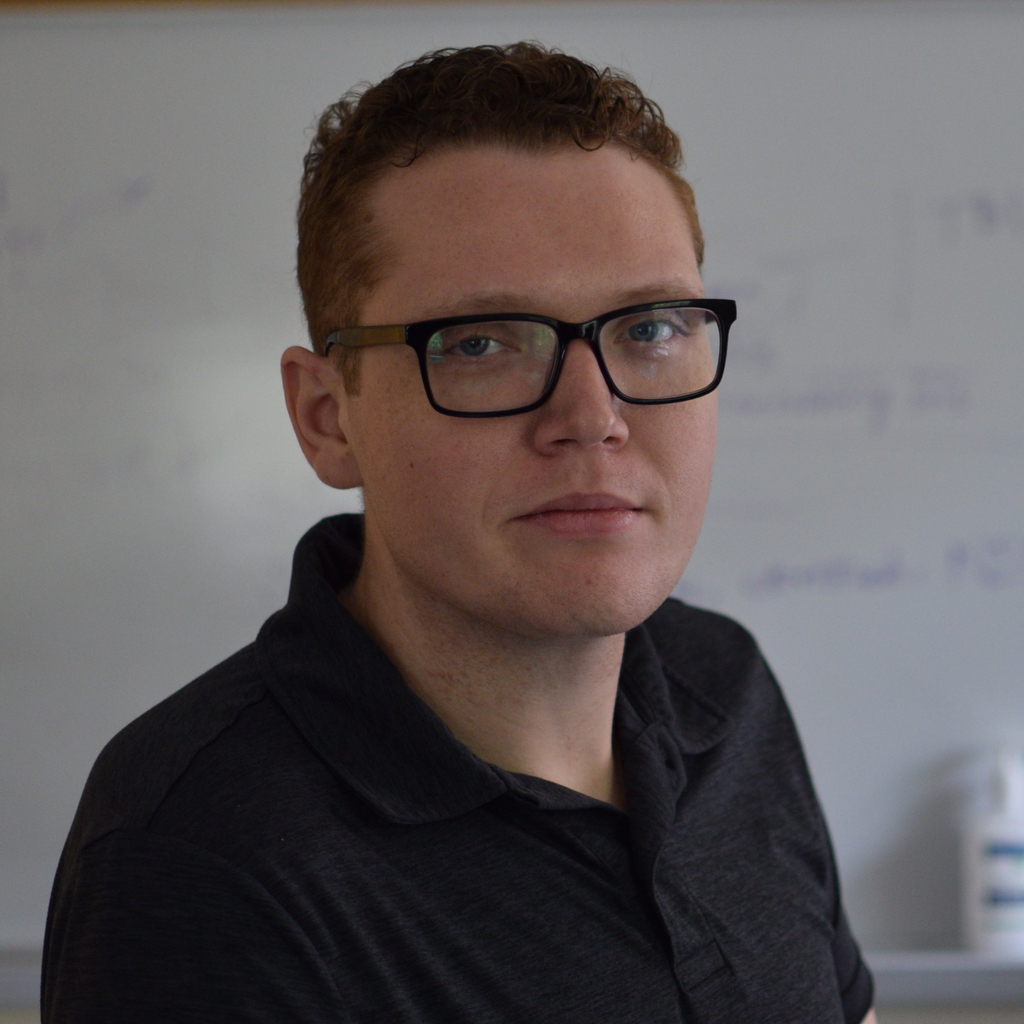}}]{Nathan Cooper} is a nerd and received a B.S. degree in Software Engineering from the University of West Florida in 2018. He is currently a Ph.D. candidate in Computer Science at William \& Mary under the mentorship of Dr. Denys Poshyvanyk and is a member of the Semeru Research group. He has research interests in Software Engineering, Machine / Deep Learning applications for Software Engineering, information retrieval, and question \& answering applications for Software Engineering. He has published in the top peer-reviewed Software Engineering venues ICSE and MSR. He has also received the ACM SIGSOFT Distinguished paper award at ICSE'20. More information is available at \url{https://nathancooper.io/#/}.
\end{IEEEbiography}

\begin{IEEEbiography}[{\includegraphics[width=1in,height=1.25in,clip,keepaspectratio]{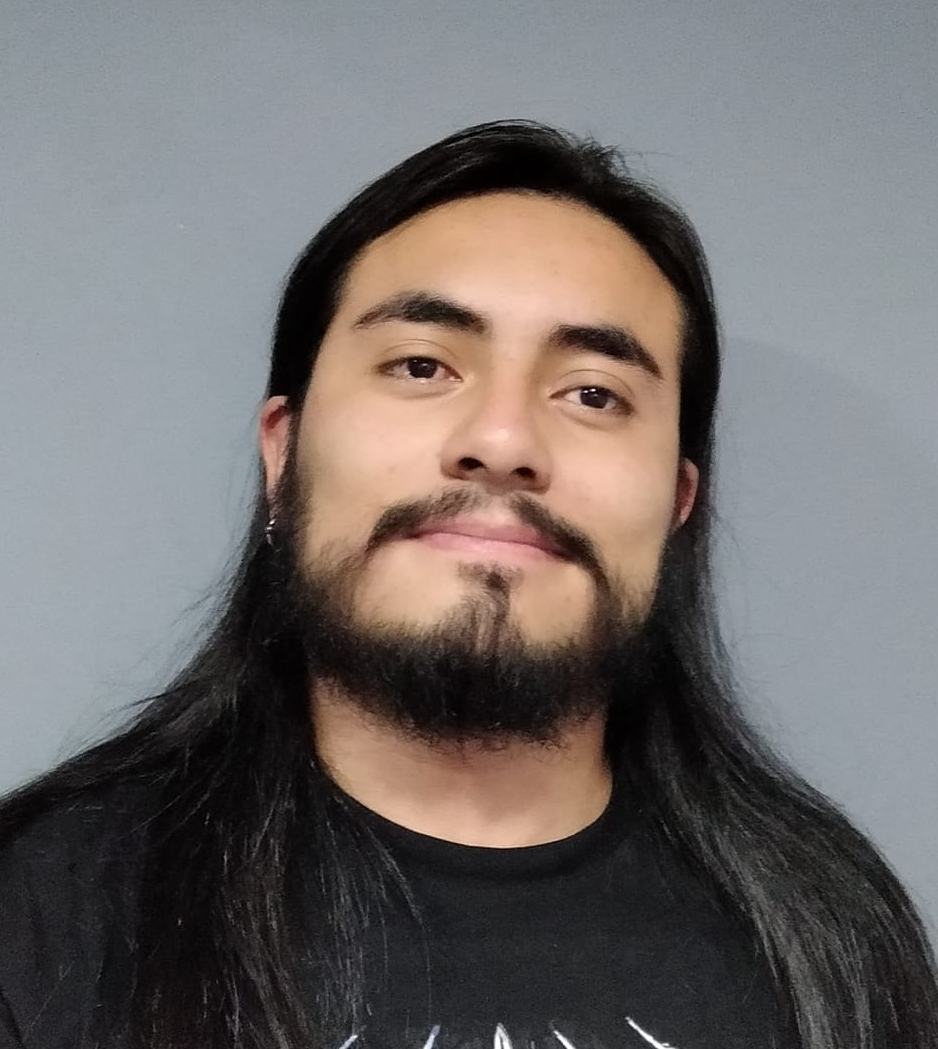}}]{Alvaro Rodriguez} is a MSc. Candidate in Computer Engineering at Universidad Nacional de Colombia (UNAL). He received his B.S degree in Computer Engineering in 2018.  His research interests include Artificial Intelligence for Software Engineering and Evolutionary Computation.
\end{IEEEbiography}

\begin{IEEEbiography}[{\includegraphics[width=1in,height=1.25in,clip,keepaspectratio]{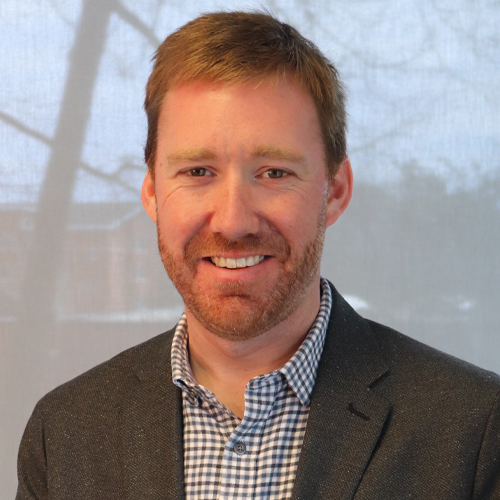}}]{Kevin Moran} Kevin Moran is an Assistant Professor of Computer Science and a member of the Cybersecurity \& Privacy (CyberSP) Cluster at UCF. He directs the SAGE research group. He graduated with a B.A. in Physics from the College of the Holy Cross in 2013, an M.S. and Ph.D. from William \& Mary in 2015 and 2018 respectively. His main research interest involves facilitating the processes of software engineering, security, and maintenance by building developer tools enhanced by machine learning. He has published over 30 papers at various software engineering and security conferences, and his research has been recognized with ACM SIGSOFT distinguished paper awards at ESEC/FSE 2019 and ICSE 2020, and a Best Paper Award at CODASPY’19. He was also recently recognized with the 2023 MOBILESoft Rising Star Award. More information is available at \url{https://www.kpmoran.com/}.
\end{IEEEbiography}

\begin{IEEEbiography}[{\includegraphics[width=1in,height=1.25in,clip,keepaspectratio]{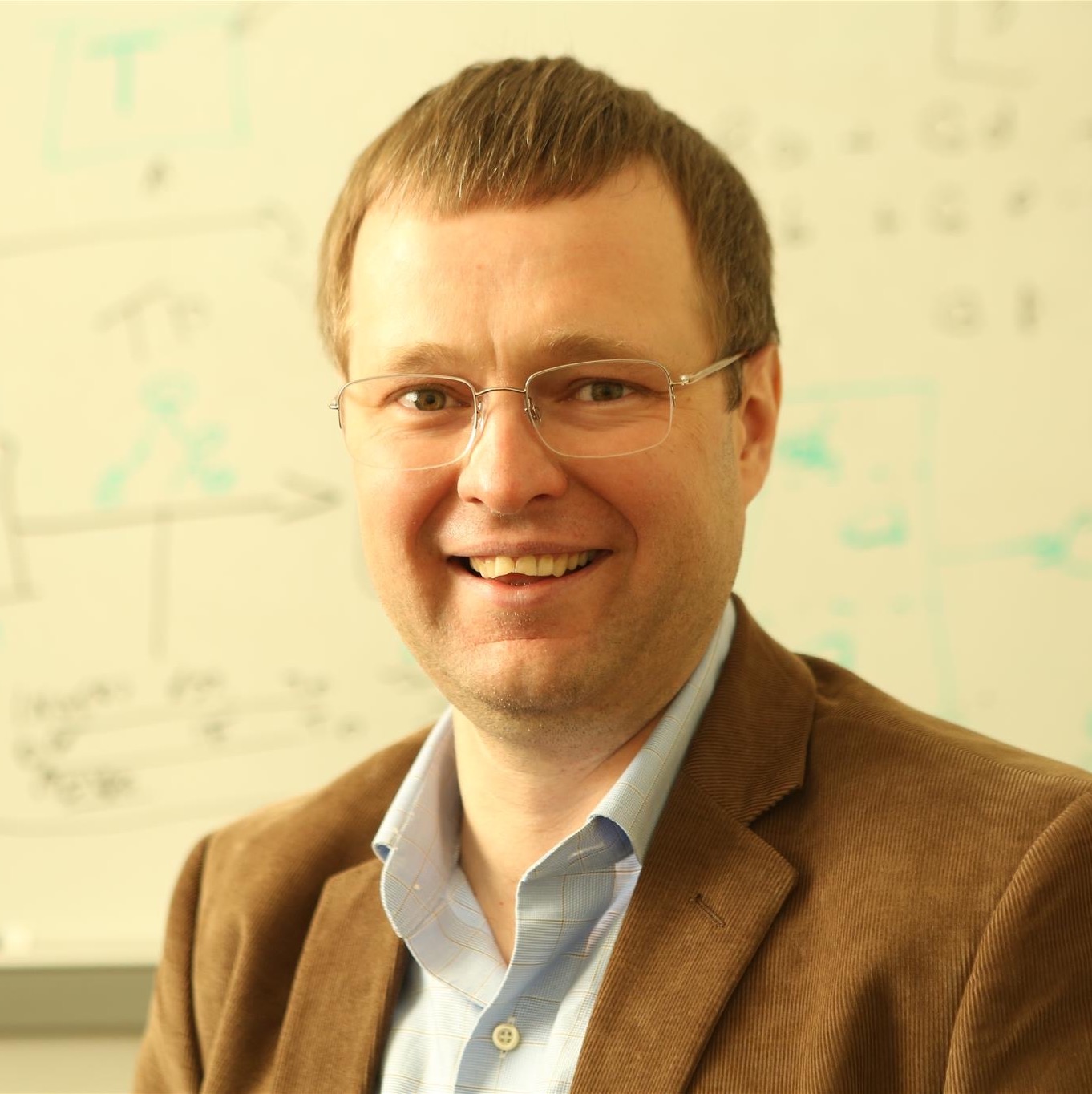}}]{Denys Poshyvanyk} is a Chancellor Professor of Computer Science at William \& Mary. He received the MS and MA degrees in Computer Science from the National University of Kyiv-Mohyla Academy, Ukraine, and Wayne State University in 2003 and 2006, respectively. He received the PhD degree in Computer Science from Wayne State University in 2008. He has served as a program co-chair for FSE'25, ASE'21, MobileSoft'19, ICSME'16, ICPC'13, WCRE'12 and WCRE'11. He currently serves as a Guest Editor-in-Chief of the AI-SE Continuous Special Section at the ACM Transactions on Software Engineering and Methodology. He served on the editorial board of IEEE Transactions on Software Engineering and ACM Transactions on Software Engineering and Methodology, Empirical Software Engineering Journal (Springer), Journal of Software: Evolution and Process (Wiley) and Science of Computer Programming. His research interests include software engineering, software maintenance and evolution, program comprehension, reverse engineering and software repository mining. His research papers received several Best Paper Awards at ICPC'06, ICPC'07, ICSM'10, SCAM'10, ICSM'13, CODAPSY'19 and ACM SIGSOFT Distinguished Paper Awards at ASE'13, ICSE'15, ESEC/FSE'15, ICPC'16, ASE'17, ESEC/FSE'19 and ICSE'20. He also received the Most Influential Paper Awards at ICSME'16, ICPC'17, ICPC'20 and ICSME'21. He is a recipient of the NSF CAREER award (2013).  He is an IEEE Fellow and an ACM distinguished member. More information is available at: \url{http://www.cs.wm.edu/~denys/}.
\end{IEEEbiography}

\end{document}